\documentclass[final, 3p, times, 10pt]{elsarticle}

\usepackage{amsmath}
\usepackage{amssymb}
\usepackage{color}
\DeclareMathAlphabet{\altmathcal}{OMS}{cmsy}{m}{n}
\usepackage{mathptmx}
\usepackage{color}
\usepackage{multirow}
\usepackage{aas_macros}

\journal{Journal of Computational Physics}

\begin{document}
\newcommand{\av}[1]{\left<{#1}\right>}
\newcommand{\DS}{\displaystyle}
\newcommand{\HALF}{\frac{1}{2}}
\newcommand{\hvec}[1]{\hat{\mathbf{#1}}}
\newcommand{\pd}[2]{\frac{\partial #1}{\partial #2}}
\newcommand{\quotes}[1]{``#1''}
\newcommand{\tens}[1]{\mathsf{#1}}
\renewcommand{\vec}[1]{\mathbf{#1}}

\newcommand{\cE}{\altmathcal{E}}
\newcommand{\cF}{\altmathcal{F}}
\newcommand{\cR}{\altmathcal{R}}
\newcommand{\cU}{\altmathcal{U}}
\newcommand{\cV}{\altmathcal{V}}

\newcommand{\cc}{ {\boldsymbol{c}} }
\newcommand{\cf}{ f }

\newcommand{\xf}{ {\mathbf{x}_f} }
\newcommand{\yf}{ {\mathbf{y}_f} }
\newcommand{\zf}{ {\mathbf{z}_f} }

\newcommand{\xe}{ {\mathbf{x}_e} }
\newcommand{\ye}{ {\mathbf{y}_e} }
\newcommand{\ze}{ {\mathbf{z}_e} }

\newcommand{\RED}{\color{red}}
\newcommand{\BLACK}{\color{black}}
\newcommand{\BLUE}{\color{blue}}
\newcommand{\GREEN}{\color{green}}

\begin{frontmatter}



\title{Systematic construction of upwind constrained transport schemes for MHD}


%\author[address1,address2]{A. Mignone, L. Del Zanna}
\author[address1]{A. Mignone}
\author[address2,address3]{L. Del Zanna}

\address[address1]
{Dipartimento di Fisica, Universit\`a di Torino, via Pietro Giuria 1, 10125 Torino, Italy}

\address[address2]
{Dipartimento di Fisica e Astronomia -- Universit\`a di Firenze e INFN -- Sez.
di Firenze, via G. Sansone 1, I-50019 Sesto F.no, Italy}

\address[address3]
{INAF, Osservatorio Astrofisico di Arcetri, Largo E. Fermi 5, I-50125 Firenze,
Italy }

%\author[A. Mignone]{A. Mignone$^{1}$\thanks{E-mail: {mignone@to.infn.it}}, L. Del
%Zanna$^{2,3}$
% \\
%$^1$ Dipartimento di Fisica, Universit\`a di Torino, via P. Giuria 1, I-10125
%Torino, Italy \\
%$^2$ Dipartimento di Fisica e Astronomia -- Universit\`a di Firenze e INFN -- Sez.
%di Firenze, 
%via G. Sansone 1, I-50019 Sesto F.no, Italy \\
%$^3$ INAF, Osservatorio Astrofisico di Arcetri, Largo E. Fermi 5, I-50125 Firenze,
%Italy }
%

\begin{abstract}
The constrained transport (CT) method reflects the state of the art numerical technique for preserving the divergence-free condition of magnetic field to machine accuracy in multi-dimensional MHD simulations performed with Godunov-type, or upwind, conservative codes.
The evolution of the different magnetic field components, located at zone interfaces using a staggered representation, is achieved by calculating the electric field components at cell edges, in a way that has to be consistent with the Riemann solver used for the update of cell-centered fluid quantities at interfaces.
Albeit several approaches have been undertaken, the purpose of this work is, on the one hand, to compare existing methods in terms of robustness and accuracy and, on the other, to extend the \emph{upwind contrained transport} (UCT) method by Londrillo \& Del Zanna (2004) and Del Zanna et al. (2007) for the systematic construction of new averaging schemes.
In particular, we propose a general formula for the upwind fluxes of the induction equation which simply involves the information available from the base Riemann solver employed for the fluid part, provided it does not require full spectral decomposition, and 1D reconstructions of velocity and magnetic field components from nearby intercell faces to cell edges.
Our results are presented here in the context of second-order schemes for classical MHD, but they can be easily generalized to higher than second order schemes, either  based on finite volumes or finite differences, and to other physical systems retaining the same structure of the equations, such as that of relativistic or general relativistic MHD.
\end{abstract}


\end{frontmatter}

% \linenumbers

%%%%%%%%%%%%%%%%%%%%%%%%%%%%%%%%%%%%%%%%%%%%%%%%%%%%%%%%%%%%%%%%%%%%%%%%
\section{Introduction}
%
%
%
%%%%%%%%%%%%%%%%%%%%%%%%%%%%%%%%%%%%%%%%%%%%%%%%%%%%%%%%%%%%%%%%%%%%%%%%

Magnetohydrodynamics (MHD) is the basic modelization framework to treat plasmas at the macroscopic level, that is neglecting kinetic effects and as a single fluid, an approximation commonly used for applications to laboratory, space and astrophysical plasmas.
Magnetic fields and currents created by the moving charges play a fundamental role in the dynamics of the fluid, which is considered to be locally neutral, and the set of hydrodynamical (Euler) equations must be supplemented by the magnetic contributions to the global energy and momentum, and by a specific prescription for the evolution of the magnetic field itself, the so-called induction equation, that is Faraday's law combined to a constitutive relation between the current and the electric field (Ohm's law). 

In extending the same numerical techniques employed for the Euler equations to multi-dimensional MHD, a major challenge dwells in preserving the divergence-free constraint of the magnetic field which inherently follows from the curl-type character of Faraday's law.
This is especially true for Godunov-type shock-capturing schemes based on the properties of the hyperbolic set of conservation laws, let us call them \emph{standard upwind procedures}, where spatial partial derivatives do not commute when discontinuities are present and spurious effects (\emph{numerical magnetic monopoles}) arise. 

The research in this field, which is crucial for building accurate and robust finite-volume (FV) or finite-difference (FD) shock-capturing numerical codes for computational MHD, started more than twenty years ago and several different methods have been proposed.
Here, for sake of conciseness, we simply refer the reader to the paper by T{\'o}th (2000) \cite{Toth2000} for a comprehensive discussion and comparison of the early schemes.
Summarizing in brief, among the proposed solution methods are schemes based on the cleaning of the numerical monopoles, either by solving an elliptic (Poisson) equation and removing the monopole contribution from the updated fields \cite{Brackbill_Barnes1980}, or by adding specific source terms and an additional evolution equation for the divergence of $\vec{B}$ itself to preserve the hyperbolic character of the MHD set \cite{Powell_etal1999, Munz_etal2000, Dedner_etal2002}.
Other methods evolve in time the vector potential \cite{Rossmanith2006, Helzel_etal2011} or use different alternatives, such as for the so-called flux distribution schemes \cite{Torrilhon2005}.

Conversely, a radically different strategy is adopted in the so-called constrained transport (CT) methods.
These schemes all rely on the curl-type nature of the induction equation and on its discretization based on Stokes' theorem (rather than on Gauss' one as needed for the equations with the divergence operator), as first realized in pioneering works where the evolution equation for the magnetic field alone was solved \cite{Yee1966, Brecht_etal1981, DeVore1991, Evans_Hawley1988}.
The CT method was later extended to the full MHD system of hyperbolic equations in the context of Godunov-type schemes \cite{Dai_Woodward1998,Ryu_etal1998,Balsara_Spicer1999}.
In FV-CT schemes, magnetic field components are stored as surface integrals at cell interfaces as primary variables to be evolved via the induction equation, while the corresponding fluxes are line-averaged electric field (namely \emph{electromotive force}, emf) components located at cell edges, to recover the discretized version of Stokes' theorem.
By doing so, the solenoidal constraint can be preserved exactly during time evolution.

A major difficulty of the CT formalism is the computation of upwind-stable emf components located at zone-edges \cite{Balsara_Spicer1999}.
This can be achieved either by properly averaging the interface fluxes computed when solving the 1D Riemann problems at zone interfaces, or by using genuine (but much more complex) 2D Riemann solvers computed directly at cell edges \cite{Balsara2010, Balsara2012, Balsara2014a, Balsara_Dumbser2015}.
In the latter case the dissipative part of the multidimensional emf can be shown to behave as a proper resistive term for the induction equation \cite{Balsara_Nkonga2017}.
As far as the former (simpler) case is concerned, in the original work by \cite{Balsara_Spicer1999} the emf was obtained as the arithmetic of  the four upwind fluxes nearest to the zone edge.
It was then recognized (e.g. \cite{Gardiner_Stone2005}) that this approach has insufficient numerical dissipation and it does not reduce to the plane-parallel algorithm for grid-align flow.
Gardiner \& Stone (2005) \cite{Gardiner_Stone2005} suggested that this issue could be solved by doubling the dissipation and introduced a recipe to construct a stable and non-oscillatory upwind emf with optimal numerical dissipation based on the direction of the contact mode.
This approach (here referred to as the CT-Contact method) is, however, mainly supported by empirical results as there is no formal justification that the emf derivative should  obey such selection rule.
In addition, the method can be at most $2^{\rm nd}$-order accurate thus making the generalization to higher-order methods not feasible.
%The difficulty in applying rigorously the CT approach is that emf components located at zone edges must result in either an average of 1D fluxes computed when solving the Riemann problems via standard upwind procedures at the neighbouring zone faces, or by using genuine but much more complex 2D Riemann solvers computed directly at cell edges \cite{Balsara2010, Balsara2012, Balsara2014a, Balsara_Dumbser2015, Balsara_Nkonga2017}.

A rigorous approach to this problem for both FV and FD Godunov-type schemes for computational MHD was originally proposed by Londrillo and Del Zanna (2004) \cite{Londrillo_DelZanna2004} with their \emph{upwind constrained transport} (UCT) method.
According to the UCT methodology, the continuity property of the magnetic field components at cell interfaces (which follows from the solenoidal constraint) is considered as a \emph{built-in} condition in a numerical scheme, enabling face-centered fields to be evolved as primary variables.
At the same time, staggered magnetic field components enter as single-state variables in the fluid fluxes at the corresponding cell interfaces, and as two-state reconstructed values at cell edges in the four-state emf for the induction equation.
Time-splitting techniques should be avoided as they prevent exact cancellation of $\nabla\cdot\vec{B}$ terms at the numerical level.
The emf components constructed using information from the four neighboring upwind states must also automatically reduce to the correct 1D numerical fluxes for plane parallel flows and discontinuities aligned with the grid directions.
According to the authors, these are the necessary conditions to preserve the divergence-free condition and to avoid the occurrence of numerical monopoles that may arise while computing the divergence of fluid-like fluxes numerically.
In the original work of \cite{Londrillo_DelZanna2004}, a second-order FV scheme based on Roe-type MHD solver (UCT-Roe) and a high-order scheme based on characteristic-free reconstruction and a two-wave HLL approximate Riemann solver (UCT-HLL) were proposed. 

The UCT-HLL scheme was further simplified by Del Zanna et al. (2007) \cite{DelZanna_etal2007} in the context of general relativistic MHD, 
%where the four-state electric fields are computed by simple averaging of 1D reconstructions along towards the edges of staggered magnetic fields and transverse velocity components.
and recipes were given to build a FD UCT schemes of arbitrary order of accuracy by testing several reconstruction methods.
%Notice that the proposed scheme was actually designed for general relativistic MHD (GRMHD), where the induction equation is a simple extension of that of classical MHD, and similar UCT methods are nowadays employed by the majority of GRMHD codes \cite{Porth2019}, employed for example to simulate accretion of plasma onto black holes, as done to compare synthetic data against the famous first image of a black hole obtained by the Event Horizon Telescope \cite{EHT2019}. 
High-order FD-CT methods were also recently proposed by \cite{Minoshima_etal2019} who, instead, constructed the emf by simply doubling the amount of numerical dissipation.
While FD approaches are based on a point value representation of primary variables and avoid multi-dimensional reconstructions, FV-CT schemes of higher than $2^{\rm nd}$-order accuracy are much more arduous to construct albeit they are likely to increase robustness, see the review by Balsara \cite{Balsara_Review2017} and references therein.
Efforts in this direction were taken by Balsara (2009) \cite{Balsara2009, Balsara_etal2009} in the context of ADER-WENO FV schemes who designed genuinely third- and fourth-order spatially accurate numerical methods for MHD.
More recently, fourth-order FV schemes using high-order quadrature rules evaluated on cell interfaces have been proposed by Felker \& Stone (2018) \cite{Felker_Stone2018} and Verma (2019) \cite{Verma_etal2019}.
Here the construction of the higher-order emf follows the general guidelines of the UCT-HLL (or Lax-Friedrichs) approach introduced by \cite{DelZanna_etal2007} and later resumed for truly 2D Riemann problem by \cite{Balsara2010}.

%Another mentioned difficulty of FV CT schemes for MHD is the extension to high-order schemes, so that a better alternative is to resort to a finite-difference (FD) approach based on point value representation of primary variables, where multi-dimensional reconstruction can be avoided and replaced by field-by-field standard 1D reconstruction, as achieved for the first time (twenty years ago!) by \cite{Londrillo_DelZanna2000}.

The goal of the present work is to systematically construct UCT schemes for classical MHD, using a variety of 1D Riemann solvers avoiding the full spectral resolution in characteristic waves, and providing the correct averaging recipes to build the four-state emf fluxes at zone edges, extending the scheme by \cite{DelZanna_etal2007} to less dissipative solvers like HLLD \cite{Miyoshi_Kusano2005}, where the Riemann fan is split into four intermediate state to include also the contact and Alfv\'enic contributions, other than simply the fast magnetosonic ones.
This novel UCT scheme is tested by performing several multi-dimensional numerical tests, and comparison is also made with other CT popular schemes based on emf averaging, including the simple arithmetic averaging method, those based on doubling the diffusive contribution of 1D fluxes, as in \cite{Gardiner_Stone2005}, the above mentioned UCT-HLL one, and a novel UCT version of the GFORCE scheme by \cite{Toro_Titarev2006}.
%All schemes are compared by either using second-order Runge-Kutta time-stepping and linear limiting in reconstruction, or third-order Runge-Kutta combined with MP5 reconstruction.

Our paper is structured as follows.
In \S\ref{sec:notations} we introduce the basic CT discretization and general notations while in \S\ref{sec:emf_averaging} we review basic averaging CT schemes.
The UCT method and the original Roe and HLL schemes are discussed in \S\ref{UCT}, while the new UCT-based composition formulae are presented in \S\ref{sec:composition}. Numerical benchmarks are introduced in \S\ref{sec:numerical_benchmarks} and a final summary is reported in \S\ref{sec:summary}.

%%%%%%%%%%%%%%%%%%%%%%%%%%%%%%%%%%%%%%%%%%%%%%%%%%%%%%%%%%%%%%%%%%%%%%%%
\section{Notations and general formalism for CT schemes}
\label{sec:notations}
%
%
%
%%%%%%%%%%%%%%%%%%%%%%%%%%%%%%%%%%%%%%%%%%%%%%%%%%%%%%%%%%%%%%%%%%%%%%%%

The ideal MHD equations are characterized by two coupled sub-systems, one for the time evolution of the set of \emph{conservative} hydro-like flow variables (mass, momentum and energy densities denoted, respectively, with $\rho$, $\rho\vec{v}$ and $\cE$):
\begin{equation}
  \pd{U}{t} + \nabla\cdot\tens{F} = 0,
  \qquad{\rm where}\qquad
   U = \left(\begin{array}{l}
     \rho          \\ \noalign{\medskip}
     \rho \vec{v}  \\ \noalign{\medskip}
     \cE      
  \end{array}\right) \,,\quad
  \tens{F} = \left(\begin{array}{c}
     \rho \vec{v}                                         \\ \noalign{\medskip}
     \rho \vec{v}\vec{v} - \vec{B}\vec{B} + \tens{I}p_t   \\ \noalign{\medskip}
     (\cE + p_t)\vec{v} - (\vec{v}\cdot\vec{B})\vec{B}
  \end{array}\right)^\intercal \,,\quad
\end{equation}
and an induction equation for the evolution of the magnetic field
\begin{equation}\label{eq:induction}
  \frac{\partial\vec{B}}{\partial t} + \nabla\times\vec{E}=0,
\end{equation}
where the curl operator appears instead of a divergence and where the electric field $\vec{E} = - \vec{v} \times \vec{B}$ is not a primary variable, depending on the flow velocity and the magnetic field, but its components have to be considered as the fluxes for the magnetic field itself.
Here $\vec{v} = (v_x,\, v_y,\, v_z)$ is the fluid velocity vector, $p_t = p+B^2/2$ is the total (thermal + magnetic) pressure, while the total energy density adds up kinetic, thermal and magnetic contributions:
\begin{equation}
 \cE = \frac{1}{2}\rho\vec{v}^2 + \frac{p}{\Gamma-1} + \frac{\vec{B}^2}{2} \,,
\end{equation}
$\Gamma$ is the specific heat ratio for an adiabatic equation of state.

Due to the commutativity of analytical spatial derivatives, the induction equation (\ref{eq:induction}) implicitly contains the solenoidal condition for the magnetic field
\begin{equation}
\nabla\cdot\vec{B} = 0,
\end{equation}
that if true for $t=0$ must be preserved during the subsequent time evolution.
The peculiar structure of the MHD system and the existence of the above non-evolutionary constraint makes it difficult to extend straightforwardly the methods developed for the Euler equations to MHD, especially for upwind schemes where Riemann solvers have to be modified in order to adapt to the curl operator and where numerical derivatives do not commute and spurious magnetic monopoles could arise.

%trim={<left> <lower> <right> <upper>}
\begin{figure}
  \centering
  \includegraphics[trim={90 10 50 10}, width=0.5\textwidth]{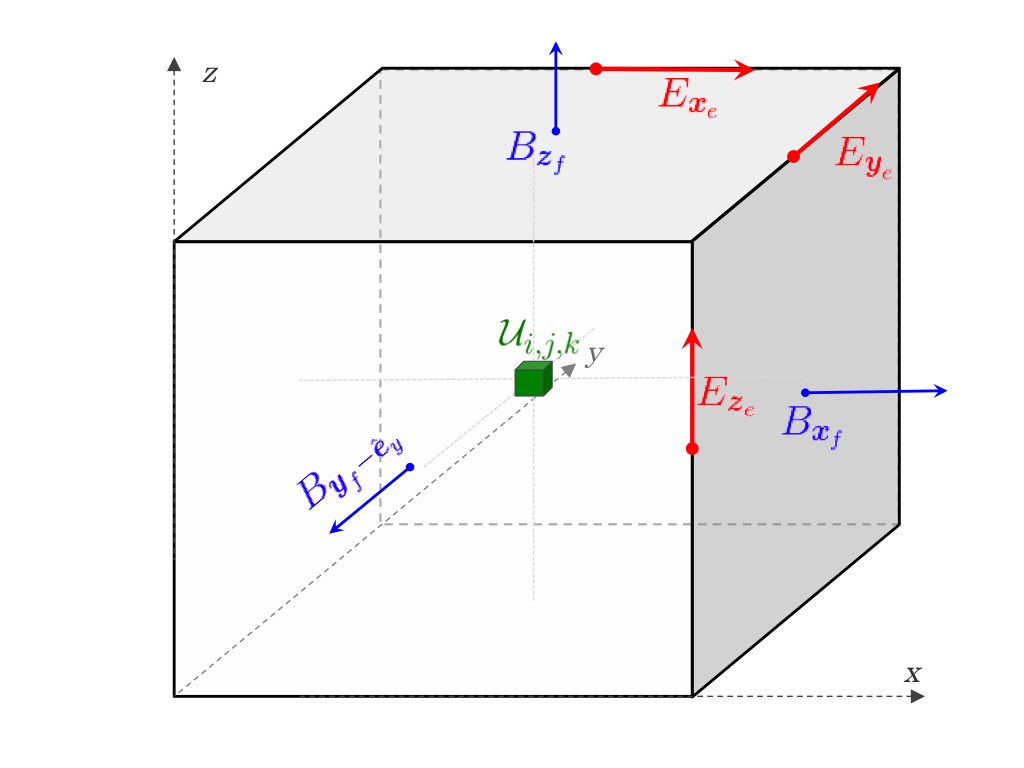}%
  \includegraphics[width=0.55\textwidth]{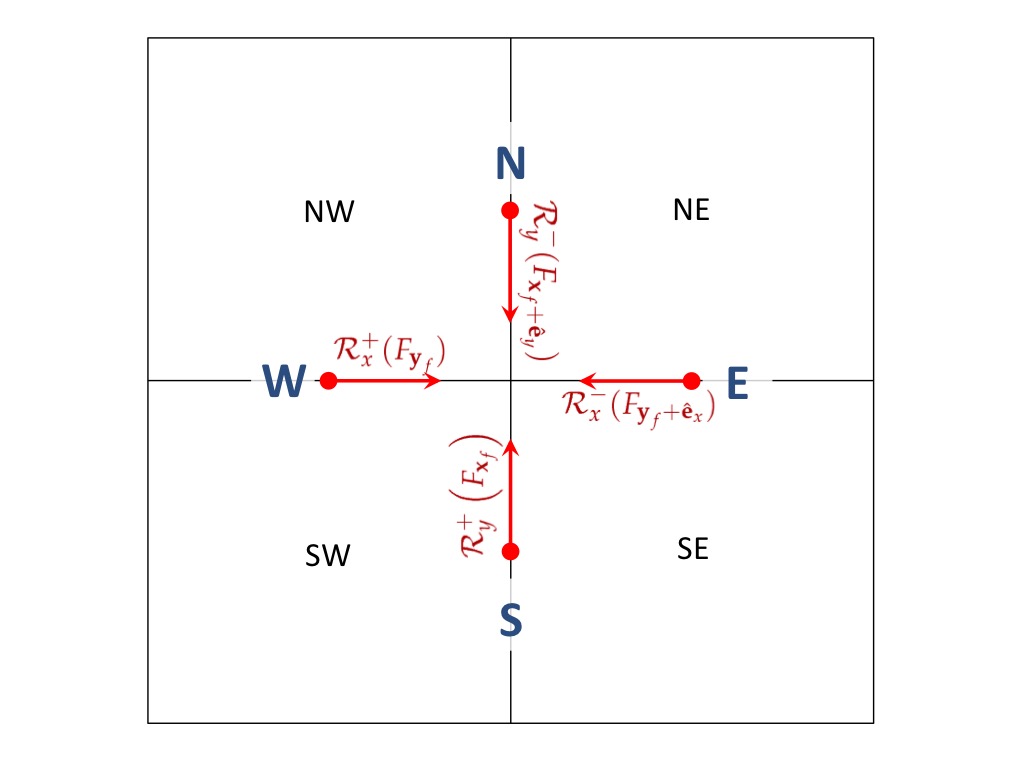}
  \caption{Right: positioning of MHD variables in the CT formalism.
   Staggered magnetic field components (blue) are face-centered,
   the electromotive force is edge-centered  (red) while remaining hydrodynamical
   quantities are located at the zone center (green).
   Right: Top view of the intersection between four neighbor zones:
    N, S, E and W indicate the four cardinal directions with respect to
    the zone edge (here represented by the intersection between four
    neighbor zones), $\cR_x(F_{\yf})$ and $\cR_y(F_{\xf})$ are
    1-D reconstruction operators
    applied to each zone face while $F$ denotes a generic flux component
    computed during the $x-$ ($F\equiv -F^{[B_y]}_{\xf}$) or
    $y-$ sweep ($F\equiv F^{[B_x]}_{\yf}$), see Eq. (\ref{eq:edge_E}).}
  \label{fig:ct}
\end{figure}

Here we adopt a Cartesian coordinate system, with unit vectors $\hvec{e}_x=(1,0,0)$, $\hvec{e}_y=(0,1,0)$ and $\hvec{e}_z=(0,0,1)$, uniformly discretized into a regular mesh with coordinate spacing $\Delta x$, $\Delta y$ and $\Delta z$.
Computational zones (or cells) are centered at $(x_i,\, y_j,\, z_k)$ and delimited by the six interfaces orthogonal to the coordinate axis aligned, respectively, with $(x_{i\pm\HALF},\, y_j,\, z_k)$, $(x_i,\, y_{j\pm\HALF},\, z_k)$ and $(x_i,\, y_j,\, z_{k\pm\HALF})$.
CT-based schemes for MHD are characterized by a hybrid collocation for primary variables, those to be evolved.
While flow variables are zone-centered, here labeled as $U_{\cc}$ where the $\cc$ subscript is a shorthand notation for $(i,j,k)$, magnetic fields have a staggered representation and are located at zone interfaces.
Numerical fluxes for flow variables are also collocated in this points, where Riemann solvers will be computed, whereas magnetic fluxes (the electric field components) are computed at zone edges, as shown in Fig. \ref{fig:ct}.
To simplify the notations, from now on these staggered electromagnetic quantities will be indicated as 
\begin{equation}
  \vec{B}_{\cf} \equiv \left( \begin{array}{l}
              B_{\xf}  \\ \noalign{\medskip}
              B_{\yf}  \\ \noalign{\medskip}
              B_{\zf} \end{array}\right)
    = \left( \begin{array}{l}
              B_{x,i+\HALF,j,k}  \\ \noalign{\medskip}
              B_{y,i,j+\HALF,k}  \\ \noalign{\medskip}
              B_{z,i,j,k+\HALF} \end{array}\right) \,, \qquad
     \vec{E}_{e} \equiv \left( \begin{array}{l}
              E_{\xe}  \\ \noalign{\medskip}
              E_{\ye}  \\ \noalign{\medskip}
              E_{\ze} \end{array}\right)
    = \left( \begin{array}{l}
              E_{x,i,j+\HALF,k+\HALF}  \\ \noalign{\medskip}
              E_{y,i+\HALF,j,k+\HALF}  \\ \noalign{\medskip}
              E_{z,i+\HALF,j+\HALF,k} \end{array}\right) \,,
\end{equation}
where the subscripts $\xf$, $\yf$ and $\zf$ identify the spatial component as well as the face-centered staggered location inside the control volume, i.e., $\xf \equiv \{x,(i+\HALF, j, k)\}$, $\yf \equiv \{y,(i, j+\HALF, k)\}$, and $\zf \equiv \{z,(i, j, k+\HALF)\}$.
Likewise, the components and corresponding positions of the different edge-centered electric field components are labeled as $\xe \equiv \{x,(i, j+\HALF, k+\HALF)\}$, $\ye \equiv \{y,(i+\HALF, j, k+\HALF)\}$, and $\ze \equiv \{z,(i+\HALF, j+\HALF, k)\}$.
This subscript notation extends also to arrays and scalar quantities in general by discarding the spatial component (e.g. \quotes{$x$} or \quotes{$y$}) when unnecessary, e.g., $U_\xf = U_{i+\HALF, j, k}$.
This should not generate confusion as its employment will be clear from the context.
%Note that, to ease up the notations, the same subscript is also be used to label the spatial component when referred to a vector, e.g., $B_\yf  = B_{y,i,j+\HALF,k}$ or $F_\xf= F_{x,i+\HALF,k}$.
We will also make frequent use of the backward difference operators $\Delta_x$, $\Delta_y$, and $\Delta_z$, defined as
\begin{equation}\label{eq:deltaOp}
    \Delta_x Q_\cc \equiv  Q_\cc - Q_{\cc-\hvec{e}_x} \,,\quad
    \Delta_y Q_\cc \equiv  Q_\cc - Q_{\cc-\hvec{e}_y} \,,\quad
    \Delta_z Q_\cc \equiv  Q_\cc - Q_{\cc-\hvec{e}_z} \,,\quad
\end{equation}
where $Q$ can be any quantity, here with cell-centered representation $Q_c$. These $\Delta$ operators can be equivalently applied to face-centered $Q_f$ or edge-centered $Q_e$ values.

In the context of a finite-volume (FV) approach, conserved variables are evolved in terms of their volume (or zone-) averages $U_{\cc}$, implying a surface-averaged representation of the fluxes at zone interface, as required by direct application of Gauss' theorem:
\begin{equation}
  \begin{array}{l}
  \DS \hat{F}_{\xf} = \frac{1}{\Delta y\Delta z}
      \int \hvec{e}_x\cdot\tens{F}\big(U(x_{i+\HALF},y,z,t)\big)  \,dy\,dz \,,
  \\ \noalign{\medskip}
  \DS \hat{F}_{\yf} = \frac{1}{\Delta z\Delta x}
      \int \hvec{e}_y\cdot\tens{F}\big(U(x,y_{j+\HALF},z,t)\big)  \,dz\,dx \,,
  \\ \noalign{\medskip}
  \DS \hat{F}_{\zf} = \frac{1}{\Delta x\Delta y}
      \int \hvec{e}_z\cdot\tens{F}\big(U(x,y,z_{k+\HALF},t)\big)  \,dx\,dy \,.
  \end{array}
\end{equation}
Conversely, magnetic field components $\vec{B}_{\cf}$, having a staggered representation, are also interpreted as face-averages and are updated using a discrete version of Stokes' theorem.
The line-averaged electric field $\vec{E}_{e}$ effectively behave as electromotive force (emf), so the numerical fluxes for the magnetic field are commonly referred to emf components in the literature:
\begin{equation}
  \begin{array}{l}
  \DS \hat{E}_{\xe} = \frac{1}{\Delta x}
      \int E_x(x,y_{j+\HALF},z_{k+\HALF},t) \,dx\,,
  \\ \noalign{\medskip}
  \DS \hat{E}_{\ye} = \frac{1}{\Delta y}
      \int E_y(x_{i+\HALF},y,z_{k+\HALF},t) \,dy\,,
  \\ \noalign{\medskip}
  \DS \hat{E}_{\ze} = \frac{1}{\Delta z}
      \int E_z(x_{i+\HALF},y_{j+\HALF},z,t) \,dz\,.
  \end{array}
\end{equation}

The semi-discrete FV version of any CT numerical scheme is then the following:
\begin{equation}
 \begin{array}{lcl}
  \DS \frac{dU_\cc}{dt} & = & \DS - 
  \left(  \frac{ \Delta_x \hat{F}_{\xf} }{ \Delta x }
        + \frac{ \Delta_y \hat{F}_{\yf} }{ \Delta y }
        + \frac{ \Delta_z \hat{F}_{\zf} }{ \Delta z }\right) \,,
 \\ \noalign{\medskip}         
  \DS \frac{dB_{\xf}}{dt} & = & \DS - 
       \left(  \frac{\Delta_{y} \hat{E}_{\ze}}{\Delta y}
              -\frac{\Delta_{z} \hat{E}_{\ye}}{\Delta z}
       \right)  \,,
 \\ \noalign{\medskip}         
  \DS \frac{dB_{\yf}}{dt} & = & \DS - 
       \left(  \frac{\Delta_{z} \hat{E}_{\xe}}{\Delta z}
              -\frac{\Delta_{x} \hat{E}_{\ze}}{\Delta x}
       \right)  \,,
\\ \noalign{\medskip}         
  \DS \frac{dB_{\zf}}{dt} & = & \DS - 
       \left(  \frac{\Delta_{x} \hat{E}_{\ye}}{\Delta x}
              -\frac{\Delta_{y} \hat{E}_{\xe}}{\Delta y}  
       \right)\,,
\end{array}
\end{equation}
Notice that no approximation has been made so far.
The condition
\begin{equation}
 \frac{d}{dt}  \left( 
           \frac{ \Delta_x B_{\xf} }{ \Delta x }
         + \frac{ \Delta_y B_{\yf} }{ \Delta y }
         + \frac{ \Delta_z B_{\zf} }{ \Delta z }\right) = 0
\end{equation}
is thus valid exactly, and at any time the discrete version of the solenoidal constraint is ensured \emph{to machine accuracy} (if so for at the initial condition).

To second-order accuracy, a midpoint quadrature rule is typically used to evaluate, e.g., $\hat{F}_{\xf}$ with its point value  obtained by means of a 1D Riemann solver at cell interfaces (the base scheme).
For higher than $2^{\rm nd}$ order schemes, $\hat{F}_{\xf}$ is obtained by suitable quadrature rules, see .e.g \cite{Corquodale_Colella2011, Felker_Stone2018, Verma_etal2019}.
Another option is to use high-order finite-difference (FD) schemes for which primary variables are stored as point-values in the same positions imposed by the CT method, and where multi-dimensional averaging is not needed \cite{Londrillo_DelZanna2000, DelZanna_etal2007}.
However, for sake of clarity and simplicity, in the following we limit our analysis to second-order schemes, so that FV and FD schemes basically coincide and the averaging operations are simply omitted in our notations.

%%%%%%%%%%%%%%%%%%%%%%%%%%%%%%%%%%%%%%%%%%%%%%%%%%%%%%%%%%%%%%%%%%%%%%%%
\subsection{Approximate Riemann solvers for the base scheme}
%
%%%%%%%%%%%%%%%%%%%%%%%%%%%%%%%%%%%%%%%%%%%%%%%%%%%%%%%%%%%%%%%%%%%%%%%%

CT schemes for MHD must be coupled to the Godunov-type method to solve the hyperbolic sub-system of Euler-like partial differential equations for $U_{\cc}$ (the base scheme).
The inter-cell fluxes $\hat{F}_{\xf}$, $\hat{F}_{\yf}$ and $\hat{F}_{\zf}$ are evaluated by solving a Riemann problem between left and right states reconstructed from the zone average to the desired quadrature point.
For the midpoint rule, left and right states can be obtained using one-dimensional upwind reconstruction techniques.
For instance, at an $x$-interface,
\begin{equation}
  U^L_\xf = \cR^+_x\left(U_{\cc}\right),\quad
  U^R_\xf = \cR^-_x\left(U_{\cc + \hvec{e}_x}\right)\, ,\quad
\end{equation}
where $\cR^\pm_x()$ is an operator in the $x$ direction giving the reconstructed value at the right ($+$) or left ($-$) interface with respect to the cell center, with the desired order of accuracy and possessing monotonicity properties.
Left and right states in the other directions are obtained similarly.
Reconstruction is best carried on primitive or characteristic variables as it is known to produce less oscillatory results.

After the reconstruction phase, one needs to solve the Riemann problem, a procedure that in modern shock-capturing schemes for MHD is hardly ever achieved using exact nonlinear solvers.
Approximate Riemann solvers provide inter-cell fluxes generally written as the sum of a centered flux terms and a dissipative term
\begin{equation}\label{eq:centered+dissipative}
  \hat{F}_{\xf} = F_{\xf} - \Phi_{\xf} \,,
\end{equation}
where $F_{\xf}$ is the centered flux term while $\Phi_{\xf}$ is the (stabilizing) dissipative term.
Consider, for instance, the Roe Riemann linear solver, based on the decomposition of variables into characteristics.
In this case the two terms are 
\begin{equation}\label{eq:RoeFlux}
  F_{\xf} = \frac{1}{2}\left(F_{\xf}^L + F_{\xf}^R\right)
  \,,\quad
  \Phi_{\xf} = \frac{1}{2}\tens{R}|\tens{\Lambda}|\tens{L}
                    \cdot\left(U^R_\xf - U^L_\xf\right) \,,
\end{equation}
where $F^{L,R}_{\xf} = \hvec{e}_x\cdot\tens{F}(U_\xf^{L,R})$ are the left and right fluxes, $\tens{R}=\tens{R}(\overline{U}_{\rm roe})$ and $\tens{L}=\tens{L}(\overline{U}_{\rm roe})$ are the right and left eigenvector matrices defined in terms of the Roe average state $\overline{U}_{\rm roe}$ (note that $\tens{R}\tens{L}=\tens{L}\tens{R}=\tens{I}$) while $|\tens{\Lambda}| = {\rm diag}(|\lambda_1|, ...,|\lambda_k|)$ is a diagonal matrix containing the eigenvalues in absolute value.

A different averaging procedure is obtained in the case of HLL schemes \cite{HLL1983}, where the inter-cell numerical flux is expressed through a convex combination of left and right fluxes plus a diffusion term 
\begin{equation}\label{eq:HLLFlux}
 \hat{F}_{\xf} =   \frac{\alpha^R_x F^L_\xf + \alpha^L_x F^R_\xf}
                         {\alpha^R_x + \alpha^L_x}
                  - \frac{\alpha^R_x\alpha^L_x(U^R_\xf - U^L_\xf)}
                         {\alpha^R_x + \alpha^L_x} \,,
\end{equation}
where $\alpha^R_x = \max (0, \lambda^R_\xf) \ge 0$ and $\alpha^L_x = -\min(0,\lambda^L_\xf)\ge 0$, with $\lambda^R_\xf$ the rightmost (largest) characteristic speed and $\lambda^L_\xf$ the leftmost (smallest with sign, not in terms of its absolute value) characteristic speed.
The HLL flux can be derived from an integral relation \cite{Toro2009} and approximates the inter-cell flux with a single intermediate state when $\lambda^R_\xf\geq 0$ and $\lambda^L_\xf\leq 0$, while retaining pure upwind properties when the two speeds have the same sign, hence $\hat{F}_{\xf} \equiv F^L_\xf$ or $ \hat{F}_{\xf} \equiv F^R_\xf$.
The HLL flux is known to be quite dissipative compared to the Roe one, especially for second order schemes, but the simplicity of component-wise resolution, possibly combined to higher-order reconstruction, has received considerable attention since \cite{Londrillo_DelZanna2000}.

The simplest solver that does not require the full characteristic decomposition is the Rusanov one, or local Lax-Friedrichs, which simply replaces the eigenvalue matrix with $|\tens{\Lambda}| = \tens{I}|\lambda_{\max}|$, where $\lambda_{\max}$ is the maximum local spectral radius.
In this case $\Phi_{\xf} = |\lambda_{\max}|(U^R_\xf - U^L_\xf)/2$.
Notice that the Rusanov flux can also be derived as a particular case of the HLL one, when $-\lambda^L_\xf =\lambda^R_\xf = |\lambda_{\max}|$, hence $\alpha_x^L = \alpha_x^R = |\lambda_{\max}|$.
Other alternatives to the Roe solver, avoiding the full spectral decomposition but still relying on the knowledge of the characteristic speeds, are the HLLC \cite{Toro2009, Gurski2004, Li2005} and HLLD \cite{Miyoshi_Kusano2005} solvers, for a better resolution with respect to HLL of contact and Alfv\'enic jumps, respectively.
These solvers will be discussed in better details in the next sections. 

It is worth noticing that also multidimensional Riemann solvers for the induction equation allow the numerical flux to be decomposed into centered and dissipative terms, as shown in the work by Balsara, see \citep{Balsara_Kappeli2017, Balsara_Nkonga2017}.
The parabolic terms arising from the dissipative terms are equivalent to a physical conductivity which makes the discretization of the induction equations numerically stable.
In this context, an attractive approach to the Riemann problem is provided by the so-called HLLI solver, accounting for multiple intermediate characteristic waves \cite{Dumbser_Balsara2016,Balsara_Nkonga2017}.

%%%%%%%%%%%%%%%%%%%%%%%%%%%%%%%%%%%%%%%%%%%%%%%%%%%%%%%%%%%%%%%%%%%%%%%%
\section{CT schemes based on emf averaging}
\label{sec:emf_averaging}
%
%%%%%%%%%%%%%%%%%%%%%%%%%%%%%%%%%%%%%%%%%%%%%%%%%%%%%%%%%%%%%%%%%%%%%%%%

The set of conserved variables $U_{\cc}$ may be extended to include the zone-centered representation of magnetic field components as well, usually provided by simple spatial average from the two neighboring faces along the relevant direction at the beginning of any timestep.
The solution to the full Riemann problem (8 equations and variables for 3D MHD) thus provides point-value upwind fluxes for the zone-centered magnetic field as well.
Indeed, indicating with a square bracket the flux component we make the formal correspondences
\begin{equation}\label{eq:edge_flux}
  \begin{array}{lll}
    F_x^{[B_x]} =  0     \,,\;  &
    F_y^{[B_x]} =  E_z   \,,\;  &
    F_z^{[B_x]} = -E_y   \,,
  \\ \noalign{\medskip}
    F_x^{[B_y]} = -E_z  \,,\;  &
    F_y^{[B_y]} = 0     \,,\;  &
    F_z^{[B_y]} = E_x   \,,
  \\ \noalign{\medskip}
    F_x^{[B_z]} =  E_y   \,,\;  &
    F_y^{[B_z]} = -E_x   \,,\;  &
    F_z^{[B_z]} = 0 \,,
  \end{array}
\end{equation}
where $F_x = \hvec{e}_x\cdot\tens{F}$ and so forth.
This formal analogy holds for the upwind fluxes $\hat{F}$ as well.
The electromotive force at cell edges can thus be obtained by taking advantage of the upwind information already at disposal during the 1D Riemann solver, thus avoiding more complex 2D Riemann problems (for a detailed discussion see the review by Balsara \cite{Balsara_Review2017} and also \cite{Balsara2009} for extensions to higher orders).
To ease up the notations, we consider a top-view of a cell edge representing the intersection of four zones (Fig. \ref{fig:ct}b) and label the left and right states along the $y$-coordinate as south ($S$) and north ($N$).
Similarly, left and right states in the $x$-direction are labeled with west ($W$) and east ($E$) with respect to the intersection point.
We then define the $S$ and $N$ states reconstructed along direction $y$ (see the vertical arrows in the cited figure) and the $W$ and $E$ states reconstructed along direction $x$ (the horizontal arrows) as
\begin{equation}\label{eq:edge_E}
  E_z^S = \cR_y^+\left(-F^{[B_y]}_{\xf}\right) ,\quad
  E_z^N = \cR_y^-\left(-F^{[B_y]}_{\xf + \hvec{e}_y}\right),\quad
  E_z^W = \cR_x^+\left( F^{[B_x]}_{\yf}\right) ,\quad
  E_z^E = \cR_x^-\left( F^{[B_x]}_{\yf + \hvec{e}_x}\right) \,,
\end{equation}
and likewise, for the dissipative part of numerical fluxes we let
\begin{equation}\label{eq:edge_diffE}
  \phi_x^S = \cR_y^+\left(\Phi^{[B_y]}_{\xf}\right),\quad
  \phi_x^N = \cR_y^-\left(\Phi^{[B_y]}_{\xf+\hvec{e}_y}\right),\quad
  \phi_y^W = \cR_x^+\left(\Phi^{[B_x]}_{\yf}\right) \,,\quad
  \phi_y^E = \cR_x^-\left(\Phi^{[B_x]}_{\yf+\hvec{e}_x}\right)\,.
\end{equation}
When not directly available, the diffusion terms at cell faces can generically be obtained as the difference between the centered contribution and the numerical flux, that is
\begin{equation}
  \Phi^{[B_y]}_{\xf} = F^{[B_y]}_{\xf} - \hat{F}^{[B_y]}_{\xf}\,,\quad
  \Phi^{[B_x]}_{\yf} = F^{[B_x]}_{\yf} - \hat{F}^{[B_x]}_{\yf}\,.
\end{equation}
The generalization to the 3D case is easily obtained by cyclic permutations. 
Different emf averaging procedures have been proposed for CT schemes, outlined in what follows.

%%%%%%%%%%%%%%%%%%%%%%%%%%%%%%%%%%%%%%%%%%%%%%%%%%%%%%%%%%%%%%%%%%%%%%%%
\subsection{Arithmetic averaging}
%%%%%%%%%%%%%%%%%%%%%%%%%%%%%%%%%%%%%%%%%%%%%%%%%%%%%%%%%%%%%%%%%%%%%%%%

Arithmetic averaging, initially proposed by \cite{Balsara_Spicer1999}, is probably the simplest CT scheme and can be trivially obtained by taking the arithmetic average of the upwind fluxes obtained at the four nearest emf sharing the same zone edge:
\begin{equation}\label{eq:Arithmetic}
  \hat{E}_{\ze} = \frac{1}{4}\left(- \hat{F}^{[B_y]}_{\xf}
                                   - \hat{F}^{[B_y]}_{\xf+\hvec{e}_y}
                                   + \hat{F}^{[B_x]}_{\yf}
                                   + \hat{F}^{[B_x]}_{\yf+\hvec{e}_x}\right)
   \,.
\end{equation}
Despite its simplicity, as pointed out by several authors \cite[see, e.g.][]{Londrillo_DelZanna2000, Gardiner_Stone2005}, this approximation suffers from insufficient dissipation and thus spurious numerical oscillations in several tests, yielding, for 1D plane-parallel flows along the grid Cartesian axes, half of the correct value.

%%%%%%%%%%%%%%%%%%%%%%%%%%%%%%%%%%%%%%%%%%%%%%%%%%%%%%%%%%%%%%%%%%%%%%%%
\subsection{The CT-Contact scheme}
%
%%%%%%%%%%%%%%%%%%%%%%%%%%%%%%%%%%%%%%%%%%%%%%%%%%%%%%%%%%%%%%%%%%%%%%%%

Gardiner \& Stone \cite{Gardiner_Stone2005} suggested that the CT algorithm could be cast as a spatial integration procedure.
The reconstruction can be operated from any one of the four nearest face centers to the zone edge.
Choosing the arithmetic average leads, in our notations, to the following expression for the zone-centered emf:
\begin{equation}\label{eq:UCT_CONTACT}
  \hat{E}_{\ze} = \hat{E}_\ze^{\rm arithm}
              +\frac{\Delta y}{8}\left[
                     \left(\pd{E_z}{y}\right)^S
                   - \left(\pd{E_z}{y}\right)^N \right]
                +\frac{\Delta x}{8}\left[
                  \left(\pd{E_z}{x}\right)^W
                - \left(\pd{E_z}{x}\right)^E \right] \,,
\end{equation}
where $\hat{E}_\ze^{\rm arithm}$ is the arithmetic average, Eq. (\ref{eq:Arithmetic}).
In the original work by \cite{Gardiner_Stone2005}, various ways for obtaining the derivatives are discussed, although an optimal expression based on the sign of the fluid velocity is suggested.
This gives an upwind-selection rule which is essentially based on the speed of the contact mode leading to stable and non-oscillatory results, yielding
\begin{equation}\label{eq:ct_contact_dEy}
  \left(\pd{E_z}{y}\right)^S =
    \frac{1 + s_\xf}{2}\left(\frac{\hat{F}^{[B_x]}_{\yf} - F^{[B_x]}_{y,\cc}}{\Delta y/2}\right)
  + \frac{1 - s_\xf}{2}\left(\frac{\hat{F}^{[B_x]}_{\yf+\hvec{e}_x}
                                    - F^{[B_x]}_{y,\cc+\hvec{e}_x}}{\Delta y/2}\right) \,,
\end{equation}
where $s_\xf = {\rm sign}(\hat{F}^{[\rho]}_\xf)$ while $F^{[B_y]}_{y,c} = (\hvec{e}_y\cdot\tens{F}(U_\cc))^{[B_y]}$ is the $B_y$ component of the flux evaluated at the cell center.
Similarly:
\begin{equation}\label{eq:ct_contact_dEx}
  \left(\pd{E_z}{x}\right)^W =
    \frac{1 + s_\yf}{2}\left(-\frac{\hat{F}^{[B_y]}_{\xf} - F^{[B_y]}_{x,\cc}}
                                   {\Delta x/2}\right)
  + \frac{1 - s_\yf}{2}\left(-\frac{\hat{F}^{[B_y]}_{\xf+\hvec{e}_y}
                                        - F^{[B_y]}_{x,\cc+\hvec{e}_y}}{\Delta x/2}\right) \,.
\end{equation}

Similar expressions can be obtained at the North (N) and East (E) edges.
Although there is no formal justification that the electric field derivative should  obey such selection rule, this algorithm has been found to yield robust and stable results in practice.
The scheme correctly reduces to the base upwind method for grid-aligned planar flows and, as also pointed out by \cite{Felker_Stone2018}, it is at most of second-order spatially accurate given the way the derivatives in Eq. (\ref{eq:ct_contact_dEy})-(\ref{eq:ct_contact_dEx}) are computed.
We rename here this method as the \emph{CT-Contact} scheme.

%An improvement was proposed by \cite{Gardiner_Stone2005}, where the previous expression for the emf was doubled, in order to double the dissipation (and half of the averaged centered part removed in order to recover the right composition rule). The formula for the flux is
%\begin{equation}\label{eq:UCT_CONTACT}
%  E_{\ze} = \frac{1}{2}\left(- \cF^{[B_y]}_{\xf} - \cF^{[B_y]}_{\xf+\hvec{e}_y}
%                             + \cF^{[B_x]}_{\yf} + \cF^{[B_x]}_{\yf+\hvec{e}_x}
%                      \right) -  \frac{1}{4}\left( {E_z}_c + {E_z}_{c + \hvec{e}_x} + 
%                      {E_z}_{c + \hvec{e}_y}  + {E_z}_{c + \hvec{e}_x + \hvec{e}_y}   \right) \,,
%\end{equation}
%and in the paper an upwinding procedure for the dissipative fluxes based on the contact mode was suggested. 
%%%%%%%%%%%%%%%%%%%%%%%%%%%%%%%%%%%%%%%%%%%%%%%%%%%%%%%%%%%%%%%%%%%%%%%%
\subsection{The CT-Flux scheme}
%
%
%
%%%%%%%%%%%%%%%%%%%%%%%%%%%%%%%%%%%%%%%%%%%%%%%%%%%%%%%%%%%%%%%%%%%%%%%%

From the previous considerations, it appears that a cost-effective and straightforward possibility is to double the weights of the dissipative flux terms.
The edge-centered emf can then be obtained by separately reconstructing the centered flux terms and dissipative contributions from cell faces to the edge, using Eq. (\ref{eq:edge_E}) and (\ref{eq:edge_diffE}).
These contributions are then added with the correct weights so as to reduce to plane-parallel flows for 1D configurations.
This leads to 
\begin{equation}\label{eq:CT_Flux}
  \hat{E}_{\ze} = \frac{1}{4}
            \left(  E_z^N + E_z^S + E_z^E + E_z^W  \right)
    + \frac{1}{2}\left(\phi_x^S + \phi_x^N\right)   
    - \frac{1}{2}\left(\phi_y^W + \phi_y^E\right)    \,,
\end{equation}
where the different $E_z$ are the centered contributions obtained using Eq. (\ref{eq:edge_E}).
This approach, named here \emph{CT-Flux}, has been recently adopted by \cite{Minoshima_etal2019} in designing high-order finite difference scheme.
Note that the staggered magnetic fields are not employed when reconstructing from the face to the corner.

%For conveniency, one can directly interpolate the centered terms plus \emph{twice} the dissipation terms obtained during the same flux component computation, e.g., 
%%
%\begin{equation}\label{eq:UCT_Flux}
%  E_z^S = -\cR^y_+\left(\cF^{[B_y]}_{\xf} + 2\Phi^{[B_y]}_{\xf}\right)
%\end{equation}
%%

%%%%%%%%%%%%%%%%%%%%%%%%%%%%%%%%%%%%%%%%%%%%%%%%%%%%%%%%%%%%%%%%%%%%%%%%
\section{The upwind constrained transport (UCT) method: original framework}
\label{UCT}
%
%%%%%%%%%%%%%%%%%%%%%%%%%%%%%%%%%%%%%%%%%%%%%%%%%%%%%%%%%%%%%%%%%%%%%%%%
The CT discretization scheme outlined so far allows to preserve exactly the solenoidal condition for the magnetic field but this is not enough to avoid the presence of spurious magnetic monopole terms when computing the magnetic forces, if fluxes are calculated by using reconstructed values of the magnetic field components as for the other fluid variables.
As first realized in \cite{Londrillo_DelZanna2000} and systematically demonstrated in \cite{Londrillo_DelZanna2004}, the only way to properly take into account the specific smoothness properties of the divergence-free $\vec{B}$ vector in Godunov-type schemes for MHD is to follow these guidelines:
\begin{enumerate}
  \item
   magnetic field components do not possess a left and right representation at
   the cell interface along the corresponding direction
   (upwind reconstruction is not needed).
   Hence a CT staggering is their appropriate discretization as primary variables,
   the time evolution must be performed for these magnetic field components
   at their staggered locations;

  \item
   only staggered field components must be present in the definition of the fluid
   numerical fluxes in the corresponding inter-cell positions, in order to avoid the
  formation of magnetic monopoles;

  \item
   a four-state Riemann solver for the induction equation is needed,
   as explained below;

  \item
  time integration must avoid time-splitting techniques.
\end{enumerate}

%%%%%%%%%%%%%%%%%%%%%%%%%%%%%%%%%%%%%%%%%%%%%%%%%%%%%%%%%%%%%%%%%%%%%%%%
\subsection{The UCT-Roe scheme}
%
%%%%%%%%%%%%%%%%%%%%%%%%%%%%%%%%%%%%%%%%%%%%%%%%%%%%%%%%%%%%%%%%%%%%%%%%

With the Roe formalism, the solution to the Riemann problem at zone interfaces is obtained from independent 1D matrices describing characteristic modes propagating as planar waves \cite{Londrillo_DelZanna2004}.
The flux components entering the induction equation can be expressed through a linear combination of 1-D upwind fluxes along the intersecting direction.
It is a main feature of the UCT method that this combination follows a proper upwind selection rule, since a same flux component at the same collocation point results to have two independent representations in terms of characteristic wave fans \cite{Londrillo_DelZanna2004}:
\begin{equation}\label{eq:UCT_Roe1}
  \hat{E}_{\ze} = \frac{1}{4}\left(  E_z^{SW} + E_z^{SE}
                                   + E_z^{NE} + E_z^{NW}\right)
            +\frac{1}{2}\left(\phi_x^S + \phi_x^N\right)
            -\frac{1}{2}\left(\phi_y^W + \phi_y^E\right)  \,,
\end{equation}
where the different $E_z$ at the intersection point are obtained by separate reconstruction of the velocity along the $x$- and $y$- directions from the nearest zone center and of the staggered fields in the transverse one from the adjacent interface, e.g., $E^{SW}_z = -v^{SW}_xB^W_y + v^{SW}_yB^S_x$.
The expression above clearly shows that the centered and dissipative terms are represented as a four-state function and two-point average in the orthogonal coordinate.
Alternatively, one may also use
\begin{equation}\label{eq:UCT_Roe2}
  \hat{E}_{\ze}
   =   -\frac{1}{2}\left[(\overline{v}_xB_y)^W + (\overline{v}_xB_y)^E\right]
        +\frac{1}{2}\left[(\overline{v}_yB_x)^S + (\overline{v}_yB_x)^N\right]
        +\frac{1}{2}\left(\phi_x^S + \phi_x^N\right)
        -\frac{1}{2}\left(\phi_y^W + \phi_y^E\right) \,.
\end{equation}
where, e.g., $\overline{v}_x = (v^L_x + v^R_x)_\yf/2$ while $\overline{v}_y = (v^L_y + v^R_y)_\xf/2$.
For multidimensional FV schemes solving the Riemann problem at the cell corners, these terms are already at disposal at desired location.
For Godunov-type schemes relying on face-centered flux computation, the evaluation of the Roe dissipative terms can become a rather consuming task since, in the UCT formalism, both zone-centered hydrodynamical variables \emph{and} staggered magnetic fields needed to be reconstructed towards at a cell edge.
This turns out to be more costly than interpolating just the dissipative terms from the interfaces using Eq. \ref{eq:edge_diffE}).
In the UCT-Roe scheme, the dissipative terms are obtained, e.g.,
\begin{equation}
  \phi^S_x = \frac{1}{2}\sum_\kappa |{\lambda_\kappa}|
                 (L_\kappa\cdot\Delta U)^S R^{S}_\kappa \,.
\end{equation}
The eigenvector matrices should be computed by properly averaging the adjacent L/R reconstructed states whereas jumps of conserved variables are split into a hydrodynamic ($U_h$) and magnetic part as $\Delta U^S = \{U_h^{SE} - U_h^{SW},\, B_y^E - B_y^W\}$.

While the Roe solver preserves all stationary wave families, it is also prone to numerical pathologies.
In this respect, the HLLI Riemann solver of \cite{Dumbser_Balsara2016} (and its multidimensional extension of \cite{Balsara_Nkonga2017}) offers an interesting alternative to overcome these problems.
This will be addressed in a forthcoming paper.
%%%%%%%%%%%%%%%%%%%%%%%%%%%%%%%%%%%%%%%%%%%%%%%%%%%%%%%%%%%%%%%%%%%%%%%%
\subsection{The UCT-HLL scheme}
%
%%%%%%%%%%%%%%%%%%%%%%%%%%%%%%%%%%%%%%%%%%%%%%%%%%%%%%%%%%%%%%%%%%%%%%%%
In the case of component-wise Riemann solvers, only magnetic field and velocity components are required at the cell edge (or a combination of them), thus reducing the amount of transverse reconstructions.
An attractive choice is the UCT-HLL scheme of \cite{Londrillo_DelZanna2004} in which the edge-centered electric field evaluates to
\begin{equation}\label{eq:UCT_HLL1}
  \hat{E}_{\ze} = \frac{  \alpha_x^+\alpha_y^+ E_z^{SW}
                  + \alpha_x^+\alpha_y^- E_z^{NW}
                  + \alpha_x^-\alpha_y^+ E_z^{SE}
                  + \alpha_x^-\alpha_y^- E_z^{SW}}
                 {(\alpha_x^+ + \alpha_x^-)(\alpha_y^+ + \alpha_y^-)}
            + \frac{\alpha_x^+\alpha_x^-}{\alpha_x^+ + \alpha_x^-}
              \left(B_y^E - B_y^W\right)
            - \frac{\alpha_y^+\alpha_y^-}{\alpha_y^+ + \alpha_y^-}
              \left(B_x^N - B_x^S\right) \,,
\end{equation}
where $\alpha_x^+$ is computed through some averaging procedure between the the north and south faces
Here we adopt $\alpha^+_x = \max(0, \lambda^R_{\xf}, \lambda^R_{\xf+\hvec{e}_y})$,
$\alpha^-_x = -\min(0, \lambda^L_{\xf}, \lambda^L_{\xf+\hvec{e}_y})$.
A variant of this scheme proposed by \citep{DelZanna_etal2007}, more economical in terms of storage and computations, employs the transverse velocities being entirely analogous to Eq. (\ref{eq:UCT_HLL1}):
\begin{equation}\label{eq:UCT_HLL2}
  \hat{E}_{\ze} = -\frac{  \alpha_x^+ (\overline{v}_xB_y)^W
                   + \alpha_x^- (\overline{v}_xB_y)^E
                   - \alpha_x^+\alpha_x^-(B_y^E - B_y^W)}
                  {\alpha_x^+ + \alpha_x^-}
            +\frac{  \alpha_y^+ (\overline{v}_yB_x)^S
                   + \alpha_y^- (\overline{v}_yB_x)^N
                   -\alpha_y^+\alpha_y^ - (B_x^N - B_x^S)}
                  {\alpha_y^+ + \alpha_y^-} \,,
\end{equation}
where, e.g., $(\overline{v}_xB_y)^W = \overline{v}_x^W\,B_y^W$ while the upwind transverse velocities at an $x$-interface are first computed as
\begin{equation}\label{eq:vt}
  \overline{v}_{t,\xf} = \frac{  \alpha_x^+ \vec{v}^L_\xf
                               + \alpha_x^- \vec{v}^R_\xf}
                              {\alpha_x^+ + \alpha_x^-}\cdot\hvec{e}_t \,,
\end{equation}
for $t=y,z$ and then properly reconstructed in the transverse directions.

The previous formalism may be further developed and extended to other Riemann solvers as well.
This is discussed in the next section.

%%%%%%%%%%%%%%%%%%%%%%%%%%%%%%%%%%%%%%%%%%%%%%%%%%%%%%%%%%%%%%%%%%%%%%%%
\section{The UCT method: a novel generalized composition formula for Jacobian-free Riemann solvers}
\label{sec:composition}
%
%%%%%%%%%%%%%%%%%%%%%%%%%%%%%%%%%%%%%%%%%%%%%%%%%%%%%%%%%%%%%%%%%%%%%%%%

We now present a general formalism for constructing emf averaging schemes with in-built upwind dissipation properties and using component-wise Riemann solver, that is, not directly employing characteristic information.
Our starting point is the definition of the inter-cell numerical flux function for which we assume the approximate form given by Eq. (\ref{eq:centered+dissipative}).
We shall also assume that the dissipative terms of the induction system can be expressed as linear combinations of the left and right transverse magnetic field components alone and that, by suitable manipulation, the induction fluxes can be arranged as
\begin{equation}\label{eq:RoeForm_Induction}
  \hat{F}^{[B_t]}_{\xf} =   a^L_xF^L_{\xf} + a^R_xF^R_{\xf} - (d^R_xB^R_t - d^L_xB^L_t) \,.
\end{equation}
where $F = v_xB_t - v_tB_x$ is the induction flux, $t=y,z$ labels a transverse component at an $x$-interface, while $a_x^L + a_x^R = 1$.
Analogous expressions are obtained at $y$- or $z$-interfaces.
The precise form of the coefficients $a^s_x$ and $d^s_x$ where $s=L,R$  depends, of course, on the chosen Riemann solver.
Consider, for instance the Rusanov Lax-Friedrichs solver; in this case one has the simple expressions
\begin{equation}\label{eq:LFcoeffs}
  a^L_x = a^R_x = \HALF \,,\qquad
  d^L_x = d^R_x = \frac{|\lambda^{\max}_\xf|}{2} \,,
\end{equation}
where $|\lambda^{\max}_\xf|$ is the largest characteristic speed (in absolute value) computed from the L/R states at the interface, e.g., $|\lambda^{\max}_\xf| = \max(|\lambda^L_\xf|, |\lambda^R_\xf|)$.
Likewise, the HLL solver (Eq. \ref{eq:HLLFlux}) can be rewritten in the form given by Eq. (\ref{eq:RoeForm_Induction}) with coefficients
\begin{equation}\label{eq:HLLcoeffs}
  a^L_x = \frac{\alpha_x^R}{\alpha^R_x + \alpha^L_x}
        = \frac{1}{2} + \frac{1}{2}\frac{|\lambda^R_\xf| - |\lambda^L_\xf|}
                                        {\lambda^R_\xf - \lambda^L_\xf}
                                               \,,\quad   
  a^R_x = \frac{\alpha_x^L}{\alpha^R_x +\alpha^L_x}
        = \frac{1}{2} - \frac{1}{2}\frac{|\lambda^R_\xf| - |\lambda^L_\xf|}
                                        {\lambda^R_\xf - \lambda^L_\xf}
                                             \,,\quad
  d^L_x = d^R_x = \frac{\alpha^R_x\alpha^L_x}{\alpha^R_x + \alpha^L_x} \,,
\end{equation}
where $\alpha_x^{L,R}$ are given after Eq. (\ref{eq:HLLFlux}).

With these assumptions, the edge-centered emf with the desired upwind properties can be constructed from (\ref{eq:RoeForm_Induction}) as
\begin{equation}\label{eq:emf2D}
  \hat{E}_\ze = 
           -\left[(a_x\overline{v}_xB_y)^W + (a_x\overline{v}_xB_y)^E\right]
           +\left[(a_y\overline{v}_yB_x)^N + (a_y\overline{v}_yB_x)^S\right]
           + \left[(d_xB_{y})^E  - (d_xB_{y})^W \right]
           - \left[(d_yB_{x})^N  - (d_yB_{x})^S \right] \,,
\end{equation}
where the transverse velocities are reconstructed from the interface values given (unless otherwise stated) by Eq. (\ref{eq:vt}) whereas flux and diffusion coefficients $a_x^{W,E}$ and $d_x^{W,E}$ are computed by combining the corresponding expressions obtained at $x$-interfaces with a 1D Riemann solver, e.g.,
\begin{equation}\label{eq:dEW}
  d^W_x = \frac{d^L_{\xf} + d^L_{\xf+\hvec{e}_y}}{2} \,,\quad
  d^E_x = \frac{d^R_{\xf} + d^R_{\xf+\hvec{e}_y}}{2} \,.\quad
\end{equation}
Similarly, we obtain $d_y^{N,S}$ by averaging the diffusion coefficients at $y$-faces:
\begin{equation}\label{eq:dSN}
  d^S_y = \frac{d^L_{\yf} + d^L_{\yf+\hvec{e}_x}}{2} \,,\quad
  d^N_y = \frac{d^R_{\yf} + d^R_{\yf+\hvec{e}_x}}{2} \,.
\end{equation}
Other forms of averaging for these coefficients - based on the upwind direction or by maximizing the diffusion terms - are of course possible.
However, for the present work, we will employ the simple averaging given by Eq. (\ref{eq:dEW}) and (\ref{eq:dSN}) for both the $a$ and $d$ coefficients.

In the next sections we derive the coefficients also for other Riemann solvers, namely, the HLLC, HLLD and the GFORCE schemes.
From now on, we specialize to an $x$-interface and drop the $\xf$ subscript for ease of notations as it should now be clear from the context.

%%%%%%%%%%%%%%%%%%%%%%%%%%%%%%%%%%%%%%%%%%%%%%%%%%%%%%%%%%%%%%%%%%%%%%%%
\subsection{The UCT-HLLC scheme}
%
%%%%%%%%%%%%%%%%%%%%%%%%%%%%%%%%%%%%%%%%%%%%%%%%%%%%%%%%%%%%%%%%%%%%%%%%
The HLLC solver (see \cite{TSS1994} for the original formulation) describes the Riemann fan in terms of three waves consisting of two outermost fast modes separated by a middle contact wave.
Extensions to ideal MHD have been developed by Gurski \cite{Gurski2004} and Li \cite{Li2005}).
Both formulations, however, fail to satisfy exactly the full set of integral relations across the Riemann fan\footnote{For a detailed mathematical analysis, see section 3.3 of \cite{Mignone_Bodo2006} and the discussion on page 82 in the book by \cite{Mignone_Bodo2008}} and the resulting numerical schemes are prone to instabilities.
A consistent formulation has been presented by Mignone \& Bodo \cite{Mignone_Bodo2008} showing that, for non-zero normal magnetic field ($B_x\ne 0$), the solution must assume continuity of the transverse fields across the middle wave.
The HLLC flux for the induction system can be written in the form (\ref{eq:RoeForm_Induction}) as
\begin{equation}\label{eq:HLLCFlux}
  \hat{F} =  \frac{1}{2}\Big[F^L + F^R 
                   - |\lambda_L|\left(B_t^{*L} - B_t^L\right)
                   - |\lambda^*|\left(B_t^{*R} - B_t^{*L}\right)
                   - |\lambda_R|\left(B_t^R    - B_t^{*R}\right) \Big] \,,
\end{equation}
where $F=v_xB_t - v_tB_x$ while $\lambda^*$ is the speed of the contact mode.
Since $B_t^{*L}=B_t^{*R}$ holds across the middle wave, it is easily verified from Eq. (\ref{eq:HLLCFlux}) that this method produces the same coefficients as the HLL solver (Eq. \ref{eq:HLLcoeffs}) and thus an equivalent amount of numerical diffusion.

However, for $B_x=0$, the HLLC solver admits a jump in the transverse magnetic field across the middle wave.
In particular (see also Sec. 4.2 of \cite{Miyoshi_Kusano2005}) it is easy to show that
\begin{equation}
  B^{*s} - B^s = B^s\chi^s \,,\qquad{\rm where} \qquad
  \chi^s = -\frac{v_x^s - \lambda^*}{\lambda^s - \lambda^*}  \,.
\end{equation}
Working out the explicit expressions leads to the following flux and diffusion coefficients,
\begin{equation} \label{eq:HLLCcoeffs}
  a^L = a^R = \HALF\,,\qquad
  d^s = \left(\frac{|\lambda^*|-|\lambda^s|}{2}\right)\chi^s + \frac{|\lambda^*|}{2}
  \,.
\end{equation}
As we shall see later, the expressions in this degenerate case will be useful to obtain the correct singular limit in the HLLD solver.

%%%%%%%%%%%%%%%%%%%%%%%%%%%%%%%%%%%%%%%%%%%%%%%%%%%%%%%%%%%%%%%%%%%%%%%%
\subsection{The UCT-HLLD scheme}
%
%%%%%%%%%%%%%%%%%%%%%%%%%%%%%%%%%%%%%%%%%%%%%%%%%%%%%%%%%%%%%%%%%%%%%%%%

The HLLD (see \cite{Miyoshi_Kusano2005} and \cite{Mignone2007} for adiabatic and isothermal MHD, respectively) approximates the Riemann fan with a five-wave pattern that includes two outermost fast shocks propagating with speed $\lambda_L$ and $\lambda_R$, two rotational waves $\lambda^{*L}$ and $\lambda^{*R}$ separated, in the adiabatic case, by a contact wave in the middle moving at speed $\lambda^*$.
Across the contact mode, when $B_x\ne0$, the transverse components of magnetic field $B_t \equiv B_y,\, B_z$ are continuous.
For our purposes, we conveniently rewrite the HLLD flux for the induction system in the form (Eq. \ref{eq:RoeForm_Induction}) as
\begin{equation}\label{eq:hlld1}
   \hat{F} = \frac{1}{2}\Big[ F^L + F^R 
   - \left|\lambda_L\right|  \left(B^{*L}_t - B^L_t\right) 
   - \left|\lambda^{*L}\right|\left(B^{**}_t  - B^{*L}_t\right)
   - \left|\lambda^{*R}\right|\left(B^{*R}_t - B^{**}_t\right) 
   - \left|\lambda_R\right| \left(B^R_t     - B^{*R}_t\right)\Big]
\end{equation}
where $F=v_xB_t - v_tB_x$ and we have assumed $B^{**R}_t = B^{**L}_t = B^{**}_t$.
The rotational modes are given by
\begin{equation}\label{eq:hlld_lambda*}
  \lambda^{*L} = \lambda^* - \frac{|B_x|}{\sqrt{\rho^{*L}}} \,,\quad
  \lambda^{*R} = \lambda^* + \frac{|B_x|}{\sqrt{\rho^{*R}}} \,,\quad
\end{equation}
where $\rho^{*s} = \rho^s(\lambda^s - v_x^s)/(\lambda^s - \lambda^*)$ and $\lambda^* = m_x^{\rm hll}/\rho^{\rm hll}$.
Here the suffix \quotes{${\rm hll}$} labels a specific component of the HLL intermediate state, obtained from the integral form of the Riemann fan:
\begin{equation}\label{eq:Uhll}
  U^{\rm hll} = \frac{\lambda^R U^R - \lambda^L U^L + F^L - F^R}
                     {\lambda^R - \lambda^L} \,.
\end{equation}

From Eq. 45 and 47 of Miyoshi \& Kusano \cite{Miyoshi_Kusano2005} together with of the expressions above, we rewrite the jumps of $B_t$ across the outermost fast waves ($s=L,R$) as 
\begin{equation}\label{eq:By_chi}
  B^{*s}_{t} - B^s_t  = B^s_t\chi^s\,,
  \qquad{\rm where}\qquad
  \chi^s = \frac{(v^s_{x}- \lambda^*)(\lambda^s - \lambda^*)}
                {(\lambda^{*s} - \lambda^s)(\lambda^{*s} + \lambda^s - 2\lambda^*)}
                \,.
\end{equation}
%
%The coefficient $\chi^s$ can be readily found from Eq. 45 and 47 of \cite{Miyoshi_Kusano2005} assuming that $\lambda^{*s} = \lambda^* \pm |B_x|/\sqrt{\rho^{*s}}$ (where the minus sign holds for $s=L$ and the plus sign holds for $s=R$) and $\rho^{*s} = \rho^s(\lambda^s - v^s_x)/(\lambda^s - \lambda^*)$.
%
The state in the $**$ region can be identified with the HLL average (Eq. \ref{eq:Uhll}) beyond the Alfv\'en modes:
\begin{equation}\label{eq:**}
  B^{**}_t = \frac{\lambda^{*R}B^{*R}_t - \lambda^{*L}B^{*L}_t
                    + F^{*L} - F^{*R}}
                  {\lambda^{*R} - \lambda^{*L}} \,,
\end{equation}
where $F^{*s}$ can be replaced with the jump conditions across the fast waves, i.e., $F^{*s} = F^s + \lambda^s(B^{*s}_t - B^s_t)$.
The dissipative terms in the HLLD flux (\ref{eq:hlld1}) can now be expressed as a linear combination of $B^s_t$ alone.
After some tedious but otherwise straightforward algebra, one finds that the coefficients needed in Eq. (\ref{eq:RoeForm_Induction}) can be written in the form
%
%\begin{equation}\label{eq:UCT_HLLD_ad}
%  a^L = \frac{1+\nu}{2}\,,\quad
%  a^R = \frac{1-\nu}{2}\,,\quad
%  d^s = \frac{1}{2}\Big[
%                 |\lambda^{*s}|    - |\lambda^s|
%                 - \nu(\lambda^{*s} - \lambda^s)\Big]\chi^s
%                 + \frac{1}{2}|\lambda^{*s}| - \frac{1}{2}\nu\lambda^{*s} 
%\end{equation}
\begin{equation}\label{eq:UCT_HLLD_ad}
  a^L = \frac{1 + \nu^*}{2}\,,\quad
  a^R = \frac{1 - \nu^*}{2}\,,\quad
  d^s =   \frac{1}{2}(\nu^s - \nu^*)\tilde{\chi}^s
        + \frac{1}{2}\left(|\lambda^{*s}| - \nu^*\lambda^{*s}\right)  \,,
\end{equation}
where $\tilde{\chi}^s = (\lambda^{*s} - \lambda^s)\chi^s$, while
\begin{equation}\label{eq:UCT_HLLD_nu}
  \nu^s = \frac{|\lambda^{*s}| - |\lambda^s|}{\lambda^{*s} - \lambda^s}
        = \frac{\lambda^{*s} +\lambda^s}{|\lambda^{*s}| + |\lambda^s|} \,,\qquad
  \nu^* = \frac{|\lambda^{*R}| - |\lambda^{*L}|}{\lambda^{*R} - \lambda^{*L}}
        = \frac{\lambda^{*R} + \lambda^{*L}}{|\lambda^{*R}| + |\lambda^{*L}|}\,.
\end{equation}

Note that the diffusion coefficients defined in Eq. (\ref{eq:UCT_HLLD_ad}) are well-behaved when $\lambda^{*s}\to\lambda^s$ which typically occurs in the limit of zero tangential field and $B^2_x\gtrsim \Gamma p$.
In this limit, in fact, $\tilde{\chi}^s \to (v_x^s - \lambda^*)/2$ and $\nu^s=\pm1$.
On the other hand, particular care must be given to the degenerate case $B_x\to 0$, in which the two rotational waves collapse onto the entropy mode: $\lambda^{*R},\lambda^{*L}\to\lambda^*$.
In this situation, $\tilde{\chi}^s \to (v_x^s - \lambda^*)$ remains regular but the coefficient $\nu^*$ given in Eq. (\ref{eq:UCT_HLLD_nu}) becomes ill-defined at stagnation points ($\nu^*\sim v_x/|v_x|$) and one should rather resort to a three-wave pattern in which the tangential field is discontinuous across $\lambda^*$.
This limit is embodied by the HLLC solver, Eq. (\ref{eq:HLLCcoeffs}), and can be recovered by setting $\nu^*=0$ in the expressions above.
In practice we switch to the degenerate case whenever the difference between the two rotational modes falls below a given tolerance:
\begin{equation}
  \nu^* = \left\{\begin{array}{ll}
    \DS \frac{|\lambda^{*R}| - |\lambda^{*L}|}{\lambda^{*R} - \lambda^{*L}}
      & \quad {\rm if} \;\;  |\lambda^{*R} - \lambda^{*L}| >
                           \epsilon|\lambda^R-\lambda^L| \,,
  \\ \noalign{\medskip}
  0 & \quad {\rm otherwise} \,.
  \end{array}\right.
\end{equation}
with $\epsilon = 10^{-9}$.
This concludes the derivation of the UCT-HLLD averaging scheme for adiabatic MHD.

A couple of remarks are worth making.
First, if the base Riemann solver is not the HLLD, the proposed emf averaging-scheme can still be employed provided that the contact and Alfv\'en velocities are locally redefined using Eq. (\ref{eq:hlld_lambda*}) and the following expressions for consistency reasons.
Second, in the case of isothermal MHD our derivation still holds although the $\chi^s$ coefficients are different.
This case is discussed in \ref{app:UCT_HLLD_isothermal}.

%%%%%%%%%%%%%%%%%%%%%%%%%%%%%%%%%%%%%%%%%%%%%%%%%%%%%%%%%%%%%%%%%%%%%%%%
\subsection{The UCT-GFORCE scheme}
%
%%%%%%%%%%%%%%%%%%%%%%%%%%%%%%%%%%%%%%%%%%%%%%%%%%%%%%%%%%%%%%%%%%%%%%%%

In its original formulation \cite{Toro_Titarev2006}, the Generalized First ORder CEntered (GFORCE) flux  is obtained from a weighted average of the Lax-Friedrichs (LF) and Lax-Wendroff (LW) fluxes,
\begin{equation}\label{eq:GFORCE1D}
  \hat{F} = \omega_g F^{\rm LW} + (1 - \omega_g)F^{\rm LF} \,,
\end{equation}
where $F^{\rm LW} = F(U^{\rm LW})$ and
\begin{equation}\label{eq:GFORCE_Fluxes}
  U^{\rm LW} =   \frac{U^L + U^R}{2}
                 - \frac{\tau}{2}(F^R - F^L) \,,\qquad
  F^{\rm LF} =   \frac{F^L + F^R}{2}
                 - \frac{1}{2\tau}\left(U^R - U^L\right) \,,
\end{equation}
are, respectively, the LWs flux and state and the LF flux.
In Eq. (\ref{eq:GFORCE1D}), $\omega_g \in [0,1]$ is a weight coefficient usually chosen to satisfy monotonicity requirements.
This yields, according to \cite{Toro_Titarev2006},
\begin{equation}
  \omega_g \le \frac{1}{1 + c_g} \,,
\end{equation}
where $0\le c_g\le 1$ is the Courant number.
The FORCE flux is recovered with $\omega = 1/2$ and it is precisely the arithmetic mean of the Lax-Friedrichs and the Lax-Wendroff fluxes.
In the original formulation $\tau=\Delta t/\Delta x$ although here, in order to minimize the amount of numerical dissipation, we choose $\tau$ as the inverse of the local maximum signal speed, $\tau = 1/|\lambda_{\max}|$.

Specializing to a magnetic flux component, we re-write the GFORCE flux at a zone interface as
\begin{equation}\label{eq:GFORCE_Bt}
  \hat{F}_x = -\overline{v}_tB_x
              + \frac{1}{2}B_t^L\left[\omega v_x^{\rm LW} + (1-\omega) v_x^L\right]
              + \frac{1}{2}B_t^R\left[\omega v_x^{\rm LW} + (1-\omega) v_x^R\right]
              - (d^RB^R_t - d^LB_t^L) \,,
\end{equation}
which now closely relates to the form given by Eq. (\ref{eq:RoeForm_Induction}) with $a^L_x = a^R_x = 1/2$, provided that we redefine the transverse velocities as
\begin{equation}\label{eq:UCT_GFORCE_vt}
  \overline{v}_t =
  \omega\left(v_t^{\rm LW} - \frac{\tau_x}{2}\Delta v_yv_x^{\rm LW}\right)
  + (1-\omega) \left(\frac{v_y^L + v_y^R}{2}\right)
\end{equation}
and the diffusion coefficients as
\begin{equation}\label{eq:UCT_GFORCE_dLR}
  d^{s} = \omega \frac{\tau_x}{2} v_x^{s} v_x^{\rm LW}
          + (1-\omega)\frac{|\lambda_x|}{2} \,.
\end{equation}
The Lax-Wendroff velocities can be obtained zone interfaces as $v_k^{\rm LW} = m_k^{\rm LW}/\rho^{\rm LW}$ where momentum components and density are calculated using the first of (\ref{eq:GFORCE_Fluxes}).
Similar quantities are obtained at $y$- and $z$-interfaces by suitable index permutation.
At the practical level, the UCT-GFORCE scheme is therefore obtained by storing, during the Riemann solver call at any given interface, the transverse velocities $\vec{v}_t$ as well as $d^{s}$ with $s=L,R$.

%%%%%%%%%%%%%%%%%%%%%%%%%%%%%%%%%%%%%%%%%%%%%%%%%%%%%%%%%%%%%%%%%%%%%%%%
\section{Numerical benchmarks}
\label{sec:numerical_benchmarks}
%
%
%
%
%%%%%%%%%%%%%%%%%%%%%%%%%%%%%%%%%%%%%%%%%%%%%%%%%%%%%%%%%%%%%%%%%%%%%%%%

In what follows we compare different emf-averaging schemes in terms of accuracy, robustness and dissipation properties.
Our selection includes:
\begin{itemize}

\item
%%%%%
Thew arithmetic averaging, given by Eq. (\ref{eq:Arithmetic});

\item
%%%%%%
the CT-Contact emf averaging, given by Eq. (\ref{eq:UCT_CONTACT});

\item
%%%%%
the CT-Flux emf, given by Eq. (\ref{eq:CT_Flux});

\item
%%%%%
the UCT-HLL scheme following our new composition formula, Eq. (\ref{eq:emf2D}) with coefficients given by Eq. (\ref{eq:HLLcoeffs}).
Extensive numerical testing (including several other additional tests not shown here) has demonstrated our formulation of UCT-HLL scheme yields essentially equivalent results to the original formulation (Eq. \ref{eq:UCT_HLL2}).

\item
%%%%%
the newly proposed UCT-HLLD scheme, given by Eq. (\ref{eq:emf2D}) together with Eq. (\ref{eq:UCT_HLLD_ad}) and Eq. (\ref{eq:UCT_HLLD_nu});

\item
%%%%%
the novel UCT-GFORCE scheme, defined by Eq. (\ref{eq:emf2D}) with transverse velocity and diffusion coefficients given by Eq. (\ref{eq:UCT_GFORCE_vt}) and (\ref{eq:UCT_GFORCE_dLR}).
\end{itemize}

%The selected emf are summarized below:
%
%\begin{itemize}
%  \item Arithmetic averaging, given by Eq. (\ref{eq:Arithmetic});
%  \item The CT-Contact scheme, Eq. (\ref{eq:UCT_CONTACT});
%  \item The CT-Flux scheme, Eq. (\ref{eq:UCT_Flux});
%  \item The UCT-HLL scheme, in the formulation given by Eq. (\ref{eq:UCT_HLL2});
%  \item The new UCT-HLLD scheme, Eq. (\ref{eq:emf2D}) together with
%        Eq. (\ref{eq:UCT_HLLD_ad}) and (\ref{eq:UCT_HLLD_nu});
%  \item The UCT-GFORCE scheme, Eq. (\ref{eq:emf2D}) with transverse velocity
%        and diffusion coefficients given by Eq. (\ref{eq:UCT_GFORCE_coeffs}).
%\end{itemize}
%
During the comparison we will employ the same base scheme for all emf averaging methods.
The base scheme is chosen to be either the $2^{\rm nd}$-order Strong Stability-Preserving (SSP) Runge Kutta scheme \cite{Gottlieb_etal2001} with piecewise linear reconstruction or the $3^{\rm rd}$-order Runge-Kutta time-stepping with the $5^{\rm th}$-order monotonicity-preserving spatial reconstruction \cite{Suresh_Huynh1997}, first introduced in the context of relativistic MHD flows by \cite{DelZanna_etal2007}.
Although both base schemes are second-order accurate (reconstructions are applied direction-wise), the latter has reduced dissipation properties when compared to the former.

Unless otherwise stated, an adiabatic equation of state with specific heat ratio $\Gamma=5/3$ is adopted.
The interface Riemann solver is either the Roe solver of \cite{Cargo_Gallice1997} or the HLLD solver of \cite{Miyoshi_Kusano2005}, depending on the test, while the CFL number is $C_a = 0.4$ in 2D and $C_a = 0.3$ in 3D, unless otherwise stated.

%%%%%%%%%%%%%%%%%%%%%%%%%%%%%%%%%%%%%%%%%%%%%%%%%%%%%%%%%%%%%%%%%%%%%%%%
\subsection{Field Loop Advection in two and three dimensions}
%
%%%%%%%%%%%%%%%%%%%%%%%%%%%%%%%%%%%%%%%%%%%%%%%%%%%%%%%%%%%%%%%%%%%%%%%%

As a first first test, we consider the advection of weakly magnetized field loop in both 2D and 3D.
In the limit of a pressure-dominated plasma, the magnetic field is essentially transported as a passive scalar and the upwind properties of any multidimensional scheme can be easily inspected.

In the 2D version, computations are carried out on the rectangle $x\in[-1,1]$ and $y\in[-1/2,1/2]$ using both the $2^{\rm nd}$ and $3^{\rm rd}$ order base-schemes covered by a uniform grid of $128\times 64$ zones.
The initial condition consists of a medium with constant density and pressure, $\rho = p = 1$ while the magnetic field is initialized through the $z$-component of the vector potential:
\begin{equation}\label{eq:fl_Az}
  A_z(x,y) = \left\{\begin{array}{ll}
    A_0(R-r) & \quad{\rm if}\quad r < R \\ \noalign{\medskip}
    0        & \quad{\rm otherwise}\,,
  \end{array}\right.
\end{equation}
where $A_0 = 10^{-3}$, $R = 0.3$ while $r=\sqrt{x^2 + y^2}$.
The velocity is constant and equal to $\vec{v} = 2\hvec{e}_x + \hvec{e}_y$ so that the system is uniformly advected along the main diagonal.
Periodic boundary conditions are imposed on all sides and the base scheme employs the Riemann solver of Roe is used for all computations.

\begin{figure*}[!ht]
  \centering
  \includegraphics[width=0.98\textwidth]{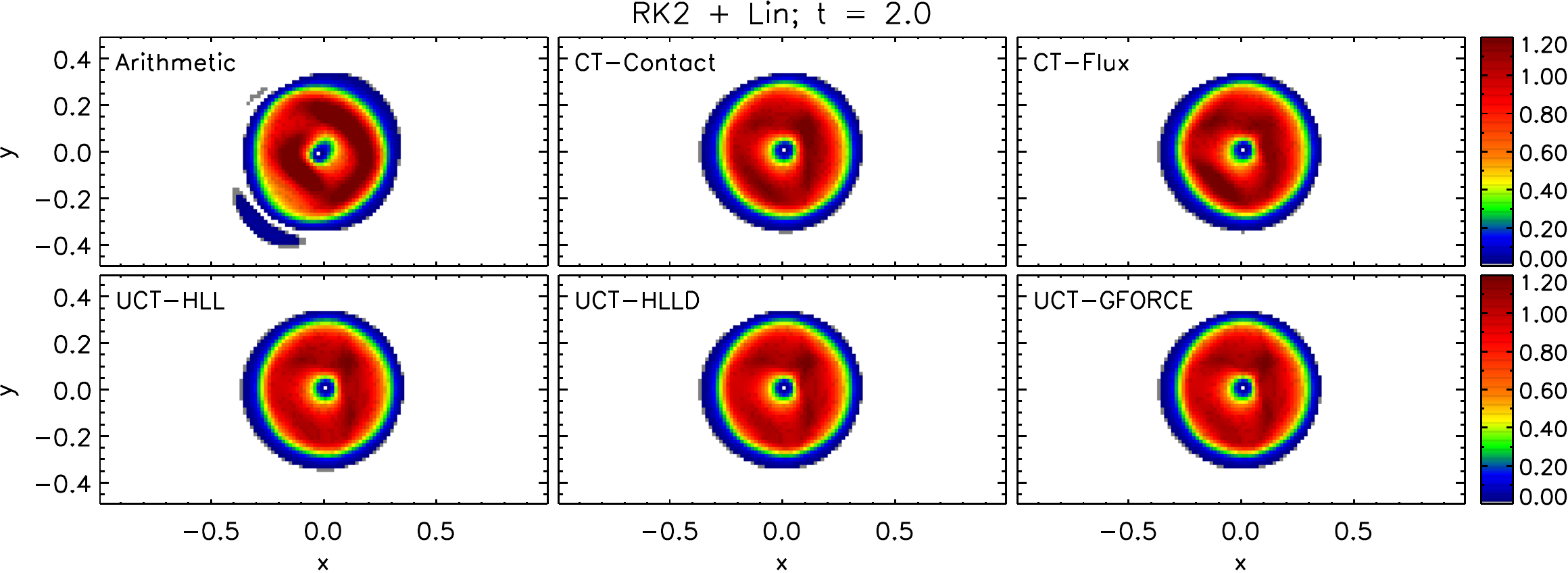}
  \caption{\footnotesize Magnetic energy (normalized to the maximum initial
           value) maps for the field loop test using the linear scheme at $t=2$.
           The six panels show the results obtained with different e.m.f. averaging
           schemes, indicated in the upper left portion of the panel.
           \label{fig:fl2D_mapsRK2}}
\end{figure*}
\begin{figure*}[!h]
  \centering
  \includegraphics[width=0.98\textwidth]{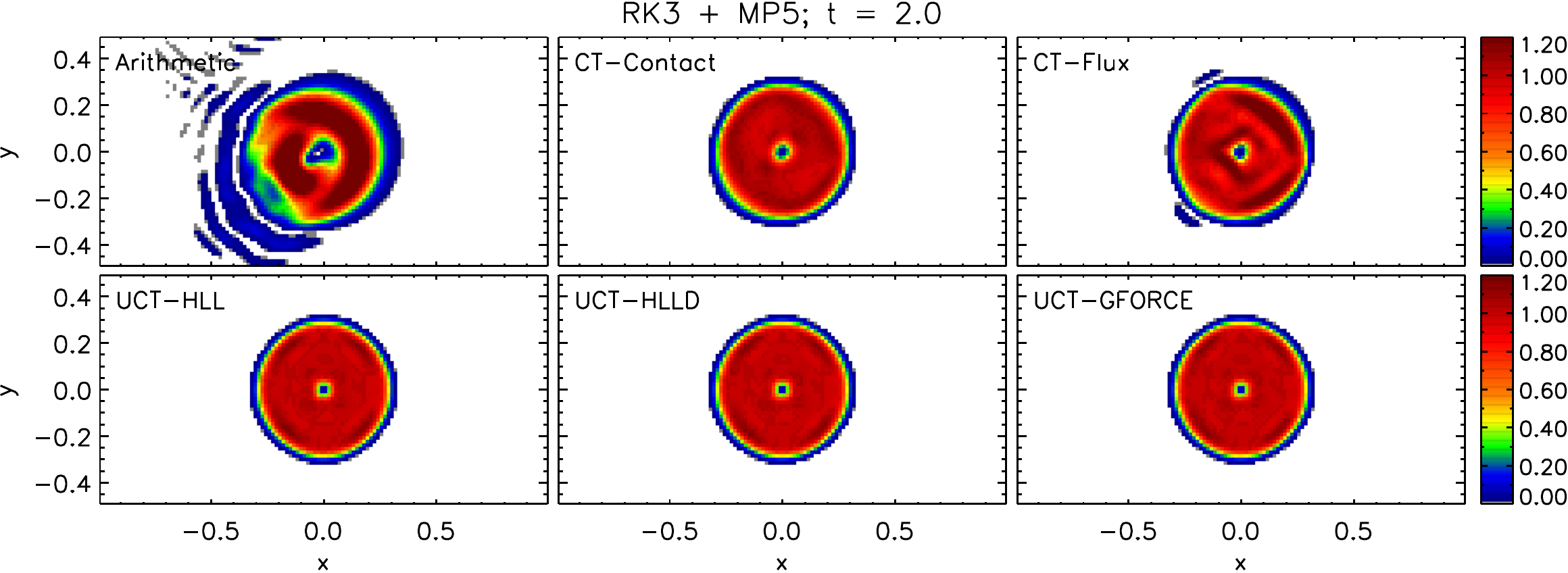}
  \caption{Same as Fig. \ref{fig:fl2D_mapsRK2} but for RK3 time stepping and
           MP5 reconstruction.
           \label{fig:fl2D_mapsRK3}}
\end{figure*}

Results are shown in Fig. \ref{fig:fl2D_mapsRK2} and \ref{fig:fl2D_mapsRK3} for the $2^{\rm nd}$-and $3^{\rm rd}$-order schemes, respectively.
The amount of numerical diffusion, mostly discernible from the smearing of the loop edges,  is primarily determined by the choice of the reconstruction scheme.
The smearing is greatly reduced with a higher than $2^{\rm nd}$-order reconstruction.
However, the choice of the averaging scheme shows striking differences in the shape of the loop.
Arithmetic averaging performs the worse, indicating large-amplitude oscillations already visible with the linear scheme and corrupting the loop shape even more when a higher-order reconstruction is employed.
Some oscillations are also present in the CT-Flux scheme while the remaining averaging methods show oscillation-free behavior, thus indicating a sufficient amount of dissipation.
\begin{figure*}
  \centering
  \includegraphics[width=0.45\textwidth]{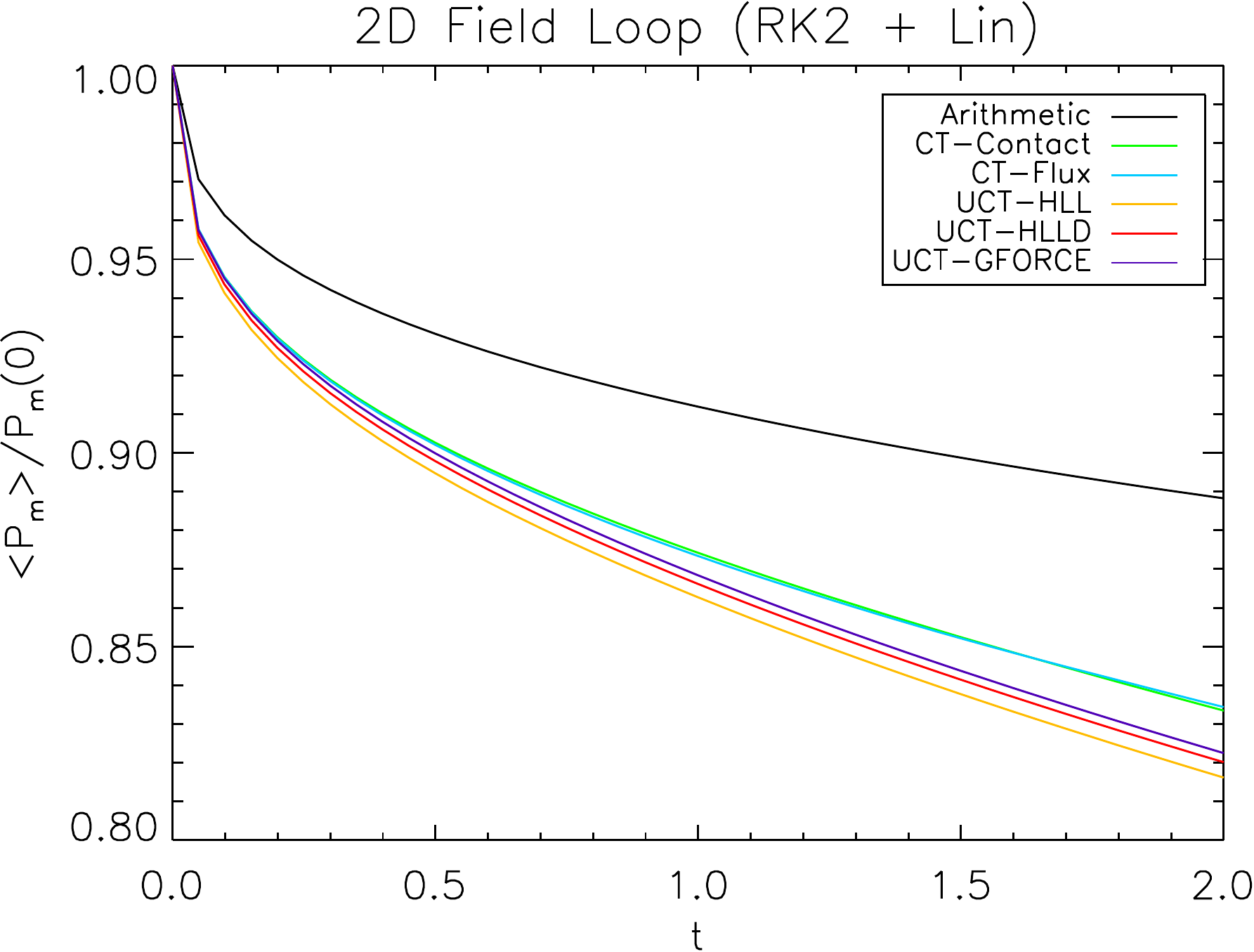}%
  \includegraphics[width=0.45\textwidth]{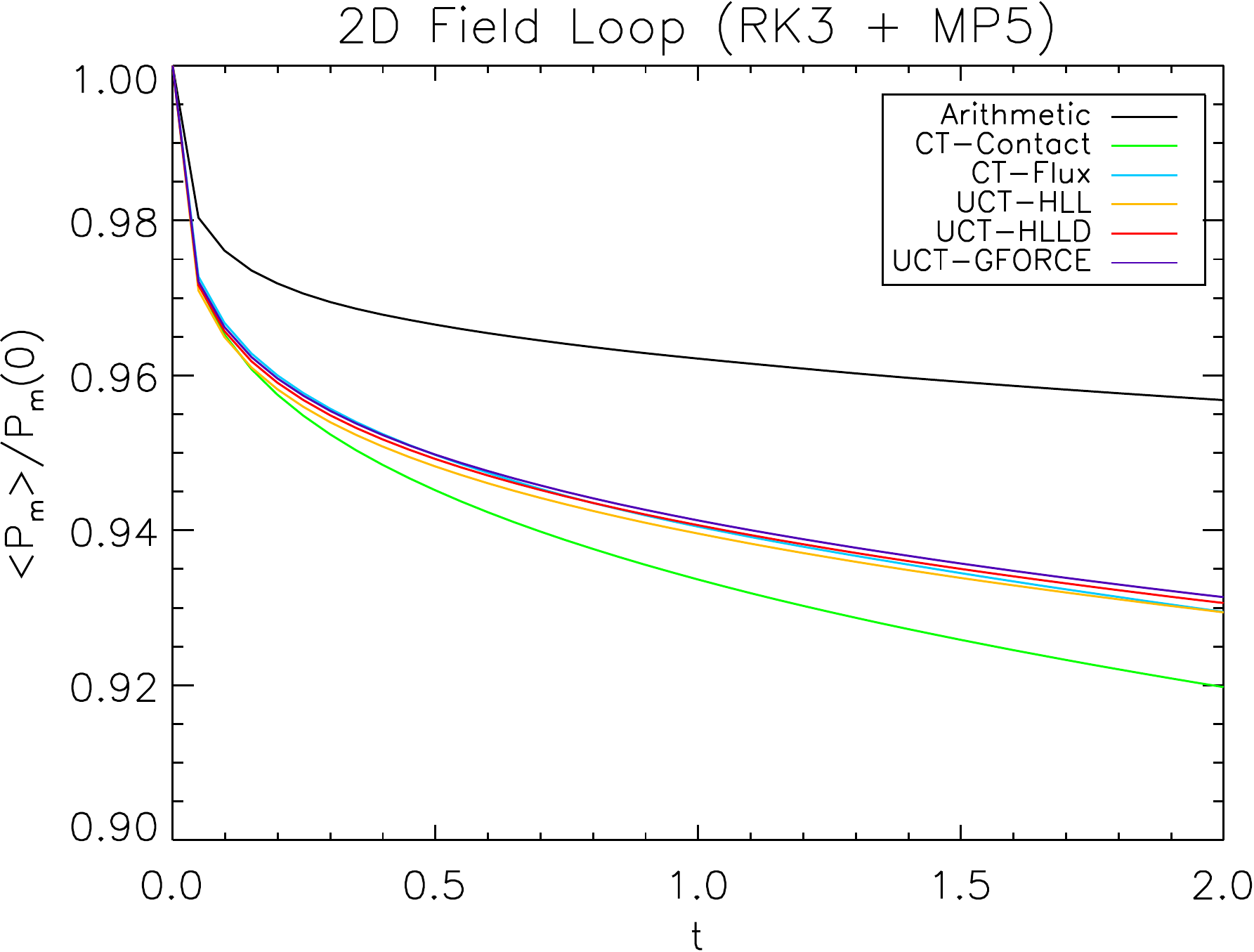}
  \caption{\footnotesize Magnetic energy decay for the 2D field loop test using
  the $2^{\rm nd}$ (left) and $3^{\rm rd}$ (right) order scheme.
  The different curves have been normalized to the initial energy and correspond
  to Arithmetic (black), CT-Contact (green), CT-Flux (cyan), UCT-HLL (orange), 
  UCT-HLLD (red) and UCT-GFORCE (purple).
  \label{fig:fl2D_decay}}
\end{figure*}
A more quantitative analysis is provided in Fig. \ref{fig:fl2D_decay} where we plot the total integrated magnetic energy as a function of time for the selected schemes.
The decay provides a measure of the scheme dissipation properties but not necessarily of its stability.
Indeed, arithmetic averaging exhibits the smallest decay rate while, at the second-order level, CT-Contact and CT-Flux provide the optimal level of dissipation.
On the other hand, with a higher-order reconstruction, the newly proposed UCT-HLLD schemes provides the lesser amount of dissipation, still ensuring stability of the integration, while CT-Contact yields slightly more diffusive results.

\begin{figure*}[!ht]
  \centering
  \includegraphics[trim=0 20 180 20, width=0.3\textwidth]{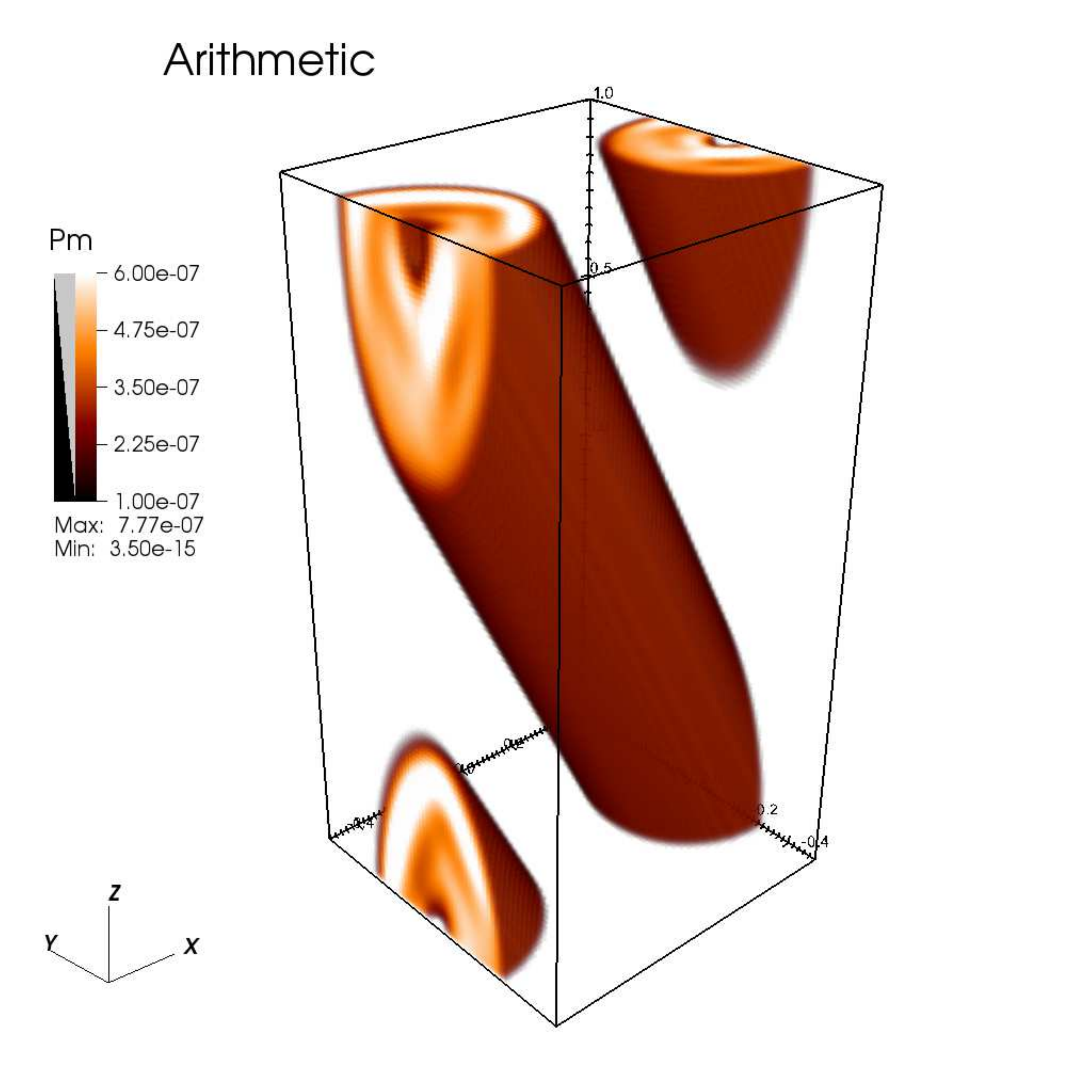}%
  \includegraphics[trim=0 20 180 20, width=0.3\textwidth]{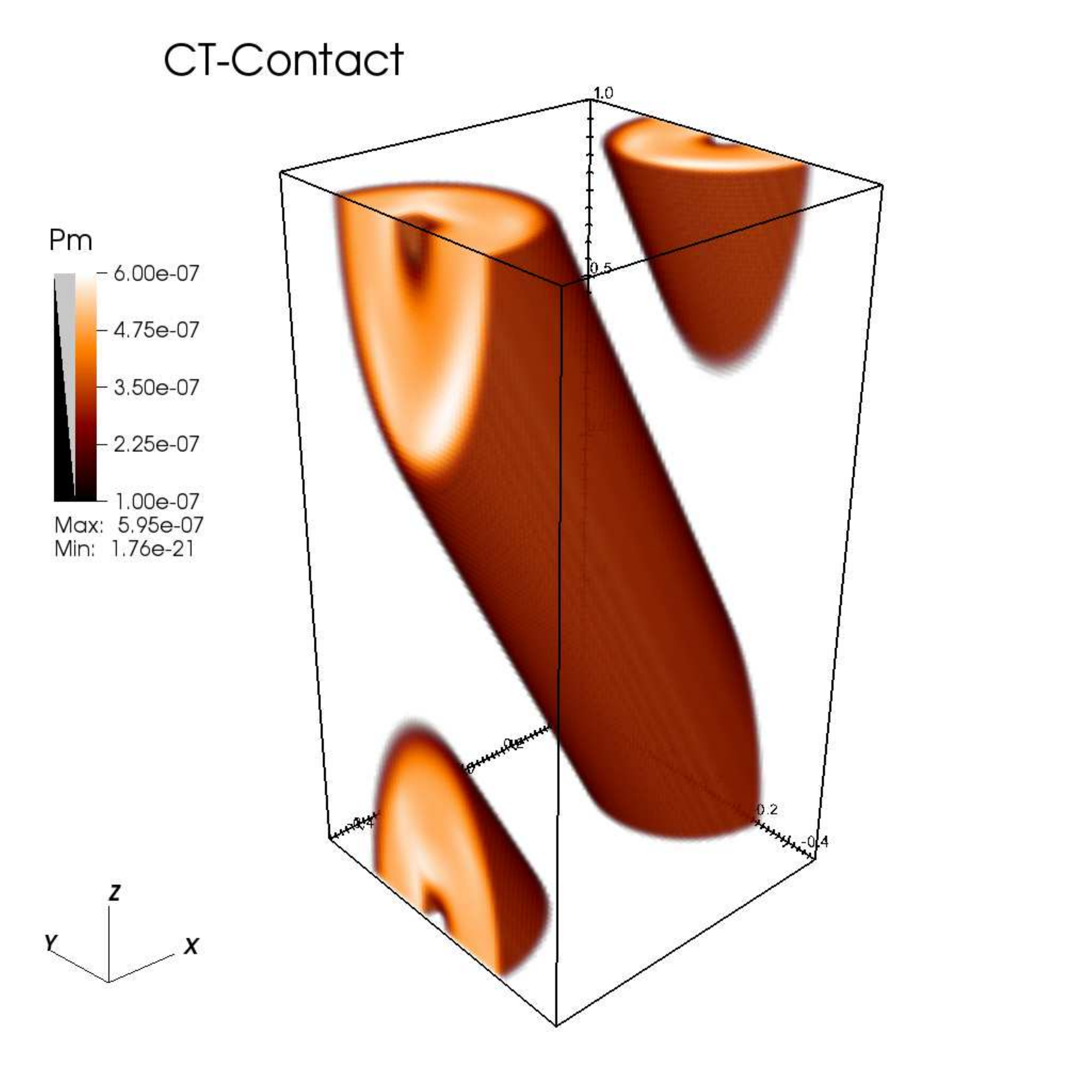}%
  \includegraphics[trim=0 20 180 20, width=0.3\textwidth]{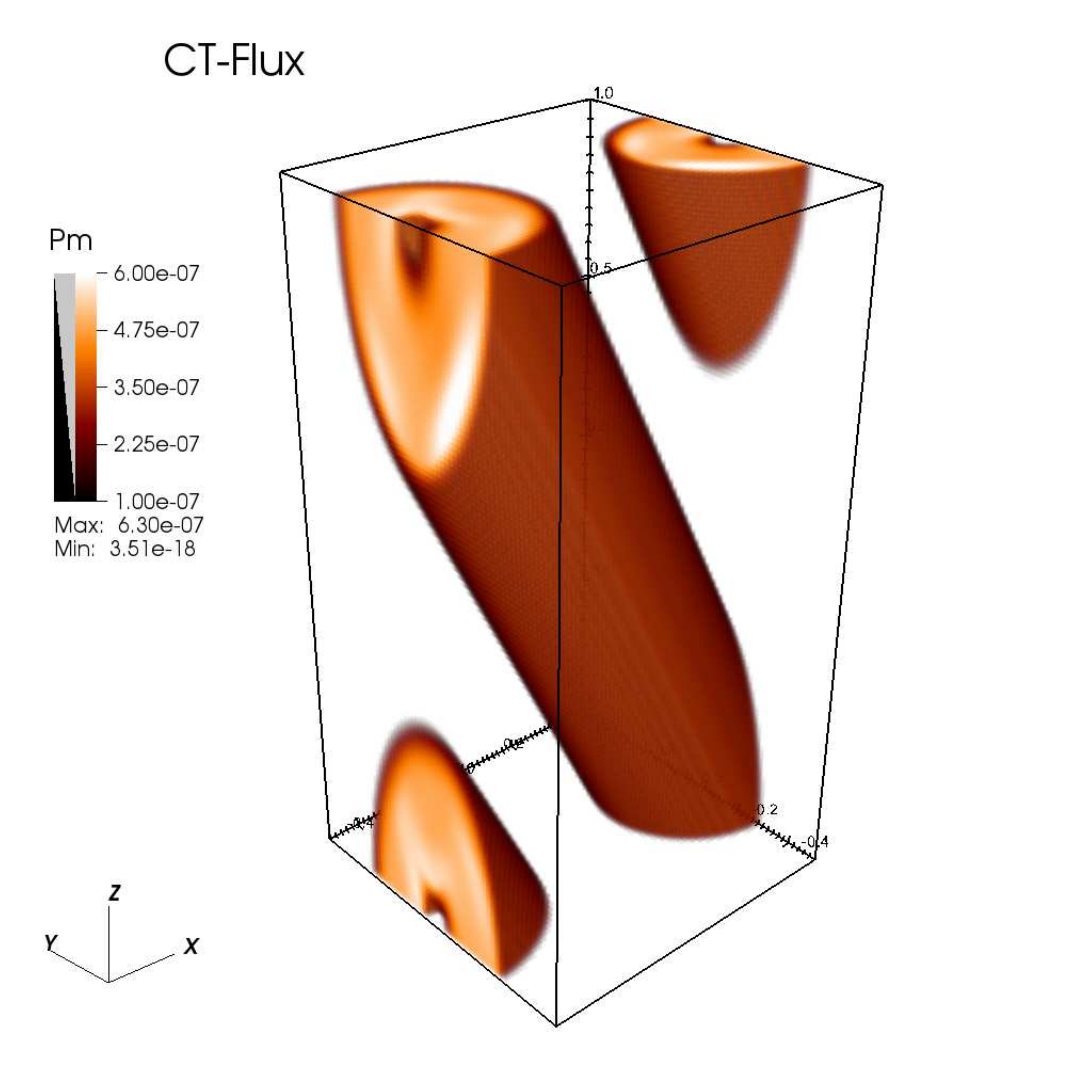}
  \includegraphics[trim=0 20 180 20, width=0.3\textwidth]{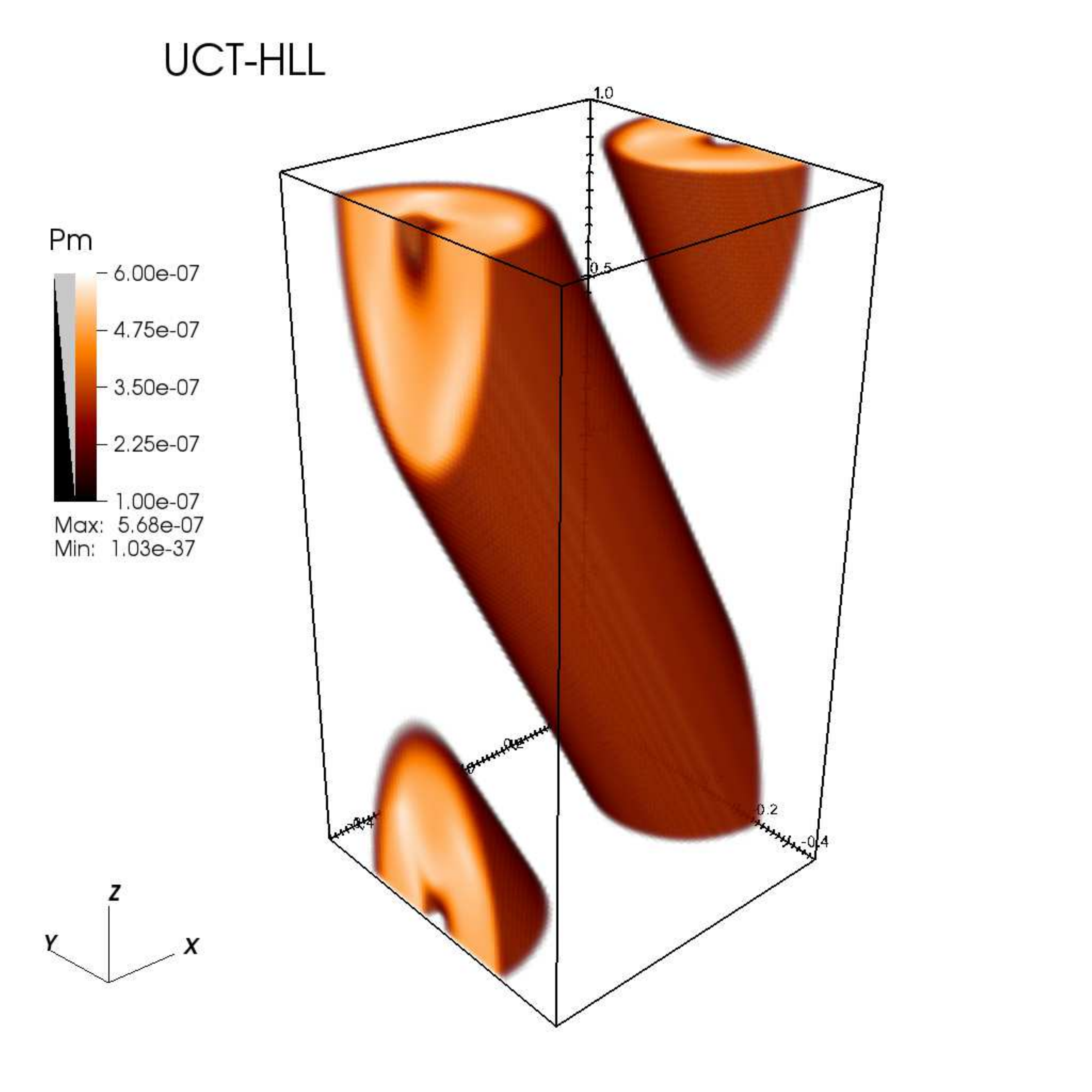}%
  \includegraphics[trim=0 20 180 20, width=0.3\textwidth]{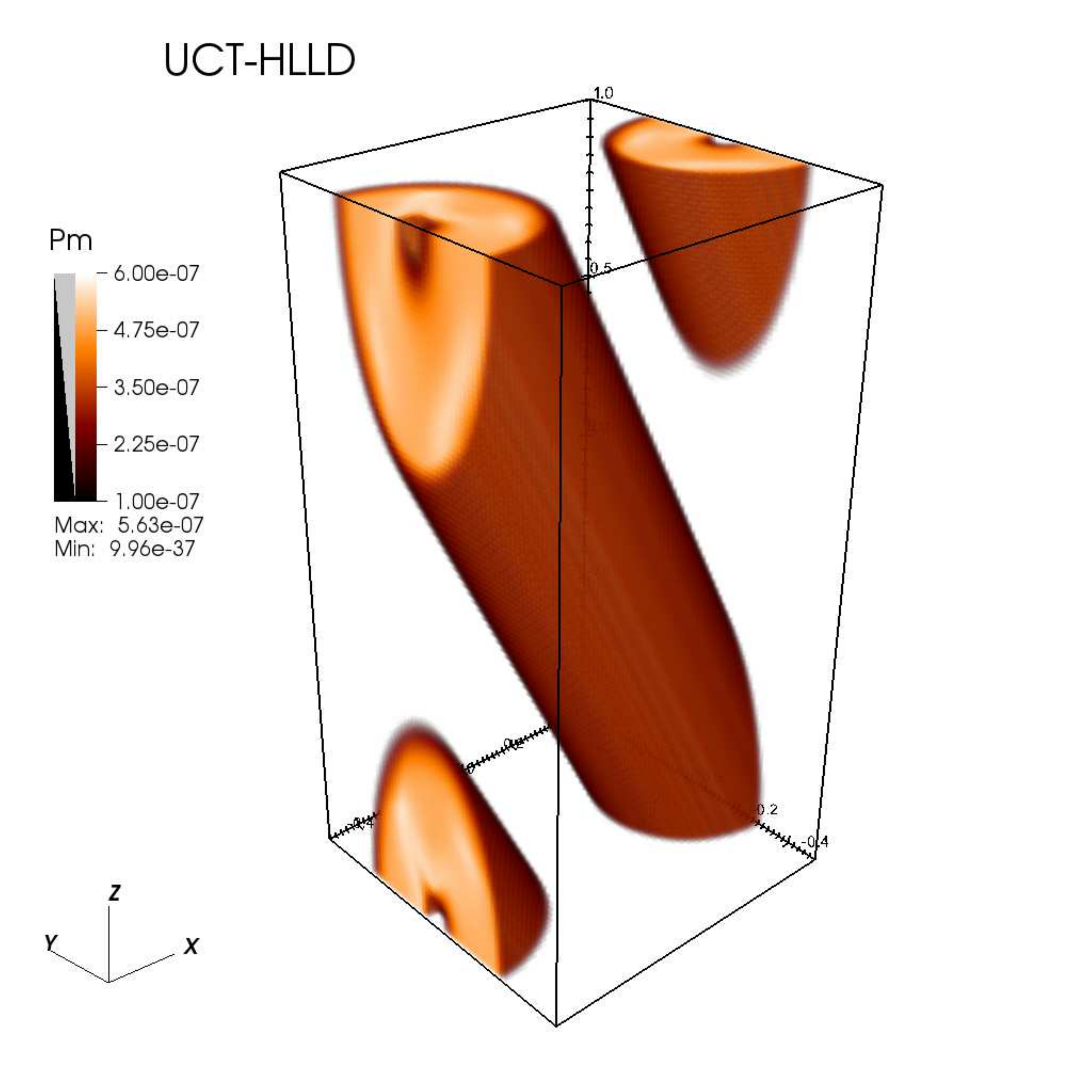}%
  \includegraphics[trim=0 20 180 20, width=0.3\textwidth]{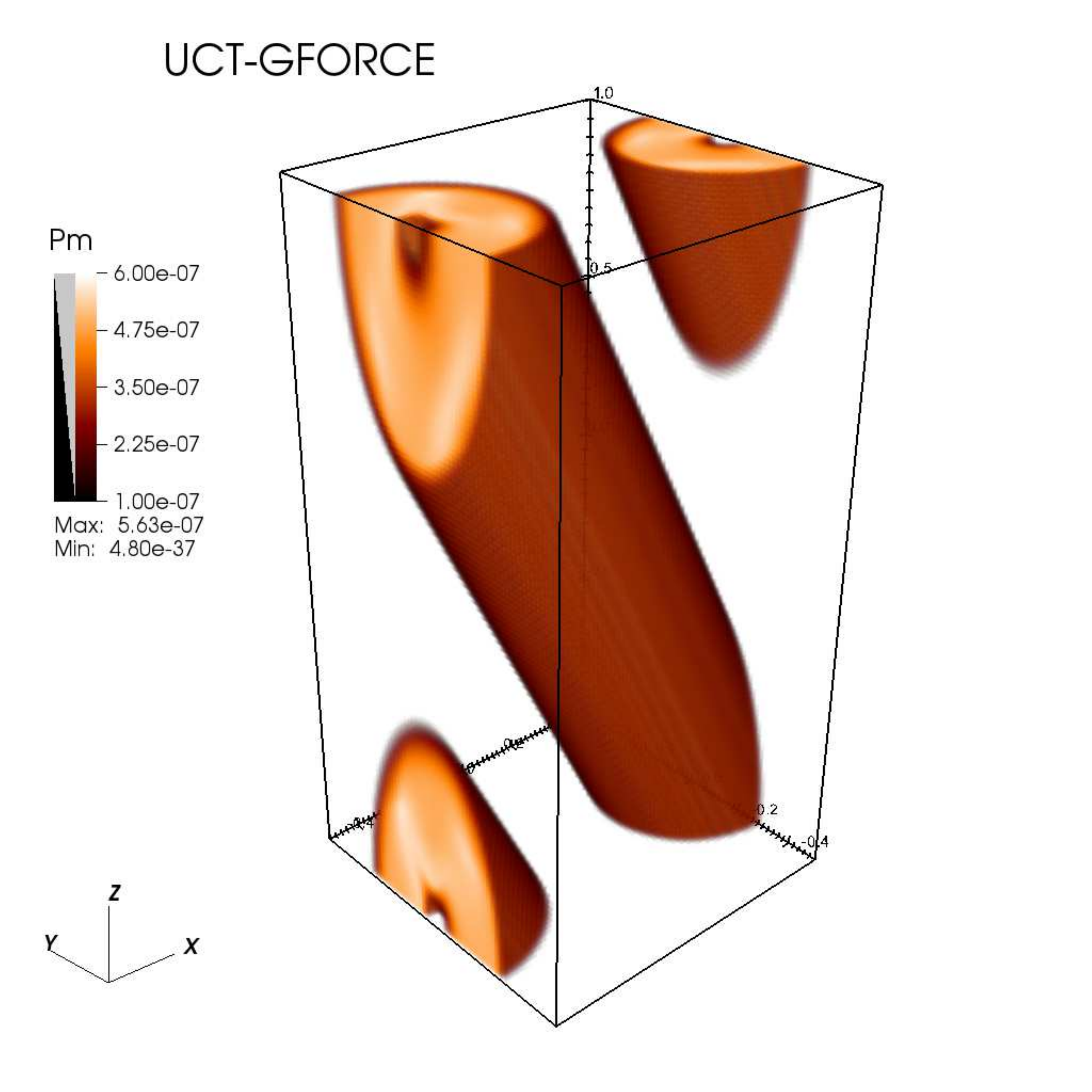}
  \caption{\footnotesize Volume rendering of the magnetic energy at $t=1$ for the 3D
  field loop test using the $2^{\rm nd}$-order scheme.
  The panel order is the same used for Fig. \ref{fig:fl2D_mapsRK2} and the
  corresponding emf-averaging scheme is reported in the top left corner of
  each panel.
  While the color-scale has been chosen to be the same for all plots,
  the maximum and minimum values are reported below the legend. 
  \label{fig:fl3D_mapsRK2}}
\end{figure*}
\begin{figure*}[!ht]
  \centering
  \includegraphics[trim=0 20 180 20, width=0.3\textwidth]{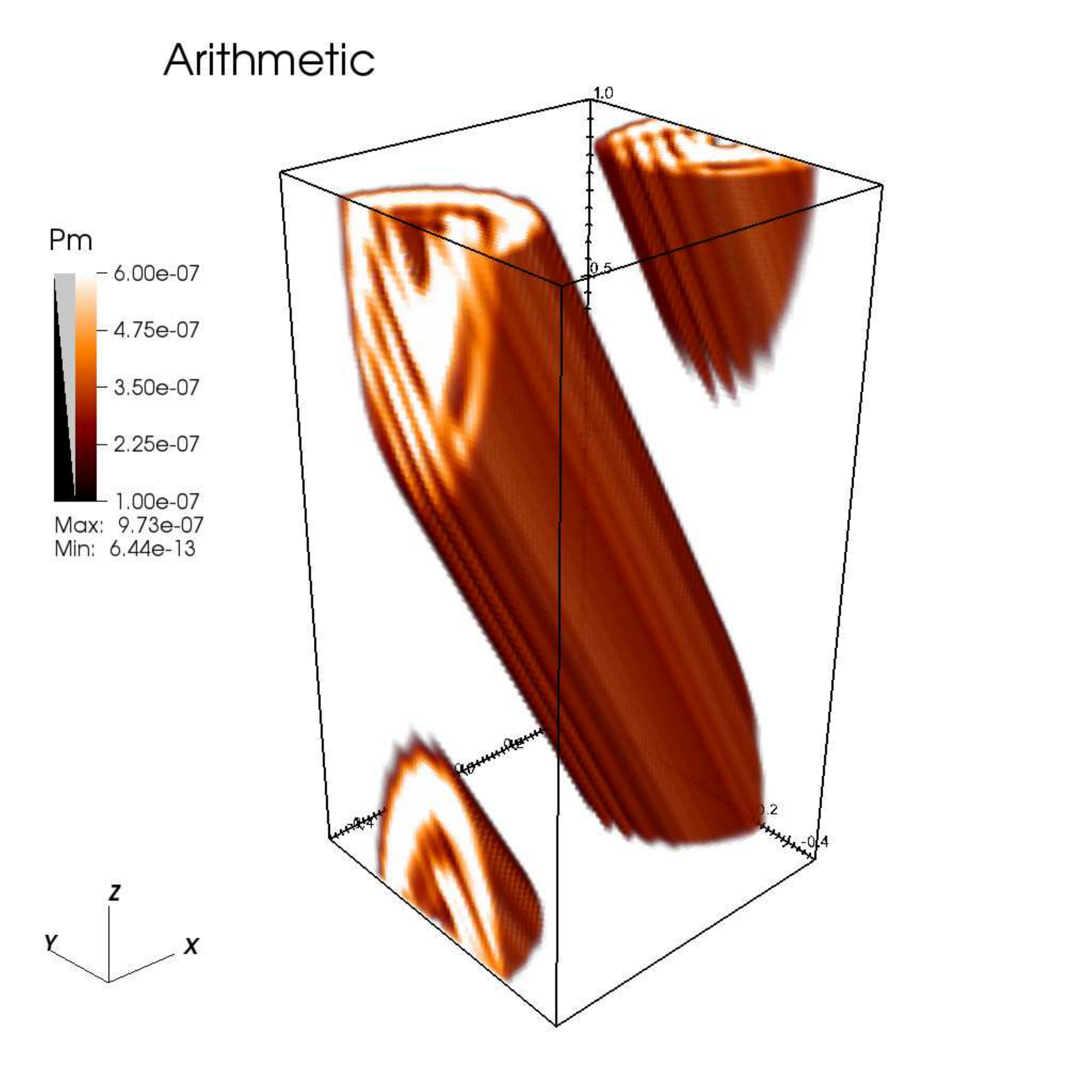}%
  \includegraphics[trim=0 20 180 20, width=0.3\textwidth]{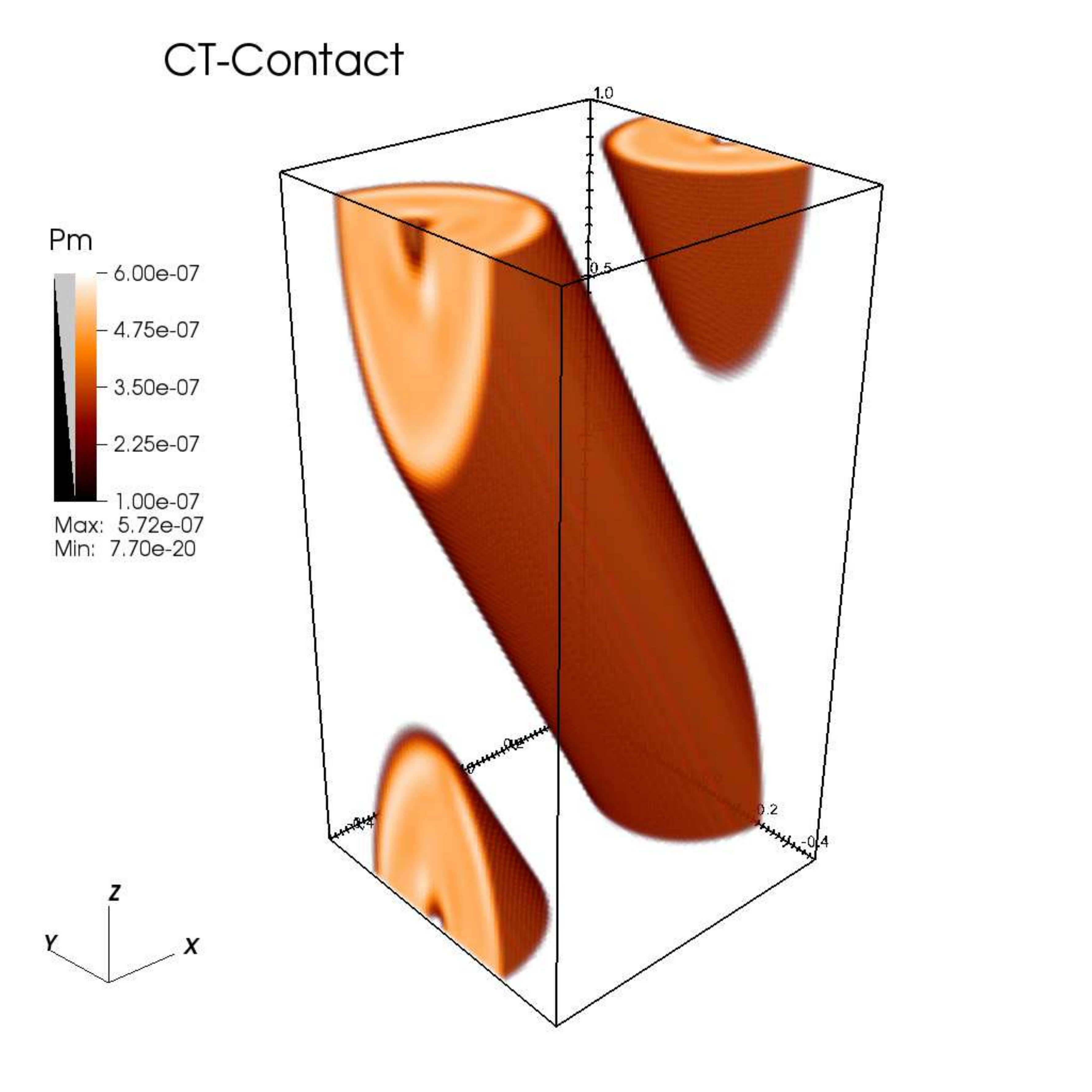}%
  \includegraphics[trim=0 20 180 20, width=0.3\textwidth]{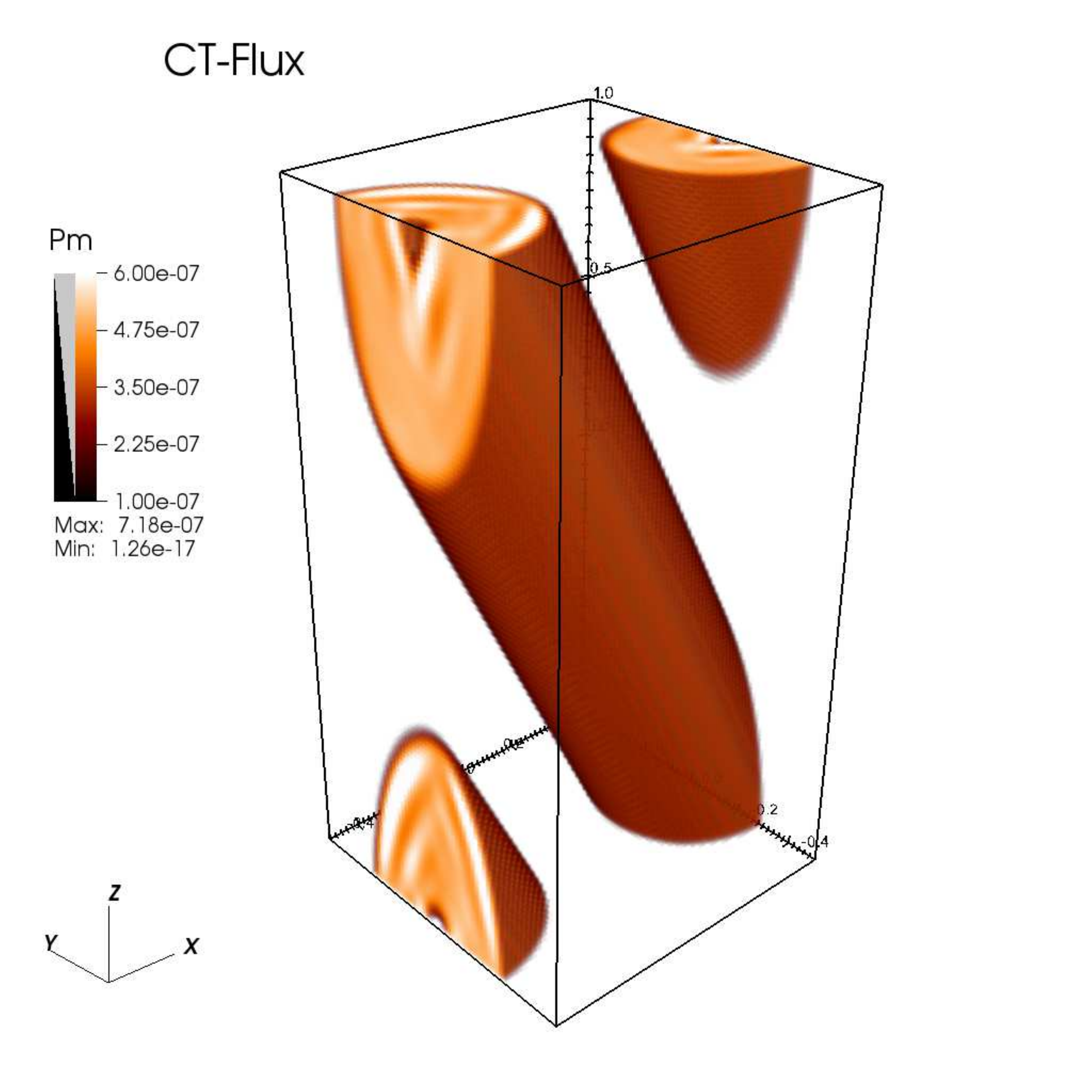}
  \includegraphics[trim=0 20 180 20, width=0.3\textwidth]{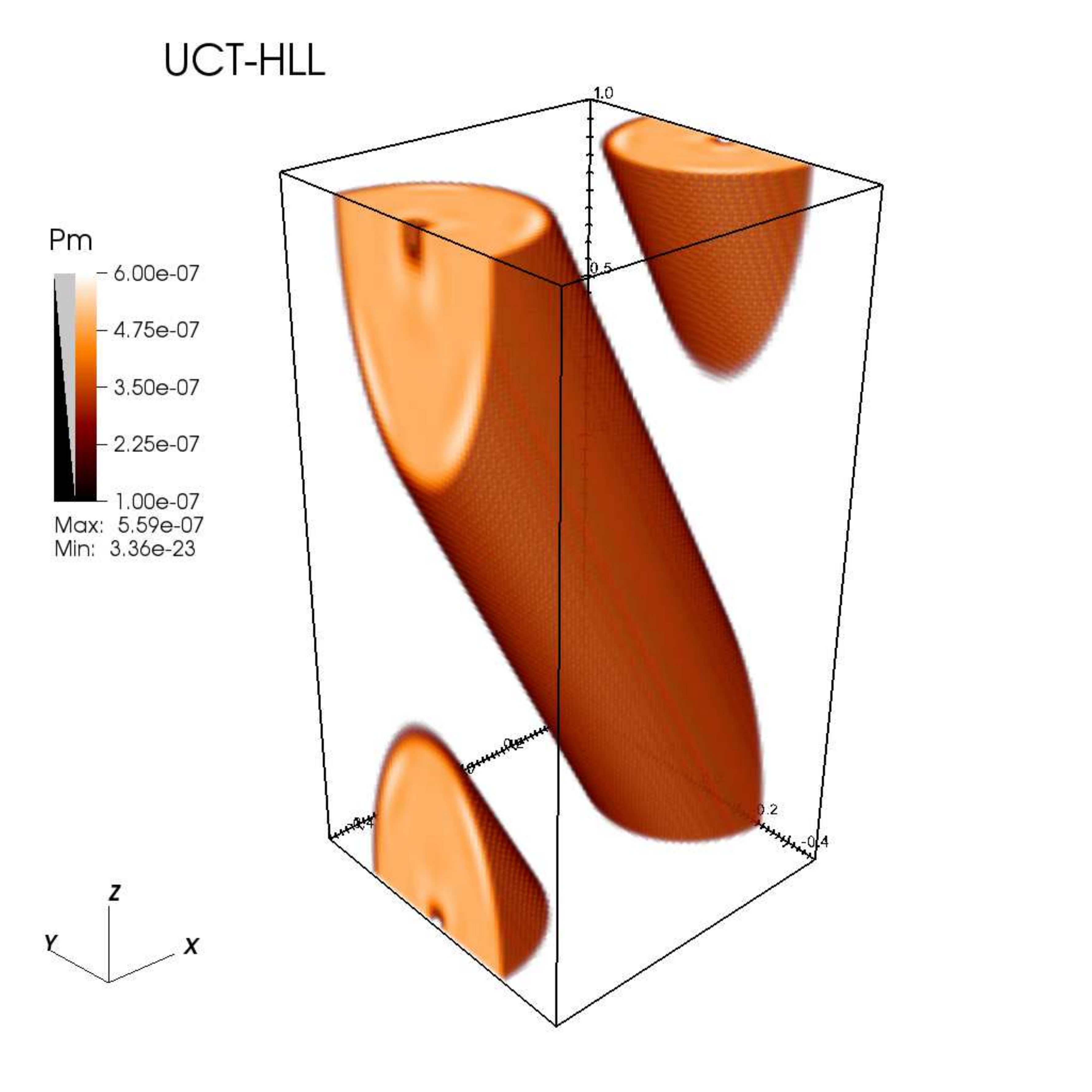}%
  \includegraphics[trim=0 20 180 20, width=0.3\textwidth]{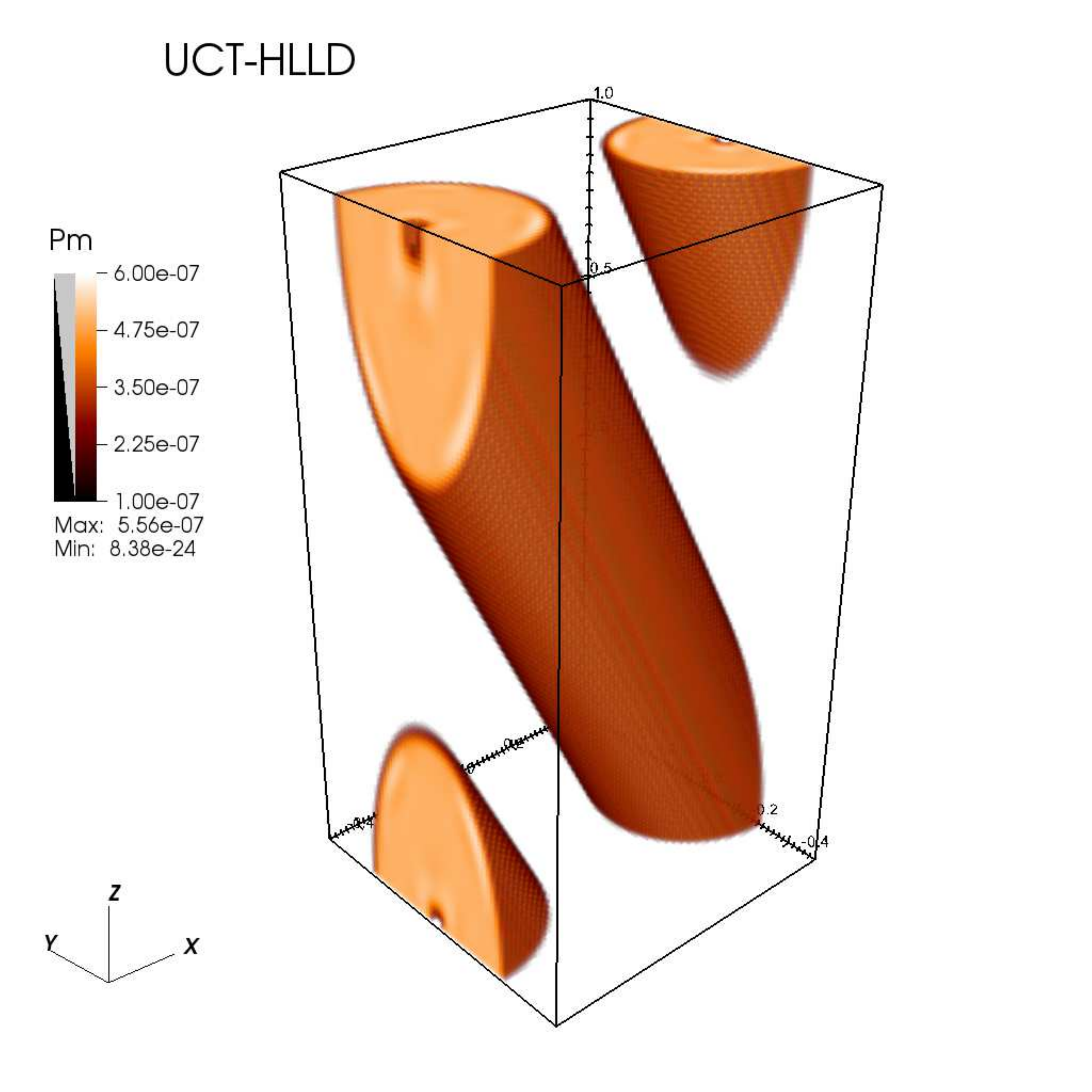}%
  \includegraphics[trim=0 20 180 20, width=0.3\textwidth]{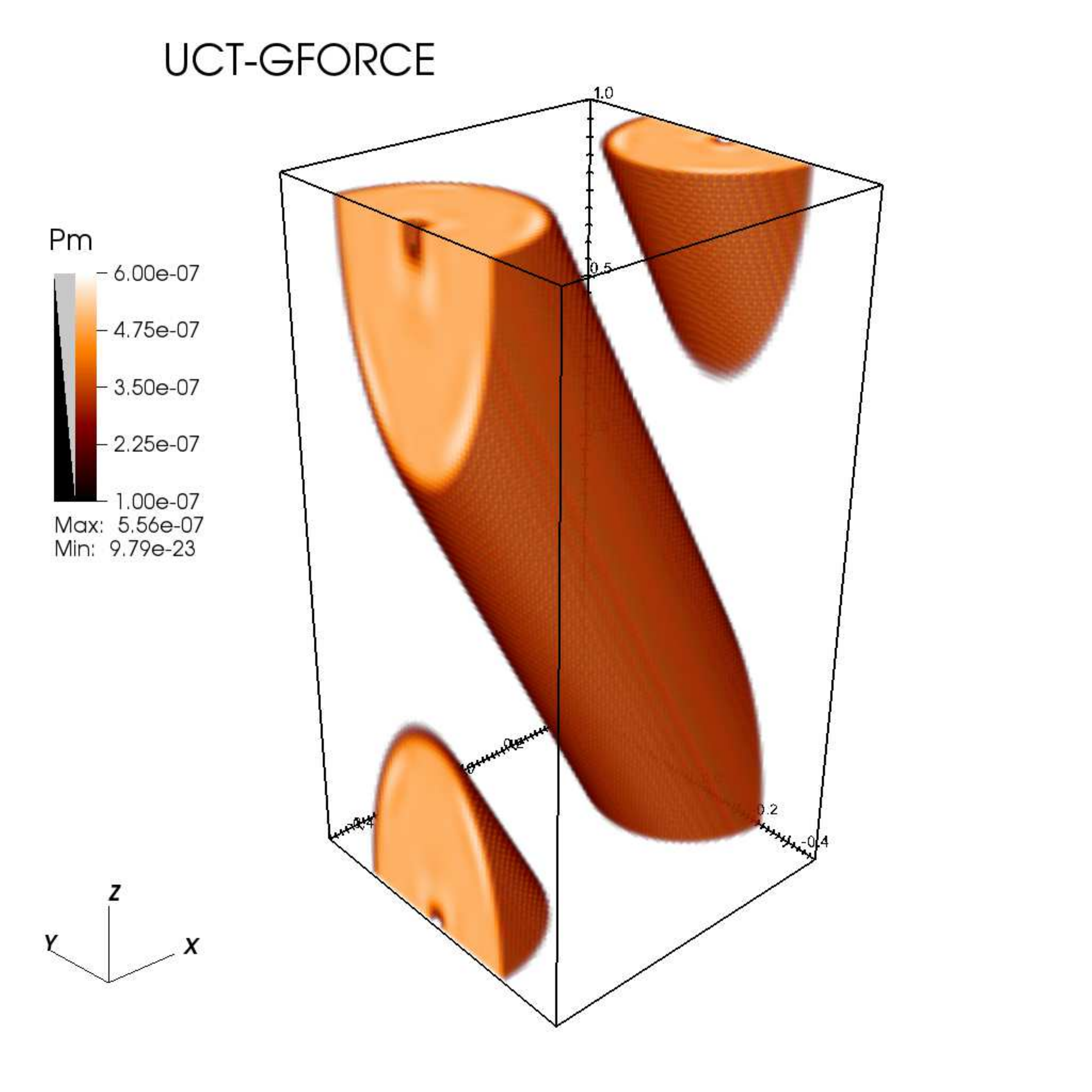}
  \caption{\footnotesize Same as Fig. \ref{fig:fl3D_mapsRK2} but for RK3
  time stepping and MP5 reconstruction.
  \label{fig:fl3D_mapsRK3}}
\end{figure*}

Computations have been repeated using a three-dimensional configuration as described in \cite{GS2008} (see also \cite{MigTze2010} and, more recently, \cite{Minoshima_etal2019}).
The domain is now chosen to be $x,y\in[-1/2,1/2]$ and $z\in[-1,1]$ with periodic boundary conditions and uniform flow velocity $\vec{v}=(1,\, 1,\, 2)$.
The initial magnetic field can be obtained by rotating the original 2D frame by an angle $\gamma$ around the $y$ axis.
The relation between the unrotated coordinates $(x',y',z')$ and the actual computational coordinates $(x,y,z)$ is given by
\begin{equation}
  \left\{\begin{array}{l}
    \DS x' =  x\cos\gamma + z\sin\gamma   \\ \noalign{\medskip}
    \DS y' = y                            \\ \noalign{\medskip}
    \DS z' = -x\sin\gamma + z\cos\gamma   \,. 
  \end{array}\right.
\end{equation}
Coordinates in the primed frame must satisfy periodicity in the range $x'/L'_x\in[-1/2,1/2]$, $y'/L'_y\in[-1/2,1/2]$.
This is achieved by modifying
\begin{equation}
  x' \leftarrow x' - L'_x\,{\rm floor}\left(\frac{x'}{L'_x} + \HALF\right) \,,\qquad
  y' \leftarrow y' - L'_y\,{\rm floor}\left(\frac{y'}{L'_y} + \HALF\right) \,,
\end{equation}
where $L'_x = 2/\sqrt{5}$, $L'_y = 1$.
We then define the vector potential in the primed frame using Eq. (\ref{eq:fl_Az}) with $x'$ and $y'$ used as arguments.
The inverse transformation is applied in order to recover the magnetic vector potential in the rotated frame:
\begin{equation}
  \vec{A} = A'_z(-\sin\gamma,\, 0,\, \cos\gamma)  \,.
\end{equation}
We choose $\gamma$ so that $\tan\gamma = 1/2$.

\begin{figure*}[!h]
  \centering
  \includegraphics[width=0.45\textwidth]{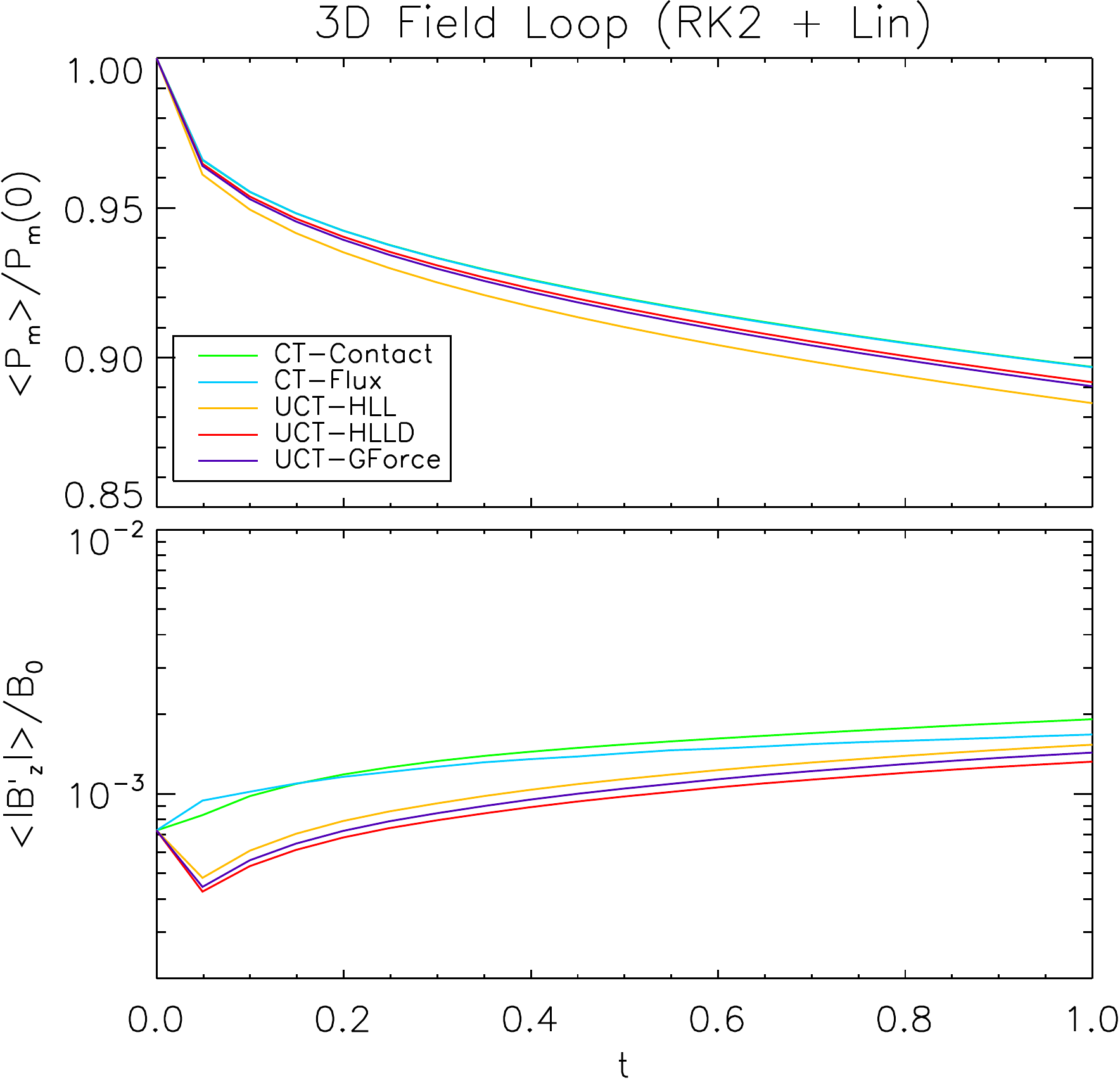}%
  \includegraphics[width=0.45\textwidth]{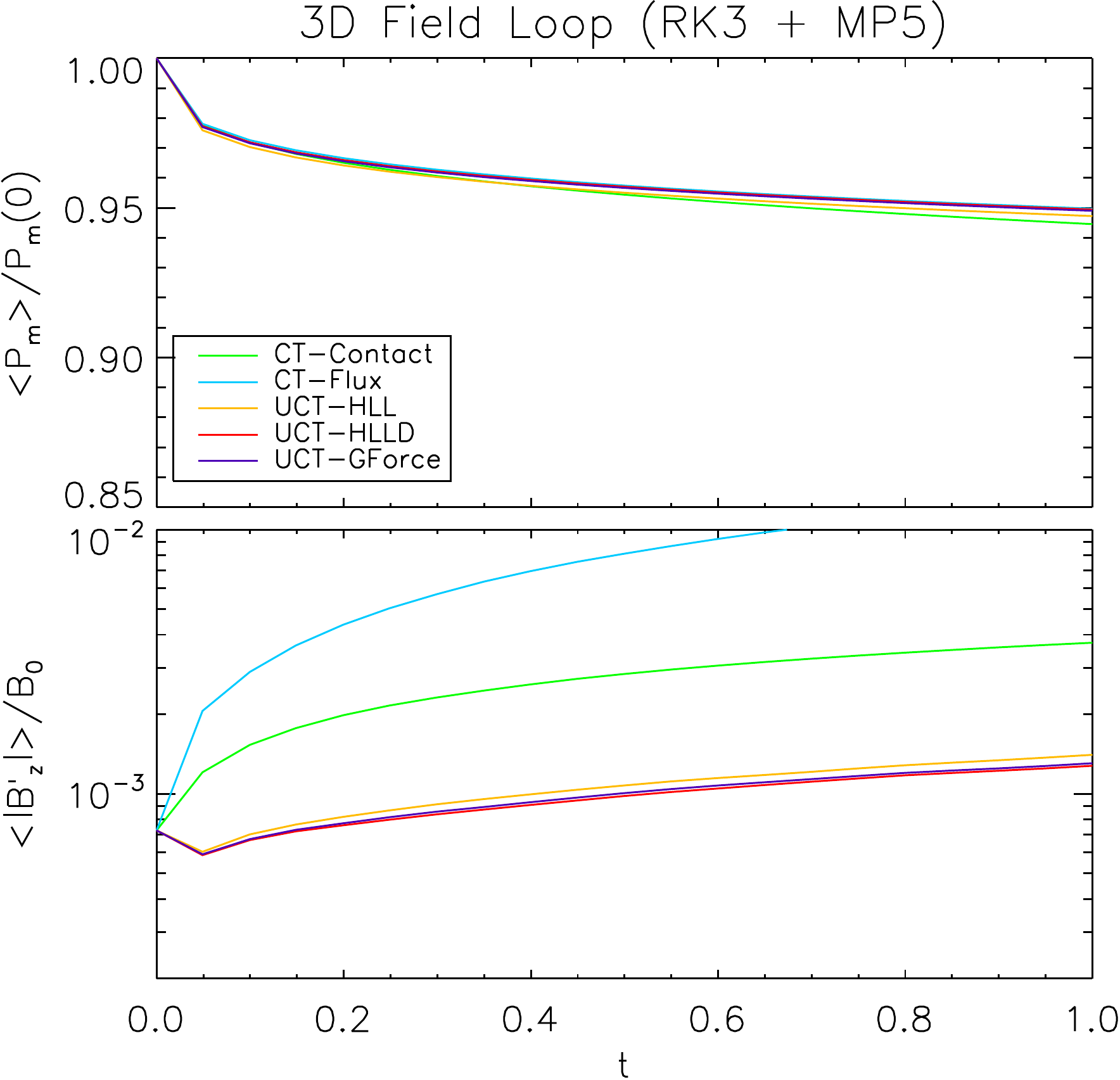}
  \caption{\footnotesize \emph{Top panels}: Magnetic energy decay (normalized to
  the initial value) as a function of time for the 3D field loop test.
  Left and right panels correspond to the $2^{\rm nd}$ and $3^{\rm rd}$ order
  schemes while color (reported in the legend) keeps the same convention used
  in Fig. \ref{fig:fl2D_decay} (arithmetic averaging has been omitted due to
  large errors).
  \emph{Bottom panels}: normalized evolution of the axial component of magnetic
  field ($B'_z$) for selected schemes.  
   \label{fig:fl3D_decay}}
\end{figure*}

Figures \ref{fig:fl3D_mapsRK2} and \ref{fig:fl3D_mapsRK3} show volume renderings of  magnetic pressure (for different emf) at $t=1$ obtained, respectively, with the $2^{\rm nd}$- and $3^{\rm rd}$-order base schemes and a resolution of $64\times64\times128$ zones.
Our results agree with the 2D expectations showing a very similar trend.
Arithmetic averaging still yields insufficient dissipation leading to severe distortions and oscillations which are amplified when switching form the $2^{\rm nd}$- to the $3^{\rm rd}$-order scheme.
This eventually leads to the formation of an unstable checkerboard pattern (top left panel of Fig. \ref{fig:fl3D_mapsRK3}) and the disruption of the loop.
Modest fluctuations are also visible with the CT-Flux, although integration  (with both the $2^{\rm nd}$- and $3^{\rm rd}$-order schemes) remain stable.
The decay of magnetic energy, shown in the top panels of Fig. \ref{fig:fl3D_decay}, confirms the same trend already established in the 2D case although discrepancies between schemes are less pronounced in the 3D case.
As pointed out by \cite{GS2008}, the component of magnetic field along the loop cylindrical axis should be zero analytically.
At the numerical level, however, this is verified only at the truncation level of the scheme as indicated by the bottom panels in Fig. \ref{fig:fl3D_decay}, where we plot the error $\av{|B'_z|}/B_0$ for the different schemes.
The CT-Contact and CT-Flux yield larger errors (apart from arithmetic averaging which yields the worse results) and discrepancies become more evident with the $3^{\rm rd}$-order scheme.
Overall, the three UCT averaging schemes (UCT-HLLD, UCT-GFORCE and UCT-HLL) perform best.
It is worth pointing out that the employment of RK time stepping adopted here avoids the complexities - generally inherent with corner-transport-upwind schemes \cite{GS2008,MigTze2010} - in the calculation of the interface states, since primitive variables need not be evolved through a separate normal predictor step and no balancing source term is required in a fully conservative evolution scheme.

%%%%%%%%%%%%%%%%%%%%%%%%%%%%%%%%%%%%%%%%%%%%%%%%%%%%%%%%%%%%%%%%%%%%%%%%
\subsection{Magnetized Current Sheet}
%
%%%%%%%%%%%%%%%%%%%%%%%%%%%%%%%%%%%%%%%%%%%%%%%%%%%%%%%%%%%%%%%%%%%%%%%%

We now consider a particularly interesting configuration where the amount of numerical dissipation introduced by either the base scheme or the emf-averaging procedure (or both) is crucial in determining the system evolution.
The computational domain is initially filled with plasma at rest ($\vec{v}=0$) having  uniform density $\rho = 1$ and a Harris current sheet is used for the magnetic field:
\begin{equation}
  \vec{B}(y) = B_0\tanh\left(\frac{y}{a}\right)\hvec{e}_x\,,\quad
\end{equation}
where $B_0=1$ and $a=0.04$ is the current sheet width.
An equilibrium configuration is constructed by counter-acting the Lorentz force with a thermal pressure gradient,
\begin{equation}
  p(y) = \frac{B_0^2}{2}\left(\beta + 1\right) - \frac{B_x(y)^2}{2} \,,
\end{equation}
%
%or by assuming a force-free configuration with vanishing Lorentz force,
%
%\begin{equation}
%  p      = \frac{B_0^2}{2}\beta \,,\quad
%  B_z(y) = B_0\sqrt{1 - \frac{B_x(y)^2}{2B_0^2}} \,,\quad
%\end{equation}
%
where $\beta = 2p_\infty/B_0^2 = 10$ is the initial plasma-beta parameter.
The equilibrium magnetic field is perturbed with
\begin{equation}
  \delta\vec{B} = \epsilon B_0\left[
   -\frac{1}{2}k_y\sin\left(\frac{1}{2}k_yy\right)\cos\left(k_xx\right)\hvec{e}_x
   + k_x\cos\left(\frac{1}{2}k_yy\right)\sin\left(k_xx\right)\hvec{e}_y
   \right]\,,
\end{equation}
where $k_x = 2\pi/L_x$, $k_y = 2\pi/L_y$ while $\epsilon = 10^{-3}$ is the initial amplitude.
In order to fulfill the divergence-free condition to machine accuracy, we differentiate the vector potential $\delta A_z = \epsilon B_0\cos(k_yy/2)\cos(k_xx)$ in order to produce the desired perturbation.
We employ a rectangular box defined by $x\in[-1,1]$ and $y\in[-\HALF,\HALF]$ with periodic boundary conditions in the $x$-direction and reflective conditions at the top and bottom boundaries.
We carry out two sets of computations at the resolution of $128\times64$ zones using the $2^{\rm nd}$-order base scheme with the Riemann solver of Roe (first set) and the HLLD solver (second set) at cell interfaces.

We begin our discussion by pointing out that, in absence of a physical resistivity, the previous (unperturbed) equilibrium is a stationary solution of the ideal MHD equations and any dissipative process should be absent.
In practice, however, the discretization process introduces a numerical viscosity/resistivity which allows the current sheet to reconnect to some extent.

\begin{figure*}
  \centering
  \includegraphics[width=0.99\textwidth]{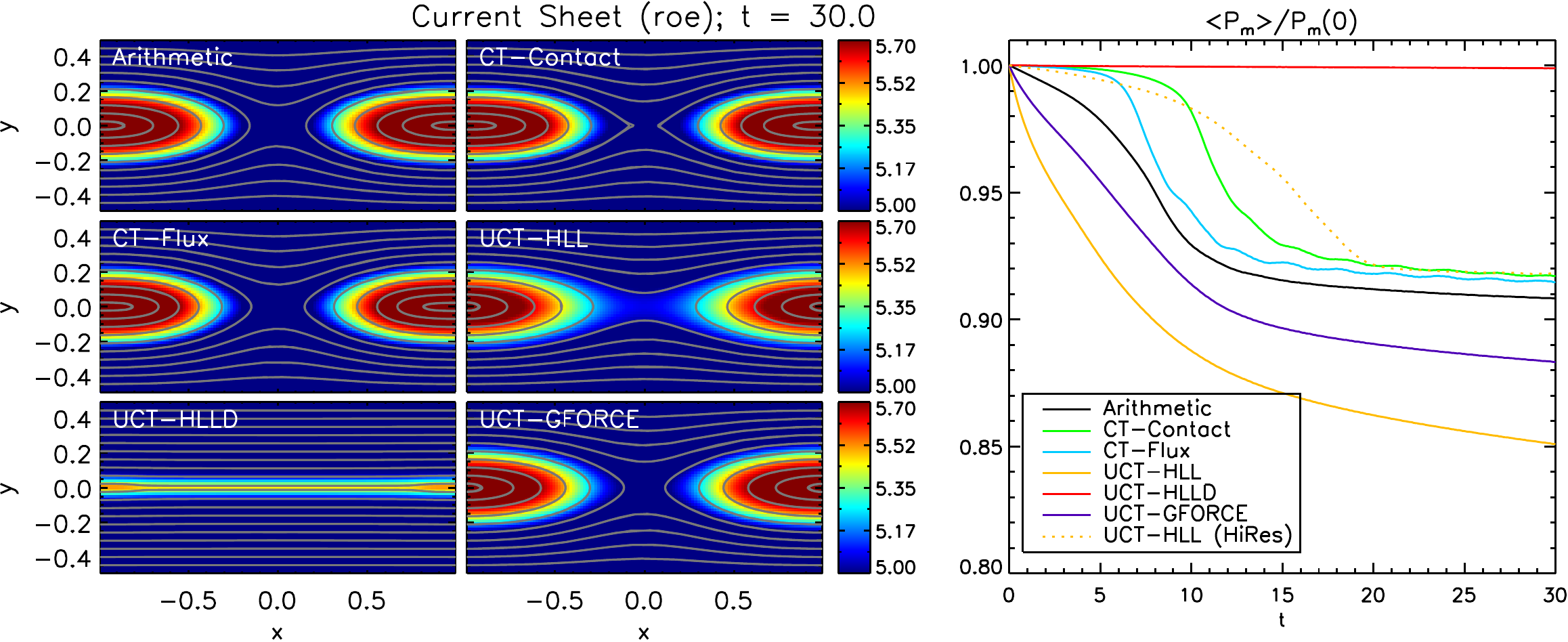}
  \caption{\footnotesize \emph{Left}: thermal pressure maps for the current sheet
  problem at $t=30$ using the different emf-averaging schemes reported
  in each panel.
  Magnetic field lines are over-plotted.
  Results have been produced using the $2^{\rm nd}$-order base scheme
  with the Roe Riemann solver and a resolution of $128\times64$ zones.
  \emph{Right}: magnetic energy (normalized to its initial value)
   as a function of time.
   The dotted orange line corresponds to the UCT-HLL scheme with resolution
   four times larger.
   The color convention (reported in the legend) is the same one adopted in
   previous plots.   
  \label{fig:cs_maps_roe}}
\end{figure*}
\begin{figure*}
  \centering
  \includegraphics[width=0.99\textwidth]{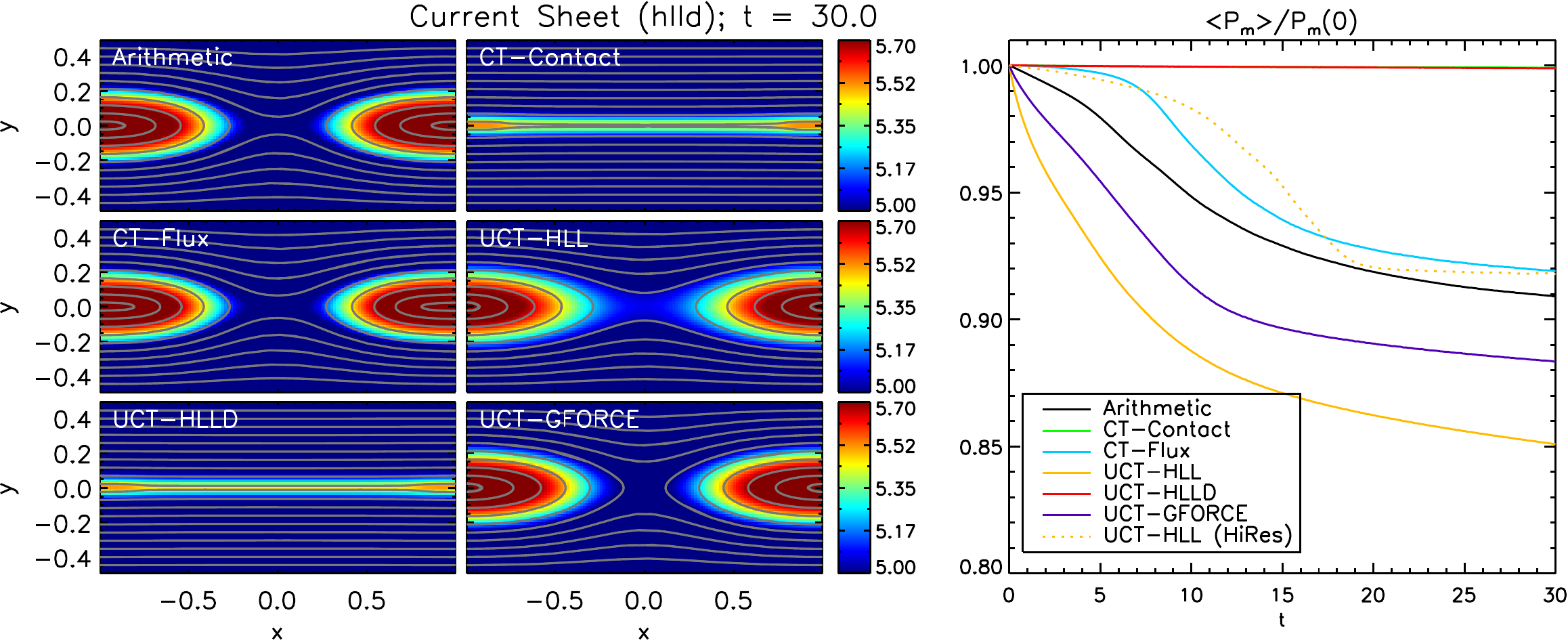}
  \caption{\footnotesize Same as Fig. \ref{fig:cs_maps_roe} but this
  time using the $2^{\rm nd}$-order base scheme with the HLLD Riemann solver.
           \label{fig:cs_maps_hlld}}
\end{figure*}
Thermal pressure maps are shown for different emf methods on the left side of Fig. \ref{fig:cs_maps_roe} at $t=30$ using the base scheme with the Roe solver.
The plot on the right side shows the corresponding volume-integrated magnetic energy as a function of time.
Magnetic reconnection takes place more rapidly for the UCT-HLL emf followed by UCT-GFORCE, Arithmetic, CT-Flux, CT-Contact eventually leading to the formation of a large magnetic island located across the vertical boundaries.
The rate at which field dissipation occurs depends on the amount of numerical viscosity diffusion: more dissipative schemes will trigger reconnection events earlier.
Results obtained with the UCT-HLLD scheme, in fact, show that the amount of dissipation is considerably reduced and the layer remains more stable, as one would expect for an ideal system.
This conclusion is also supported by a high-resolution run ($512\times256$ zones) with the UCT-HLL method (dotted orange line on the right side) indicating that magnetic field dissipation takes place at later times.

When the base Riemann solver is switched to HLLD (Fig. \ref{fig:cs_maps_hlld}), no significant change is found for the UCT schemes.
However, the solution obtained with the CT-Contact (and, to a lesser extent, with the CT-Flux) emf-averaging is now considerably different, bearing closer resemblance with the UCT-HLLD scheme.
The magnetic energy now remains nearly constant not only for the UCT-HLLD scheme (red curve) but also with the CT-Contact emf (green curve).
This apparently odd behavior may be understood by inspecting the amount of numerical dissipation inherited by the CT-Contact (or CT-Flux) scheme from the base 1D solver.
When the Roe Riemann solver is employed, contributions to the diffusion term are given by jumps in magnetic field \emph{and} thermal pressure when sweeping along the $y$-direction.
These contributions enter in the momentum flux \emph{and} the induction system as well.
Conversely, with the 1D HLLD Riemann solver, dissipation terms are proportional to the jump in magnetic field only and this contribution is confined to the electric field alone.
The CT-Contact scheme will therefore carry different amount of numerical viscosity depending on which 1D Riemann solver is selected.
Conversely, the UCT-HLLD introduces the same amount of numerical viscosity regardless of the 1D base Riemann solver.
From the discussion after Eq. (\ref{eq:UCT_HLLD_nu}), the order of magnitude of the dissipation term is $\tilde{\chi} \approx (v_x - \lambda^*) \approx O(\epsilon)$ and thus smaller when compared to the Roe dissipation matrix. 

This test clearly substantiates that the choice of the emf averaging scheme is as crucial as the choice of the interface Riemann solver in the evolution of magnetized systems.

%%%%%%%%%%%%%%%%%%%%%%%%%%%%%%%%%%%%%%%%%%%%%%%%%%%%%%%%%%%%%%%%%%%%%%%%
\subsection{Orszag Tang Vortex}
%
%%%%%%%%%%%%%%%%%%%%%%%%%%%%%%%%%%%%%%%%%%%%%%%%%%%%%%%%%%%%%%%%%%%%%%%%

The Orszag-Tang is a standard numerical benchmark in the context of the ideal MHD equations and although an exact solution does not exist, its straightforward implementation has made it an attractive numerical benchmark for inter-scheme comparison.
The problem consists of a doubly-periodic square domain, $x,y \in[0,2\pi]$ with uniform density and pressure, $\rho = 25/9$ and $p=5/3$ with velocity and magnetic field vectors given by
\begin{equation}
  \vec{v} = -\sin(y)\hvec{e}_x + \sin(x)\hvec{e}_y \,,\qquad
  \vec{B} = -\sin(y)\hvec{e}_x + \sin(2x)\hvec{e}_y \,.\qquad
\end{equation}
Albeit most numerical schemes gives comparable results at time $t = \pi$, the subsequent evolution has been discussed by few authors (see, e.g., \cite{Balsara1998, Lee_Deane2009, Mignone_etal2010_ho, Waagan_etal2011}).
Here we carry out such investigation by considering a later evolutionary time, $t=2\pi$ and by comparing different sets of emf-averaging schemes with grid resolution of $128^2$ or $256^2$ grid zones.
We employ both the $2^{\rm nd}$- and $3^{\rm rd}$-order base schemes using the HLLD Riemann solver.
As we shall see, the choice of the numerical method can appreciably impact the evolution of the system at this later time.

% 
% References to papers with late evolution:
% - Balsara 1998a
% - Lee & Deane (2009)
% - Mignone et al (2010)
% - Mignone et al (2007) Isothermal
% - Waagan et al (2011)

\begin{figure*}[!ht]
  \centering
  \includegraphics[width=0.92\textwidth]{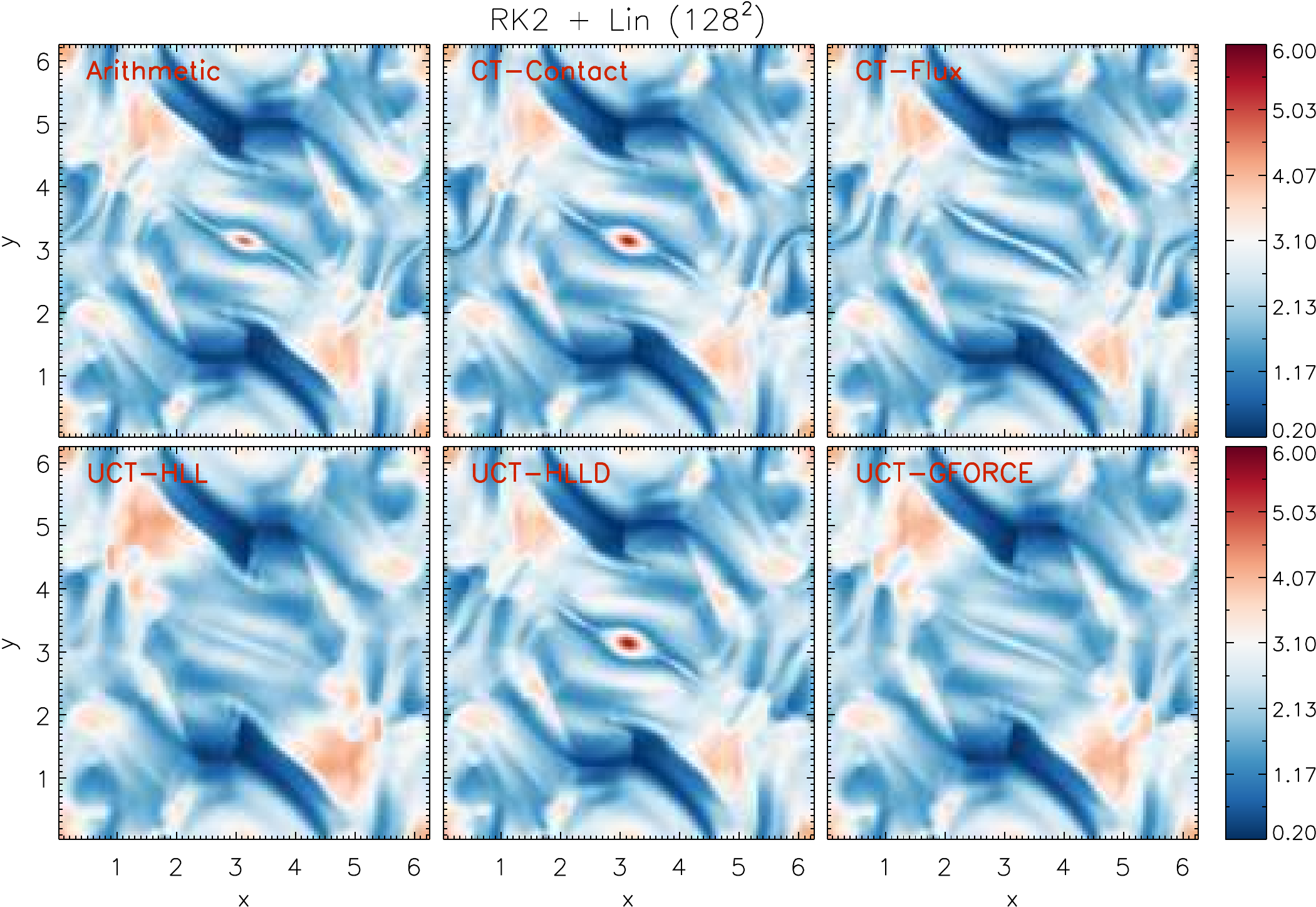}\\
  \caption{\footnotesize Pressure map distributions for the Orszag-Tang vortex
  at $t=2\pi$ using the $2^{\rm nd}$-order scheme with a grid resolution of
  $128\times 128$ zones.
  The top (bottom) row shows the results obtained with
  Arithmetic, CT-Contact and CT-Flux (UCT-HLL, UCT-HLLD and UCT-GFORCE) schemes.
   \label{fig:ot_maps_RK2_128}}
\end{figure*}
\begin{figure*}[!h]
  \centering
  \includegraphics[width=0.92\textwidth]{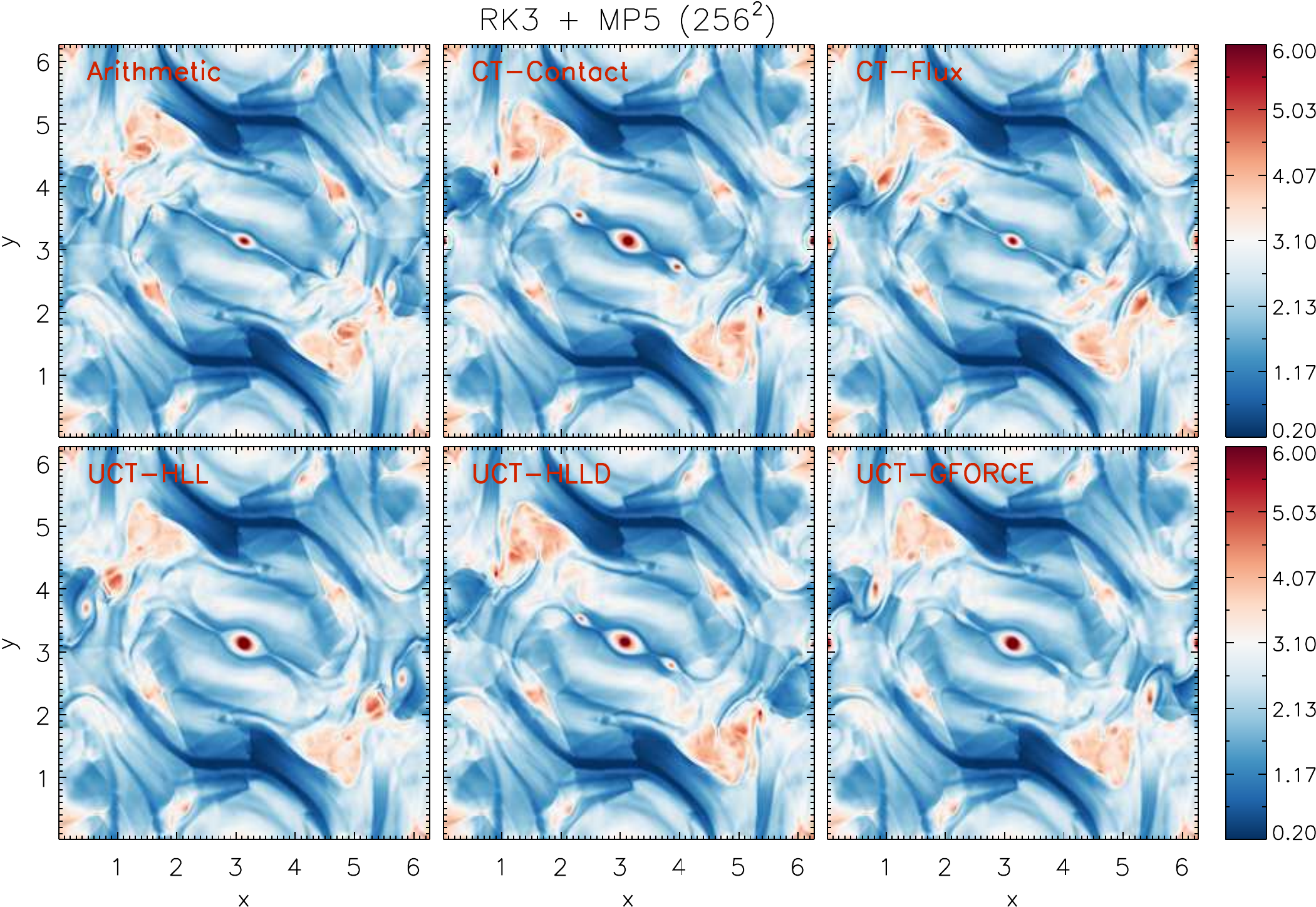}\\
  \caption{\footnotesize Same as Fig. \ref{fig:ot_maps_RK2_128} but for
  the $3^{\rm rd}$-order scheme with $256\times 256$ grid zones.
   \label{fig:ot_maps_RK3_256}}
\end{figure*}

The initial vorticity distribution spins the fluid clockwise leading to the steepening of density perturbations into shocks around $t \sim 0.8$.
The dynamics is then regulated by multiple shock-vortex interactions leading, by $t\sim 2.6$ to the formation of a horizontal current sheet at the center of the domain. 
Here magnetic energy is gradually dissipated and the current sheet twists leading, at $t=2\pi$ to the structures observed in Fig. \ref{fig:ot_maps_RK2_128} and \ref{fig:ot_maps_RK3_256} for the $2^{\rm nd}$-order and $3^{\rm rd}$-order base schemes with resolution of $128^2$ and $256^2$ zones, respectively.

The most noticeable difference lies at the center of the computational domain where the formation of a magnetic island (an O-point) can be discerned when using the Arithmetic, CT-Contact or UCT-HLLD averaging schemes while it is absent from the other solvers.
The presence of the central island may be attributed to the amount of numerical resistivity that can trigger tearing-mode reconnection episodes across the central current sheet, resulting in a final merging in this larger island.
For sufficiently low numerical dissipation, all schemes should eventually exhibits such a feature.
This may appear in contradiction with what required for the magnetized sheet test, where the initial equilibrium was expected to be stable in ideal MHD. 
Here, however, the situation is highly dynamic with the central current sheet undergoing a fast thinning process (induced by converging shock fronts), and it is known that only in the presence of a sufficiently high local Lundquist number (i.e. low numerical dissipation in the ideal MHD case of this test) the tearing instability is expected to develop on the \emph{ideal} (Alfv\'enic) timescales, see \cite{Landi_etal2015} and \cite{Papini_etal2019}.

%The most noticeable difference lies at the center of the computational domain where the formation of a magnetic island (an \quotes{O} point) can be discerned when using the Arithmetic, CT-Contact or UCT-HLLD averaging schemes while it is absent from the other solvers.
%The presence of the central island may be attributed to the amount of numerical resistivity that can trigger tearing-mode reconnection episodes across the central current sheet.
%For sufficiently low numerical dissipation, all schemes should eventually exhibits such a feature.

In our computations we found that Arithmetic averaging, CT-Contact and UCT-HLLD show the formation of the central islands for the two resolutions considered here with both the $2^{\rm nd}$- and $3^{\rm rd}$-order schemes.
A quantitative comparison is given Table \ref{tab:ot_compare} where we list the values of $\mu_p = p_{\max}/\av{p}$ for different computations.
Here $p_{\max}$ is the maximum pressure value in the region $4\pi/10 < x,\, y < 6\pi/10$ while $\av{p}$ is the average value in the entire computational domain.
For $\mu_p \lesssim 2$ no island is formed, while for $\mu_p\gtrsim2$ the island extent is roughly proportional to $\mu_p$.
CT-Contact and UCT-HLLD perform similarly yielding the smallest amount of dissipation while retaining numerical stability (no negative pressure has been encountered).
The UCT-HLL scheme is the most diffusive scheme showing the formation of the central O-point only with $256^2$ zones and MP5 reconstruction.
The UCT-GFORCE scheme performs similarly to the CT-Flux average and it is superior to the UCT-HLL scheme.
Note that the value of $\mu_p$ increases by more than $\sim 30\%$ when doubling the resolution for all methods except for arithmetic averaging which, on the other hand, yields insufficient dissipation as witnessed by several pressure fixes with the $3^{\rm rd}$-order scheme at the highest resolution employed.

\begin{table}
\centering
\begin{tabular}{ lccccccc } 
 \hline
 Scheme    & Res. &  Arithmetic  & CT-Contact & CT-Flux & UCT-HLL & UCT-HLLD & UCT-GFORCE \\
 \hline
RK2+Lin  & $128^2$ & 2.16 & 2.49 & 1.40 & 1.21 & 2.54 & 1.25 \\ \noalign{\smallskip}
RK2+Lin  & $256^2$ & 2.78 & 2.78 & 2.27 & 1.44 & 2.90 & 2.13 \\ \noalign{\smallskip}
RK3+MP5  & $128^2$ & 2.52 & 3.45 & 2.09 & 1.30 & 3.33 & 2.03 \\ \noalign{\smallskip}
RK3+MP5  & $256^2$ & 2.89 & 4.29 & 3.29 & 3.75 & 4.31 & 3.87 \\
\hline
 \end{tabular}
 \caption{\footnotesize Normalized pressure maximum $\mu_p=p_{\max}/\av{p}$ for
  the Orszag-Tang vortex at $t=2\pi$ using selected CT averaging techniques
  (columns) and different base-schemes and resolution (rows).
  An island forms only when $\mu_p \gtrsim 2$.}
 \label{tab:ot_compare}
\end{table}

Finally, it is worth to mention that the employment of the Roe Riemann solver (instead of HLLD) in the $3^{\rm rd}$-order base scheme lead to integration failures with the Arithmetic and CT-Flux averaging schemes.
No sign of numerical instability was discerned with the other emf-solvers.

%%%%%%%%%%%%%%%%%%%%%%%%%%%%%%%%%%%%%%%%%%%%%%%%%%%%%%%%%%%%%%%%%%%%%%%%
\subsection{Three-Dimensional Blast Wave}
%
%%%%%%%%%%%%%%%%%%%%%%%%%%%%%%%%%%%%%%%%%%%%%%%%%%%%%%%%%%%%%%%%%%%%%%%%

To assess the robustness of the proposed averaging schemes in a strongly magnetized plasma, we now analyze the blast wave problem in three dimensions.
Despite its simplicity, the blast wave problem is a particularly effective benchmark in testing the solver ability at handling MHD wave degeneracies parallel and perpendicularly to the field orientation.
Our configuration recalls the original paper of Balsara \& Spicer \cite{Balsara_Spicer1999} where it was first introduced and it consists of a unit cube filled with constant density $\rho_0=1$ and pressure $p_0=0.1$, threaded by a uniform magnetic field
\begin{equation}
  \vec{B} = B_0\left( \sin\theta\cos\phi,\,
                      \sin\theta\sin\phi,\ ,
                      \cos\theta \right) \,,
\end{equation}
where $B_0 = 100/\sqrt{4\pi}$, $\theta = \pi/2$ and $\phi=\pi/4$.
The plasma $\beta$ in the ambient medium is therefore $\beta \approx 2.5\times 10^{-4}$ making this test particularly challenging.
A sphere with radius $r=0.1$ is filled with a much higher pressure, $p_1 = 10^3$, and the system is evolved until $t=0.01$.
We adopt an adiabatic equation of state with specific heat ratio $1.4$ and set outflow boundary conditions everywhere.
The MHD equations are solved on the unit cube $[-1/2,1/2]^3$ using the $2^{\rm nd}$-order base scheme with the Roe Riemann solver and different emf-averaging schemes with a resolution of $192^3$ zones.
As in the original paper by \cite{Balsara_Spicer1999}, in order to avoid the occurrences of negative pressures\footnote{No scheme preserves energy positivity without energy correction for this test, not even with a minmod limiter.}, the total energy density after each step has to be locally redefined by replacing, in the magnetic term, the zone-centered magnetic field (updated using a standard Godunov-step) with the arithmetic average of the staggered fields, $E \leftarrow E - (\vec{B}_{\cc}^2 - \overline{\vec{B}}_f^2)/2$ where, e.g., $\overline{B}_x = (B_{\xf} + B_{\xf-\hvec{e}_x})/2$.

\begin{figure*}
  \centering
  \includegraphics[width=0.95\textwidth]{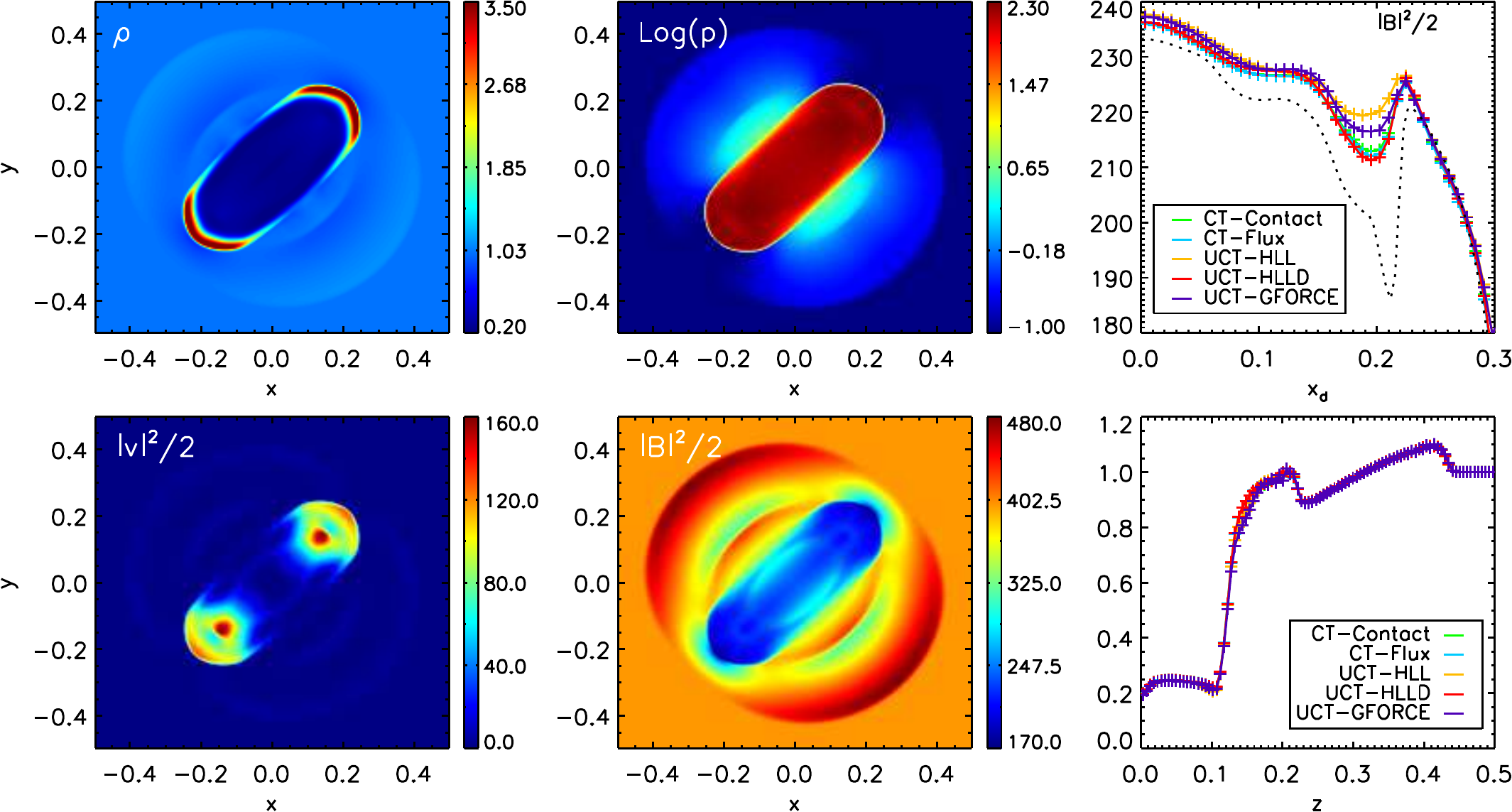}
  \caption{\footnotesize Results for the 3D blast wave problem at $t=0.01$
   using $192^3$ grid zones.
   In the four leftmost panels we show 2D colored maps at $z=0$ of density and
   log pressure (top), specific kinetic energy and magnetic energy (bottom)
   obtained with the UCT-HLLD scheme.
   In the top right panel we plot the magnetic energy densities along the
   main diagonal ($x_d=x\sqrt{2}$) at $z=0$ obtained with different emf solvers
   (solid colored lines reported in the legend).
   The dotted black line indicate the solution obtained with the
   UCT-HLLD solver at twice the resolution.
   In the bottom right panel we plot 1D slices of density along the vertical
   axis.
   The color pattern is the same already used in other figures.
  \label{fig:blast_compare}}
\end{figure*}
The four leftmost panels of Fig. \ref{fig:blast_compare} show 2D maps, taken as $xy$ slices at $z=0$, of various gas-dynamical quantities obtained with the UCT-HLLD solver.
The explosion is delimited by an outer fast forward shock and the presence of a magnetic field makes the propagation highly anisotropic by compressing the gas in the direction parallel to the field.
In the perpendicular direction the outer fast shock becomes magnetically dominated with very weak compression.
Results reproduced with the other emf schemes are also very similar.

A more careful comparison, shown in the top rightmost panel, reveals some differences in the magnetic energy plot along the main diagonal in the $z=0$ plane.
Here a dip is formed around $x_d \approx 0.2$, where $x_d$ is the distance from a point on the diagonal to the coordinate origin.
The dip becomes more pronounced as the numerical dissipation of scheme is reduced.
Indeed, UCT-HLL yields the highest minimum followed by UCT-GFORCE, CT-Contact, CT-Flux and UCT-HLLD. 
This trend is confirmed at twice the grid resolution ($384^2$ zones, dotted black line), leading to the formation of an even deeper sag.
Finally, in the bottom rightmost panel we compare the density profiles along the $z$-axis for different schemes, showing only minor differences. 

%%%%%%%%%%%%%%%%%%%%%%%%%%%%%%%%%%%%%%%%%%%%%%%%%%%%%%%%%%%%%%%%%%%%%%%%
\subsection{Kelvin-Helmholtz Instability}
%
%%%%%%%%%%%%%%%%%%%%%%%%%%%%%%%%%%%%%%%%%%%%%%%%%%%%%%%%%%%%%%%%%%%%%%%%

\begin{figure*}
  \centering
  \includegraphics[width=0.8\textwidth]{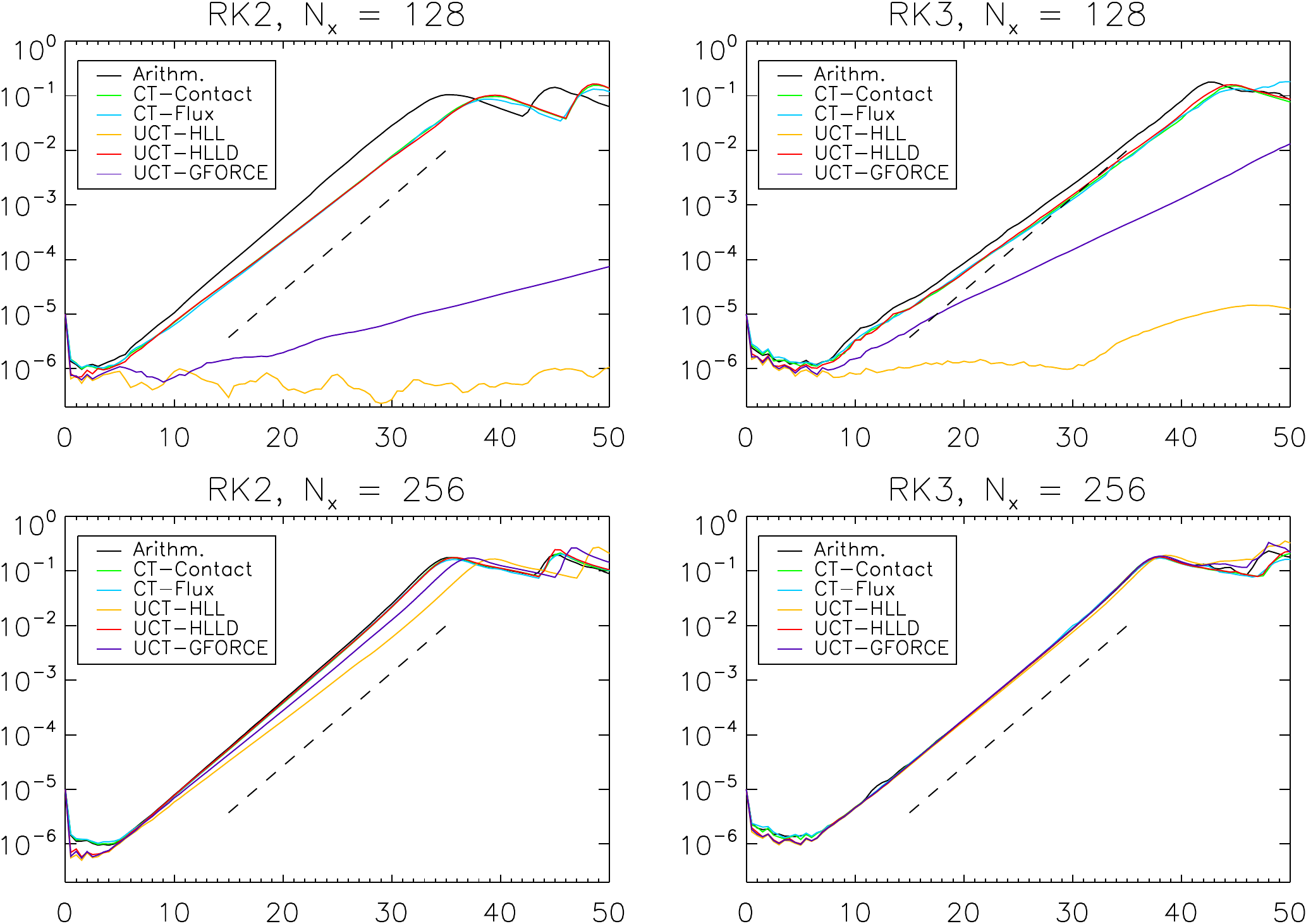}
  \caption{\footnotesize
  Growth rates, measured as $\delta v_y = (\max(v_y) - \min(v_y))/2$, for the
  2D KH instability.
  \emph{Top panels:} $\delta v_y$ for the $2^{\rm nd}$- and $3^{\rm rd}$-order
  schemes using the different emf solvers at the resolution of $128\times 256$ zones.
  \emph{Bottom panels:} $\delta v_y$ for the $2^{\rm nd}$- and $3^{\rm rd}$-order
  schemes using $256\times 512$ zones.  
  \label{fig:kh_growth}}
\end{figure*}

The Kelvin-Helmholtz instability (KH) is driven by the relative motion between two fluids.
In the presence of a magnetic field aligned with the flow direction, the instability is typically reduced by the stabilizing action of magnetic tension.
In the following test, we consider a 2D Cartesian domain $x\in[0,1]$ , $y\in[-1,1]$ initially setup to contain a velocity shear layer,
\begin{equation}
  v_x = \frac{M}{2}\tanh(y/a) \,,
\end{equation}
where $M$ is the sonic Mach number while $a=0.01$ is the shear width.
We normalize velocities to the speed of sound ($\rho=1$, $p=1/\Gamma$) while the magnetic field is aligned with the flow direction, $\vec{B} = B_0\hvec{e}_x$ while the field strength is parametrized using the Alfv\'en velocity, $B_0 = v_A\sqrt{\rho}$.
A random perturbation is applied to seed the instability,
\begin{equation}
  v_y = \epsilon M\exp\left[-(y/20a)^2\right] \,,
\end{equation}
where $\epsilon\in [-10^{-5},10^{-5}]$ is a random number.
The boundary conditions are periodic in the $x$-direction while a reflective boundary is applied at $y=\pm1$.
We tune up the parameters by choosing $M=1$, $v_A = 1/2$ and by integrating the MHD equations until $t=50$ using the HLLD Riemann solver in the base scheme.
This choice of parameters makes the test particularly severe since the configuration is only weakly unstable.
The maximum growth rate that can fit in the computational box, indeed, has been found by repeating the linear analysis of Miura \& Pritchett (1982) \cite{Miura_Pritchett1982} yielding ${\rm Im}(\omega a/c_s) \approx 0.395\times 10^{-2}$, where $\omega$ is the growth rate.
Computations are repeated using two different grid resolutions, $N_x = 128$ and $N_x = 256$ ($N_y=2N_x$) so that the shear width is resolved on $\sim 4$ and $\sim 8$ zones, respectively.

Fig. \ref{fig:kh_growth} shows the perturbation growth, measured as $\delta v_y = (\max(v_y) - \min(v_y) )/2$, as a function of time for the $2^{\rm nd}$ and $3^{\rm rd}$-order base schemes and the selected emf averaging solvers.
Perturbations grow linearly until the system turns into a nonlinear phase around $t\sim 37$.
At low resolution, the UCT-HLL scheme fails to evolve into a fully developed unstable state, the UCT-GFORCE yields a reduced instability growth while only with the CT-Contact, CT-Flux and UCT-HLLD perturbations follow a linear amplification phase in closer agreement with the analytical prediction.
Here we find ${\rm Im}(\omega) \approx 0.35$ and ${\rm Im}(\omega) \approx 0.31 - 0.33$ for the $2^{\rm nd}$- and $3^{\rm rd}$-order schemes, respectively.
Growth rates slightly lower when switching to higher-order reconstruction owing, at this poor resolution, to the excitation of short-wavelength perturbation modes with smaller growth rates.
Better convergence is achieved at twice the resolution where, for the $2^{\rm nd}$-order scheme, we find $0.38\lesssim {\rm Im}(\omega) \lesssim 0.40$ whereas, for the $3^{\rm rd}$-order method, values move closer yielding a reduced inter-scheme dispersion with $0.38\lesssim {\rm Im}(\omega) \lesssim 0.39$.
The converged growth rate, at the resolution of $512\times1024$, is ${\rm Im}(\omega) \approx 0.393$.

\begin{figure*}[!ht]
  \centering
  \includegraphics[width=0.9\textwidth]{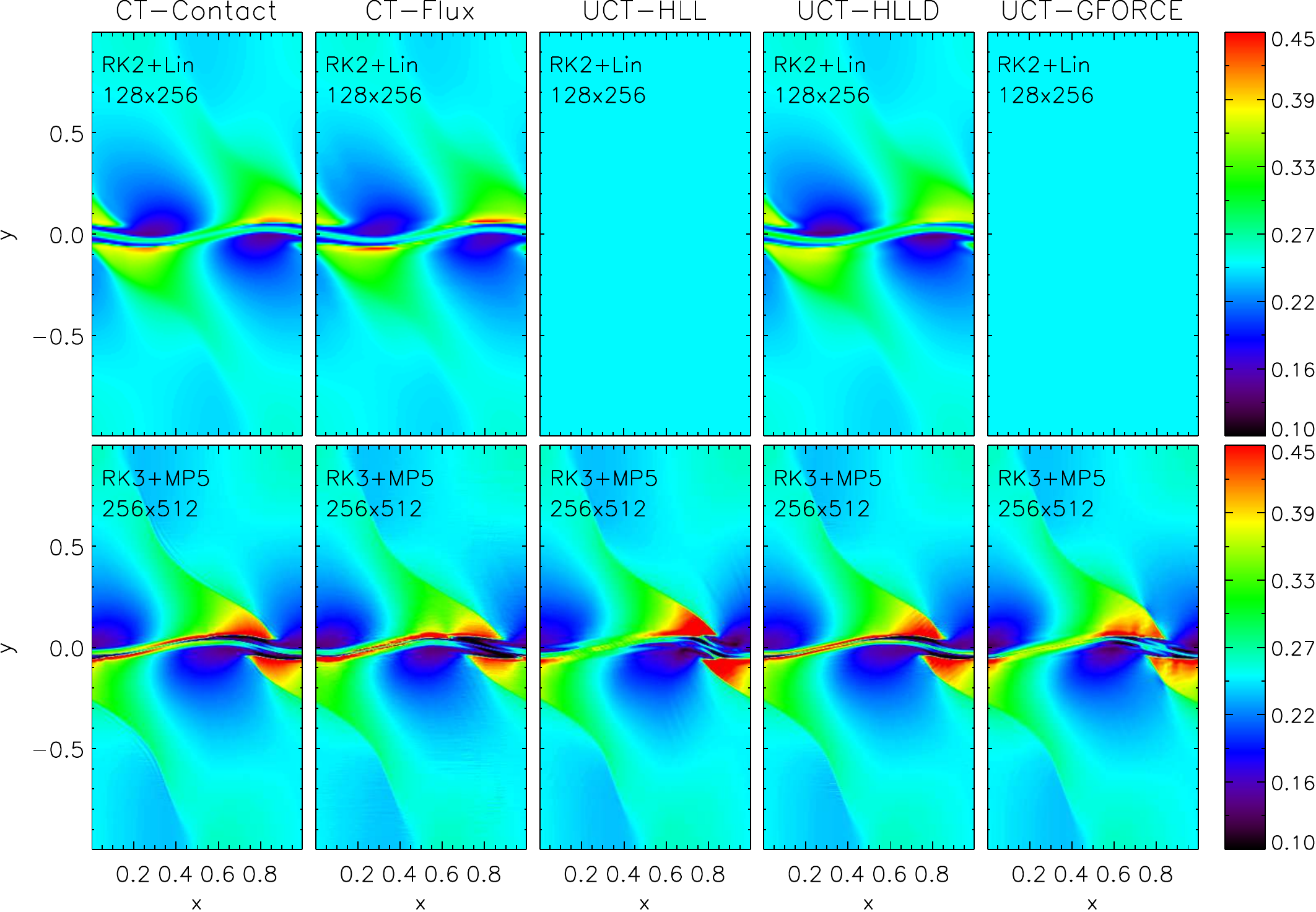}
  \caption{\footnotesize
   Final evolutionary stage ($t=50$) for the 2D Kelvin-Helmholtz test.
   Colored maps show $B^2/\rho$ for selected emf with the $2^{\rm nd}$-order
   scheme using $128\times 256$ zones (top) and the $3^{\rm rd}$-order
   scheme with $256\times 512$ grid points (bottom).
   \label{fig:kh_maps}}
\end{figure*}

Finally, in Fig. \ref{fig:kh_maps}, we show colored maps of $B^2/\rho$ for the $2^{\rm nd}$-order (low resolution, top panels) and the $3^{\rm rd}$-order (high resolution, bottom panels) computations at $t=50$.
Note that no sign of instability is visible at low resolution for the UCT-HLL and UCT- GFORCE schemes.
On the other hand, the instability is fully developed and comparable results are obtained at twice the resolution with the $3^{\rm rd}$-order base scheme for all emf methods.

%%%%%%%%%%%%%%%%%%%%%%%%%%%%%%%%%%%%%%%%%%%%%%%%%%%%%%%%%%%%%%%%%%%%%%%%
\subsection{Magnetorotational Instability in the ShearingBox model}
%
%%%%%%%%%%%%%%%%%%%%%%%%%%%%%%%%%%%%%%%%%%%%%%%%%%%%%%%%%%%%%%%%%%%%%%%%

As a final application we compare the different emf solvers by investigating the nonlinear evolution of the magneto-rotational instability (MRI) in the shearing-box approximation model.
The implementation of the shearing-box equations (which provides a local Cartesian description of a differentially rotating disk) for the PLUTO code may be found in \cite{Mignone_etal2012}.
The initial background state consists of a uniform density distribution, $\rho_0=1$, and a linear velocity shear $\vec{v} = -q\Omega x\hvec{e}_y$.
Here $\Omega=1$ is the orbital frequency while $q=3/2$ - typical of a Keplerian profile - gives a local measure of the differential rotation.
A net magnetic flux threads the computational domain and it is initially aligned with the vertical direction, $\vec{B}=B_0\hvec{e}_z$ with strength
\begin{equation}
  B_0 = c_s\sqrt{\frac{2\rho_0}{\beta}} \,.
\end{equation}
We choose $c_s=4.88$ (isothermal sound speed) and $\beta=8\times 10^3$ so that we fit approximately one most unstable MRI wavelength in the vertical direction, \cite{Pessah_etal2007, Bodo_etal2011}.
The shearing-box equations are solved on a 3D Cartesian box with $x,y\in[-2L,2L]$ and $z\in[-L/2,L/2]$ with periodic boundary conditions in the vertical and azimuthal ($y$) directions while shearing-sheet conditions are imposed at the $x$-boundaries.
Integration are carried at the resolution of $144\times144\times 36$ zones for $\approx 100$ rotations ($t_{\rm stop} = 628$) using the Roe Riemann solver with an isothermal equation of state.
We employ the orbital advection scheme to subtract the linear shear contribution from the total velocity so that the MHD equations are evolved only in the residual velocity, thus giving a substantial speedup of the algorithm, see \cite{Mignone_etal2012}.

\begin{figure*}[!ht]
  \centering
  \includegraphics[width=0.33\textwidth]{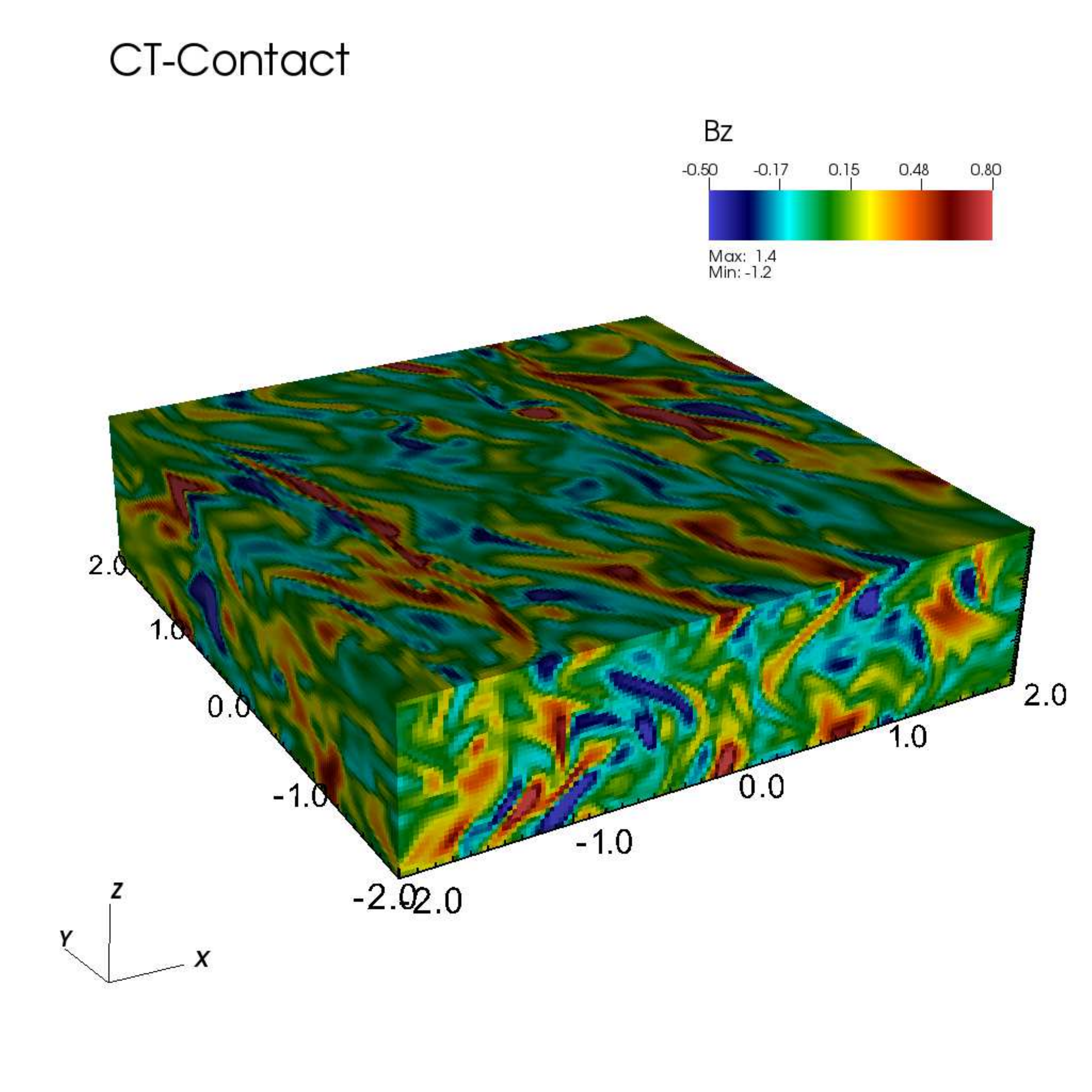}
  \includegraphics[width=0.33\textwidth]{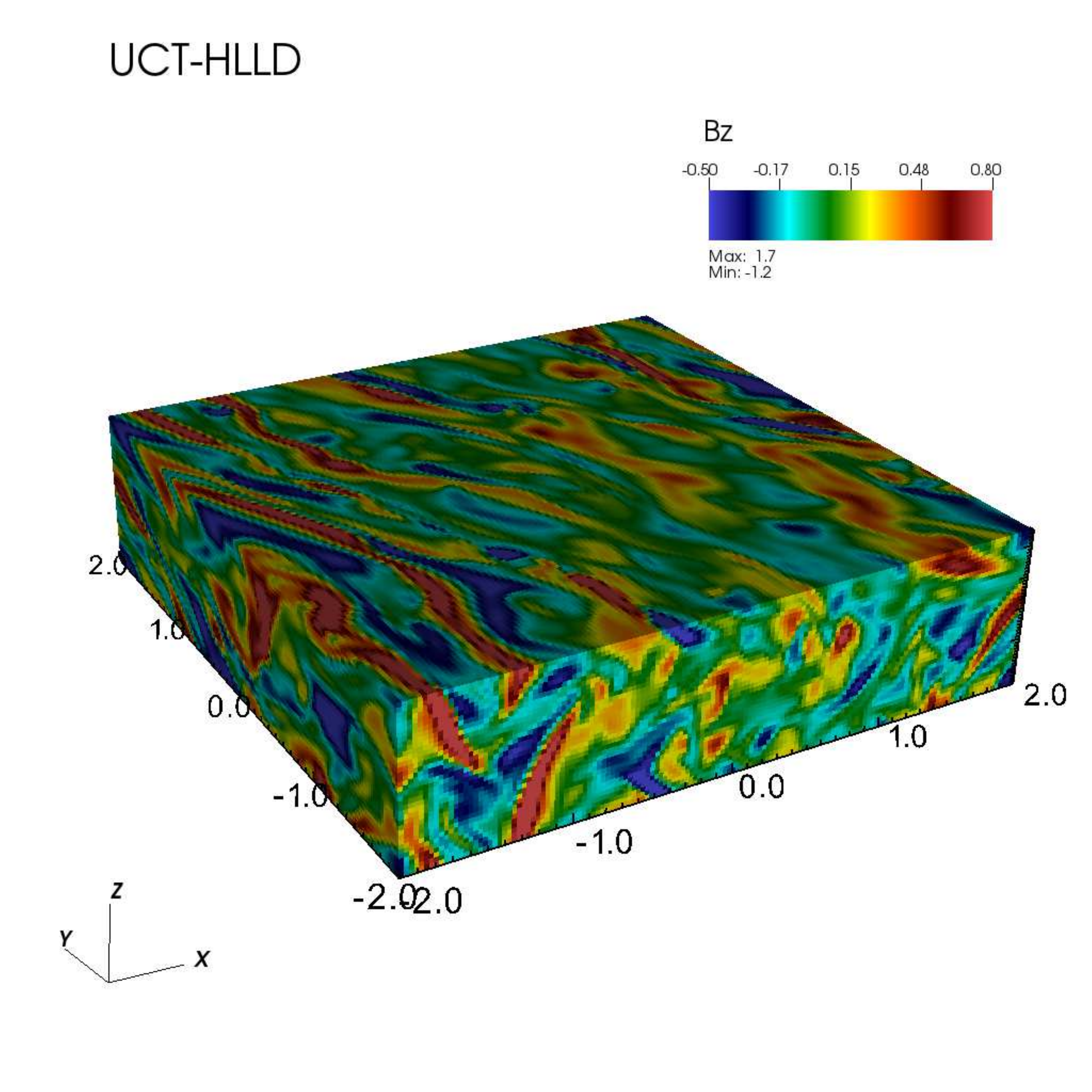}
  \includegraphics[width=0.33\textwidth]{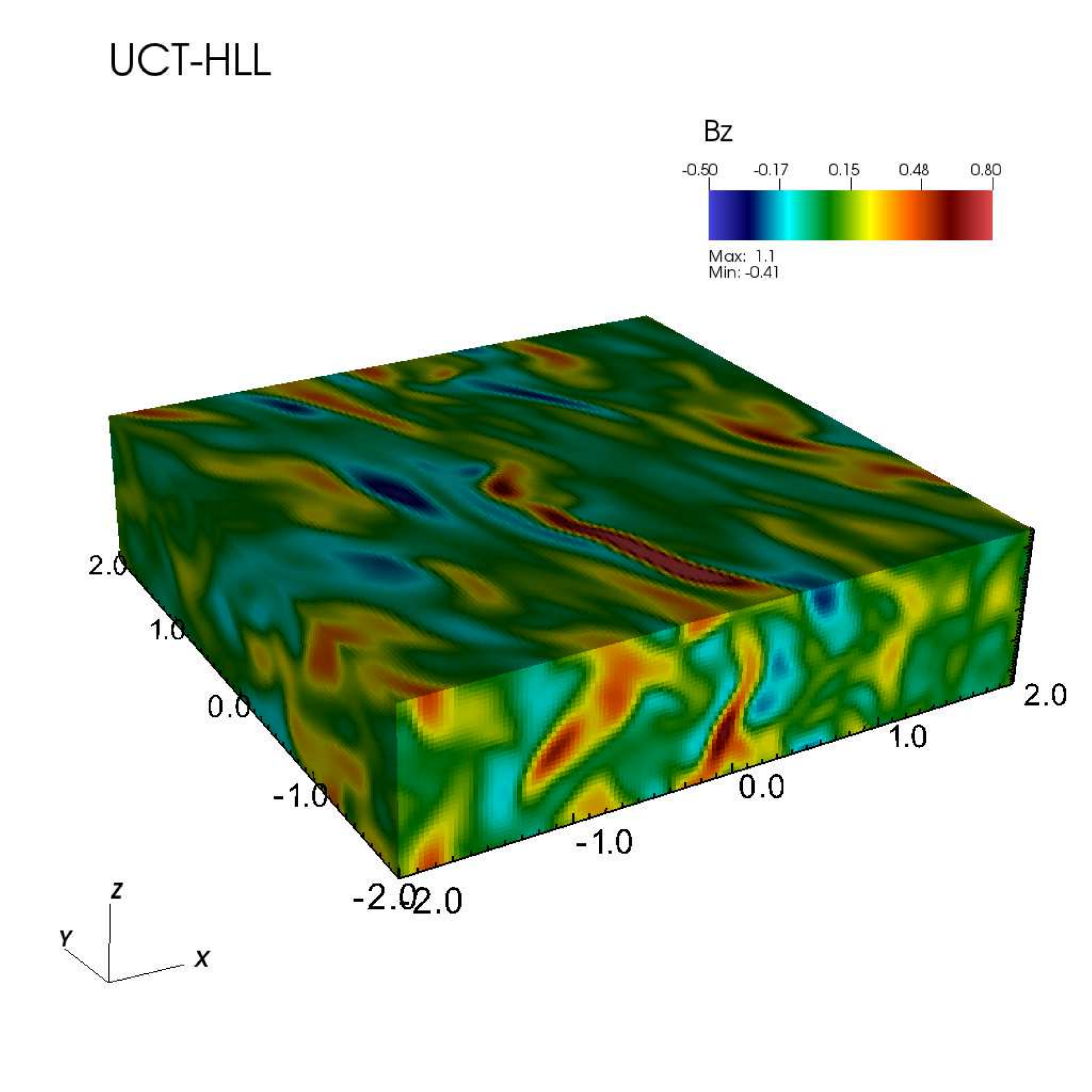}
  \caption{\footnotesize Three-slice cut of $B_z$ at $t = 628$ for the shearingbox
           test problem, using the CT-Contact (left), UCT-HLLD (middle) and
           UCT-HLL (right) schemes.
           \label{fig:sb_maps}}
\end{figure*}
\begin{figure*}[!h]
  \centering
  \includegraphics[width=0.9\textwidth]{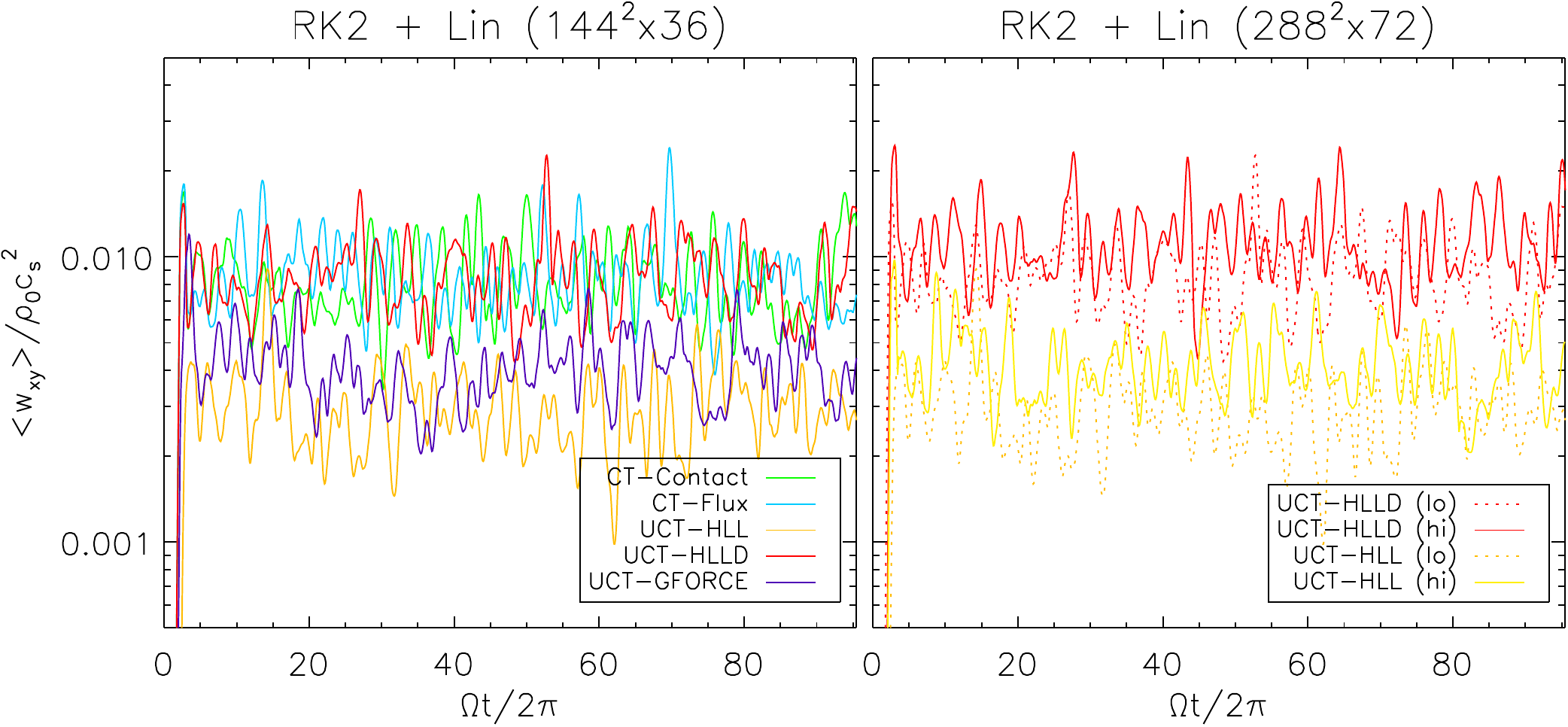}
  \caption{\footnotesize Maxwell stresses (normalized to $\rho_0c_s^2$) as
   a function of time for the the 3D shearingbox model.
   In the left panel we compare five different averaging schemes at the nominal
   resolution of $144^2\times36$ zones while in the right panel the comparison
   is repeated with results obtained at twice the resolution with UCT-HLL and
   UCT-HLLD.
   Different colors correspond to different emf-averaging schemes
   (as reported in the legend).
   \label{fig:sb_growth}}
\end{figure*}

The initial transient phase is accompanied by an exponential growth of the magnetic field followed by a transition to a nonlinear turbulent state.
The vertical component of magnetic field at $t=628$ is shown in Fig. \ref{fig:sb_maps}, comparing the results obtained with the CT-Contact, UCT-HLLD and UCT-HLL schemes.
The maps indicate qualitative larger amount of fine-scale structure with the formers with respect to the latter.
A more quantitative measure is provided by the plots of the volume-integrated Maxwell stress $w_{xy} = -\av{B_xB_y}$ - normalized to $\rho_0 c_s^2$ - as a function of time shown in the left panel of Fig. \ref{fig:sb_growth}.
The time-averaged stress value, in the range $\Omega t/2\pi \in[10,100]$, is $\overline{w}_{xy}\sim 0.21$ for UCT-HLLD, $\sim 0.20$ for the CT-Contact and CT-Flux schemes.
Lower values, respectively equal to $\sim 0.07$ and $\sim 0.10$, are found when using the UCT-HLL or UCT-GFORCE methods.

Previous studies (see, e.g., \cite{Bodo_etal2011}) indicate that stresses increase with resolution and should eventually converge as the mesh spacing becomes sufficiently fine.
In this sense, lower values of $w_{xy}$ imply larger numerical diffusion.
This conclusion is supported by computations carried out at twice the resolution ($288\times288\times 72$) using the UCT-HLL and UCT-HLLD emf-averaging schemes, for which the stresses are plotted in the right panel of Fig. \ref{fig:sb_growth}.
In this case the time-averaged value of $w_{xy}$ increases to $\sim 0.10$ (for UCT-HLL) and to $\sim 0.27$ (for UCT-HLLD).

%%%%%%%%%%%%%%%%%%%%%%%%%%%%%%%%%%%%%%%%%%%%%%%%%%%%%%%%%%%%%%%%%%%%%%%%
\section{Summary}
\label{sec:summary}
%
%
%
%
%%%%%%%%%%%%%%%%%%%%%%%%%%%%%%%%%%%%%%%%%%%%%%%%%%%%%%%%%%%%%%%%%%%%%%%%

The systematic construction and comparison of averaging schemes to evaluate the electromotive force (emf) at zone edges in constrained transport MHD has been the subject of this work.
The upwind constrained transport (UCT) formalism, originally developed by \cite{Londrillo_DelZanna2004}, has been reconsidered under a more general perspective where the edge-averaged electric field can be constructed using the information available from 1D face-centered, component-wise Riemann solvers.
This approach offers enhanced flexibility allowing new upwind techniques to be incorporated in CT-MHD schemes at the modest cost of storing transverse velocity, weight coefficients for the fluxes and diffusion terms for the magnetic field.

Four popular schemes, namely Arithmetic, CT-Contact, CT-Flux and UCT-HLL, together with two novel algorithms have been presented and compared in terms of accuracy, robustness and dissipation properties.
Among the newly introduced schemes, the UCT-HLLD and UCT-GFORCE schemes build into the UCT framework the proper combination of upwind fluxes derived, respectively, from the HLLD Riemann solver of \cite{Miyoshi_Kusano2005} and the GFORCE scheme of \cite{Toro_Titarev2006}.
Through a series of 2D and 3D numerical tests and benchmark applications, our conclusions can be summarized as follows:
\begin{itemize}

\item
%%%%%%
The choice of emf averaging procedure at zone edges can be as crucial as the choice of the Riemann solver at zone interfaces of the underlying base scheme.
This becomes particularly true in problems where the magnetic field has a dominant role on the system dynamics.

\item
%%%%%%
Averaging schemes with insufficient dissipation (arithmetic averaging) may easily corrupt the solution leading to spurious numerical artifacts or to the occurrences of nonphysical values such as negative pressures.

\item
%%%%%%
The simple recipe of doubling the dissipation term (e.g. CT-Flux and CT-Contact), suggested by the need recovering the proper directional biasing for grid-aligned configurations, still represents an efficient and valid option, although not strictly compliant with the UCT formalism.

\item
%%%%%%
The newly proposed UCT-HLLD scheme presents low-diffusion and excellent stability properties when used in conjunction with the HLLD Riemann solver as well as the Roe solver.
The amount of numerical dissipation is comparable to or even lower to the CT-Contact scheme with the advantage that UCT-HLLD can be extended to higher than $2^{\rm nd}$-order schemes.

\item
%%%%%%
The UCT-GFORCE scheme, also introduced here for the first time, has dissipation properties intermediate between the UCT-HLL and UCT-HLLD (or CT-Contact) averaging schemes.

\end{itemize}

Contrary to non-UCT schemes, where the amount of numerical dissipation is inherited from the base Riemann solver applied at zone-interfaces, our formulation yields a one-valued independent and continuous numerical flux function with stable upwind properties along each direction.
Whether this is an advantage or not should be discerned for the particular application at hand.

Albeit our results have been presented in the context of $2^{\rm nd}$-order schemes in which the Riemann solver is applied at cell interfaces, our formulation can be naturally extended to higher than $2^{\rm nd}$-order finite-volume of finite-difference schemes.

Obviously, the component-wise Riemann solvers employed dimension-by-dimension as in the present work, in spite of their simplicity, may not be the optimal choice (as opposed to truly multidimensional solvers) in situations requiring unstructured triangular or geodesic meshes to treat geometrically complex MHD flows \cite[e.g.][]{Balsara_Dumbser2015,Balsara_etal2019}.

Forthcoming works will extend this formalism to the relativistic case as well, along the lines of \cite{DelZanna_etal2003,DelZanna_etal2007,Mignone_Bodo2006,Mignone_etal2009}, aiming at improving on the simple HLL choice which is nowadays the adopted standard \cite[e.g.][]{Porth_etal2019}.

\vspace*{2ex}\par\noindent
{\bf Acknowledgments.}
The authors wish to thank P. Londrillo and G. Bodo for useful discussions.
We also acknowledge the OCCAM supercomuting facility available at the Competence Centre for Scientific Computing at the University of Torino.

\appendix

%%%%%%%%%%%%%%%%%%%%%%%%%%%%%%%%%%%%%%%%%%%%%%%%%%%%%%%%%%%%%%%%%%%%%%%%
\section{Derivation of the UCT-HLLD scheme for Isothermal MHD}
\label{app:UCT_HLLD_isothermal}
%
%
%
%%%%%%%%%%%%%%%%%%%%%%%%%%%%%%%%%%%%%%%%%%%%%%%%%%%%%%%%%%%%%%%%%%%%%%%%

The wave pattern emerging from the solution of the Riemann problem in isothermal MHD differs from its adiabatic counterpart for the absence of the contact or entropy mode.
The HLLD flux can still be written in the Roe-like form (\ref{eq:hlld1}) but the jump conditions are different, as shown by Mignone (2007) \cite{Mignone2007}.
Eq. (\ref{eq:UCT_HLLD_ad}) and (\ref{eq:UCT_HLLD_nu}) still have the same form but the coefficient $\chi^s$ needed in our Eq. (\ref{eq:By_chi}) is different.
In particular, from Eq. 32-33 of \cite{Mignone2007}, we find the following expression:
\begin{equation}
  \chi^s = \frac{\rho^s(\lambda^s - v^s_x) - B_x^2}
                {\rho^*(\lambda^s - \lambda^{*L})(\lambda^s - \lambda^{*R})} - 1
  \,,
\end{equation}
where
\begin{equation}\label{eq:hlld_iso_lambda*}
  \lambda^{*L} = u^* - \frac{|B_x|}{\sqrt{\rho^*}}  \,,\qquad
  \lambda^{*R} = u^* + \frac{|B_x|}{\sqrt{\rho^*}}  
\end{equation}
are the Alfv\'en velocities, $u^* = F^{\rm hll}/\rho^{\rm hll}$ and $\rho^* = \rho^{\rm hll}$ is the density inside the Riemann fan.

From Eq. (\ref{eq:hlld_iso_lambda*}), one clearly has that $\lambda^{*R} + \lambda^{*L} = 2u^*$ and $B_x^2 = \rho^*(\lambda^{*s} - \lambda^*)^2$.
This allows us to write the $\tilde{\chi}^s$ coefficients needed in Eq. (\ref{eq:UCT_HLLD_ad}) as
\begin{equation}
  \tilde{\chi}^s = \frac{(v_x^s - u^*)(\lambda^s - u^*)}
                        {\lambda^{*s} + \lambda^s - 2u^*} \,.
\end{equation}

%  
%  
%
%\end{equation}
%
%/*
%if (fabs(chiL-chi1L) > 1.e-4 || fabs(chiR-chi1R) > 1.e-4){
%  print ("! CT_Flux: chi different\n");
%  print ("  chiL,  chiR  = %8.3e, %8.3e\n",chiL, chiR);
%  print ("  chi1L, chi1R = %8.3e, %8.3e\n",chi1L, chi1R);
%  print ("  waves = %12.6e  %12.6e  %12.6e  %12.6e  %12.6e\n",
%            SL[i], SaL[i], Sc[i], SaR[i], SR[i]);
%  QUIT_PLUTO(1);
%}
%*/
%chiL = chi1L;
%chiR = chi1R;
%

\clearpage

\bibliographystyle{elsarticle-num}
\bibliography{paper}
%%%%%%%%%%%%%%%%%%%%%%%%%%%%%%%%%%%%%%%%%%%%%%%%%%%%%%%%%%%%%%%%%%%%%%%%
%
%   Figures Here
%
%%%%%%%%%%%%%%%%%%%%%%%%%%%%%%%%%%%%%%%%%%%%%%%%%%%%%%%%%%%%%%%%%%%%%%%%
\end{document}